# Diamond Surface Functionalization via Visible Light-Driven C–H Activation for Nanoscale Quantum Sensing

Lila V. H. Rodgers[1,*], Suong T. Nguyen[2,*], James H. Cox[2], Kalliope Zervas[1], Zhiyang Yuan[1], Sorawis Sangtawesin[3], Alastair Stacey[4,5], Cherno Jaye[6], Conan Weiland[6], Anton Pershin[7,8], Adam Gali[7,8], Lars Thomsen[9], Simon A. Meynell[10], Lillian B. Hughes[11], Ania C. Bleszynski Jayich[10], Xin Gui[2], Robert J. Cava[2], Robert R. Knowles[2,†], Nathalie P. de Leon[1,†]

*[1]Princeton University, Department of Electrical and Computer Engineering, Princeton, New Jersey, 08540, USA*
*[2]Princeton University, Department of Chemistry, Princeton, New Jersey, 08540, USA*
*[3]School of Physics and Center of Excellence in Advanced Functional Materials, Suranaree University of Technology, Nakhon Ratchasima 30000, Thailand*
*[4]School of Physics, University of Melbourne, Parkville, VIC 3010, Australia*
*[5]School of Science, RMIT University, Melbourne, VIC 3000, Australia*
*[6]Material Measurement Laboratory, National Institute of Standards and Technology, Gaithersburg, Maryland 20899, USA*
*[7]Wigner Research Centre for Physics, POB 49, Budapest, H-1525, Hungary*
*[8]Department of Atomic Physics, Budapest University of Technology and Economics, Műegyetem rakpart 3., Budapest, H-1111, Hungary*
*[9]Australian Synchrotron, Australian Nuclear Science and Technology Organisation, 800 Blackburn Road, Clayton, VIC 3168, Australia*
*[10]Physics Department, University of California Santa Barbara, Santa Barbara, California 93106, USA*
*[11]Materials Department, University of California, Santa Barbara, Santa Barbara, California 93106, USA*

**these authors contributed equally to the work*
*† corresponding authors: rknowles@princeton.edu, npdeleon@princeton.edu*

**Abstract:**

Nitrogen-vacancy (NV) centers in diamond are a promising platform for nanoscale nuclear magnetic resonance (NMR) sensing[1]. Despite significant progress towards using NV centers to detect and localize nuclear spins down to the single spin level[2,3], NV-based spectroscopy of individual, intact, arbitrary target molecules remains elusive. NV molecular sensing requires that target molecules are immobilized within a few nanometers of NV centers with long spin coherence time. The inert nature of diamond[4] typically requires harsh functionalization techniques such as thermal annealing[2,5] or plasma processing[6–8], limiting the scope of functional groups that can be attached to the surface. Solution-phase chemical methods can be more readily generalized to install diverse functional groups, but they have not been widely explored for single-crystal diamond surfaces. Moreover, realizing shallow NV centers with long spin coherence times requires highly ordered single-crystal surfaces, and solution-phase functionalization has not yet been shown to be compatible with such demanding conditions. In this work, we report a versatile strategy to directly functionalize C–H bonds on single-crystal diamond surfaces under ambient conditions using visible light. Hydrogen atom abstraction from surface C–H bonds generates carbon-centered radicals that can either intercept various radical acceptors or be oxidized to carbocations to form

C–F, C–Cl, C–S, and C–N bonds at the surface. We verify covalent bond formation using lab and synchrotron-based surface analysis. This functionalization method is compatible with charge stable NV centers within 10 nm of the surface with spin coherence times comparable to the state of the art[5]. As a proof of principle, we use shallow ensembles of NV centers to detect nuclear spins from functional groups attached to the surface. Our approach to surface functionalization based on visible light-driven C–H bond activation opens the door to deploying NV centers as a broad tool for chemical sensing and single-molecule spectroscopy.

**Main:**

Nitrogen-vacancy (NV) centers in diamond are point defects in the diamond lattice that enable room temperature nanoscale nuclear magnetic resonance (NMR) of small ensembles down to the interrogation of individual molecules[1]. In these experiments, an NV center is located within a few nanometers of a target molecule on the diamond surface, and the magnetic signal from spins in the target molecule perturbs the phase of the NV spin, which is detected optically (Figure 1A). In recent years, NV centers have been used to detect magnetic noise arising from nuclear spins in a single ubiquitin protein[2] and from single electron spins at the diamond surface[9], perform NMR spectroscopy of microscale volumes of liquid with around 1 Hz spectral resolution[10], and spatially map individual $^{13}C$ nuclear spins in the diamond lattice under cryogenic conditions[3]. Despite this impressive progress, nanoscale NMR of arbitrary target molecules external to the diamond lattice remains elusive.

In order to sense a target molecule, the molecule must be immobilized within nanometers of an NV center with long spin coherence. Recent efforts to install such target molecules have used a thin layer (1-2 nm) of $Al_2O_3$ as a chemically functionalizable interface to attach sensing targets to diamond surfaces[11,12]. However, this approach comes at the expense of increased distance between NV sensor and target and has been shown to have some adverse effects on NV spin coherence. Directly functionalizing the diamond surface to attach sensing targets would eliminate the need for a separate material interface, but the chemical inertness and small lattice constant of diamond make this approach challenging.

Conventional diamond surface functionalization methods for NV experiments operate under harsh conditions, relying on the use of plasma[6–8], thermal annealing[2,5], or oxidizing acids[13]. These approaches severely limit the scope of functional groups that can be attached and can damage the diamond surface, destabilizing shallow NV centers[5]. Wet chemical methods allow for more controllable and mild surface modification methods, which enable covalent attachment of diverse functional groups to the surface. Thus far, wet chemical functionalization has been largely demonstrated on nanodiamonds[14], micro-crystalline diamonds[15,16], or poly-crystalline diamonds[17–19], but these materials have a wide range of crystal faces and high defect concentrations at the surface[20], making it difficult to translate these techniques to high-purity, single-crystal diamonds required for NV-based nanoscale NMR sensing. Specifically, the spin coherence of shallow NV centers can vary by over an order of magnitude because of the presence of electronic defects at the surface arising from subtle differences in surface disorder[5]. Prior solution-phase functionalization chemistries have not been demonstrated on pristine surfaces with coherent, shallow NV centers. Recent work has demonstrated that plasma termination followed by subsequent functionalization of amide groups can be compatible with shallow NV centers[8], and prior work has shown that defect-mediated photoexcitation of free carriers in the diamond via sub-bandgap ultraviolet light can initiate alkylation of hydrogen-terminated (H-terminated) single crystal diamond surfaces[21,22] and allows for grafting molecules with exceptional biochemical functionality[23], motivating a broader search for single-crystal surface functionalization methods and a study of their compatibility with shallow NV centers.

In this study, we present a new, generalizable photochemical strategy to functionalize high-quality, single-crystalline diamond surfaces, enabling the direct covalent attachment of diverse functional motifs, and we demonstrate the compatibility of this chemistry with shallow NV centers with long spin coherence. Inspired by recent developments in visible-light photocatalysis for activating C–H bonds in small molecules[24,25], we set out to activate inert chemical bonds on diamond surfaces using photoactivated hydrogen atom transfer (HAT) reagents (Figure 1B). These HAT reagents homolyze C–H bonds on a hydrogen-terminated (H-terminated) diamond surface to generate carbon-centered radicals, which can be intercepted with various radical acceptors to form new carbon–heteroatom bonds on the surface (Figure 1C). These solution-phase photochemical reactions operate at room temperature and enable the formation of various chemical bonds with surface coverages up to 10% (1.6 molecules/$nm^2$), including C–F, C–Cl, C–S, and C–N bonds. We establish evidence for covalent bond formation on single-crystal surfaces using multimodal surface spectroscopies and develop a mechanistic understanding of the surface reactivity of single-crystal diamond. Finally, we show that functionalized surfaces are compatible with NV centers within 10 nanometers of the surface with spin coherence times comparable to the state of the art, and we use ensembles of shallow NV centers to detect nuclear spins ($^{19}F$) contained in installed functional groups at the surface. This chemical strategy opens the door to a broad range of functionalization targets, thus overcoming a major hurdle for NV nanoscale molecular sensing.

We began by preparing high-quality, ultra-smooth, H-terminated diamond surfaces from oxygen-terminated surfaces (Figure 1C and see SI for more details) and establishing a general procedure to reliably evaluate the outcome of functionalization reactions. After a given reaction, the samples were thoroughly cleaned in a mixture of solvents to remove physisorbed contaminants. To evaluate whether or not a reaction is successful, we cannot rely on traditional solution-phase chemical analysis tools because they do not have sufficient sensitivity to probe surfaces. Thus, we sought to functionalize the surfaces with functional groups containing identifiable heteroatoms (e.g., N, S, halogens) and used surface-sensitive techniques such as X-ray photoelectron spectroscopy (XPS) to detect their presence on the surface. Since non-specific binding can also lead to spurious XPS signals (Figure S6), we iteratively cleaned our samples, checked for the consistency of the heteroatom signal in XPS, and verified that other contaminants from the reaction mixture were not present. Additionally, since physisorbed contamination is unlikely to result in a well-ordered monolayer, we used atomic force microscopy (AFM) to check that the surface was smooth (Figure S9). XPS and AFM allow for rapid feedback for reaction discovery but do not provide conclusive evidence of covalent bond formation. To confirm our results, we performed more detailed spectroscopy using synchrotron near-edge X-ray absorption fine-structure spectroscopy (NEXAFS). A strong polarization dependence of the NEXAFS signal associated with the heteroatom of interest indicates that the attached functional group has a well-defined orientation relative to the diamond surface, providing compelling evidence of covalent bond formation[26].

As an example, we first developed an HAT method for fluorination of the H-terminated surface. This reaction involved the irradiation of the H-terminated diamond in acetonitrile solvent with tetrabutylammonium decatungstate (TBADT), a fluorinating reagent such as Selectfluor or *N*-fluorobenzenesulfonimide (NFSI), and sodium bicarbonate with 390 nm light at room temperature for two days (Figure 2A). XPS analysis of the sample after the reaction and cleaning showed a clear F *1s* signal (Figure 2B and S11) and verified the absence of heteroatoms associated with other contaminants from the reaction mixture (Figure S7). Running the reaction in the dark did not result in an F *1s* peak, indicating that light is required to achieve surface functionalization (Figure 2C). Notably, the F *1s* peak did not diminish after multiple cycles of cleaning with organic solvents or after boiling in a mixture of concentrated sulfuric, nitric, and perchloric acids ('triacid cleaning') (Figure 2B). AFM reveals that the surface morphology is smooth after the functionalization reaction and that it is free of precipitates (Figure 2E). NEXAFS at the F *K*-edge shows a clear angular dependence, indicating that the C–F bonds are well-oriented on the surface (Figure 2D and S10). Low-energy electron diffraction (LEED) of a fluorinated surface shows a clear 2×1 pattern, which is consistent with an F/H mixed termination and demonstrates that the resulting surface is highly ordered (Figure 2F). Quantitative analysis of XPS spectra reveals that the newly formed C–F bonds comprise (6 to 10)% surface coverage ((0.9 to 1.6) molecules/$nm^2$), which is slightly lower than but comparable to the efficiencies of other mild processes for diamond surface functionalization[21,27]. We hypothesize that this moderate yield arises from a self-deactivation mechanism, in which the formation of C–F bonds increases the bond strength and electrophilicity of adjacent C–H bonds due to the strong electron-withdrawing effect of fluorine. Such deactivation effects are well-documented in C–H fluorination studies with small molecule substrates[28,29].

Using this HAT method and reaction discovery pipeline, we were also able to install other functional groups on the surface. Inspired by the work of Alexanian and co-workers[30], we found that treating the surface with *N*-chloroamide **1** in benzene under 456 nm light irradiation for 24 h resulted in chlorination of the surface with ≈ 4% coverage (0.6 molecules/$nm^2$), which was confirmed by the appearance of a Cl *2p* XPS signal (Figure 2G, bottom right, SI). Similarly, we installed xanthate groups with ≈ 2% coverage (0.3 molecules/$nm^2$) by irradiating the sample in the presence of *N*-xanthylamide **2** for 24 h with 456 nm light, as evidenced by the appearance of an S *2p* XPS signal (Figure 2G, bottom left)[31]. To the best of our knowledge, this reaction represents the first reported example of a xanthylation process on a diamond surface.

To further diversify the scope of structural motifs that can be attached to the surface, we sought to install a versatile functional handle that is amenable to a wide range of subsequent derivatization reactions. Serendipitously, we discovered that conducting the fluorination reaction with either Selectfluor or NFSI in the absence of the TBADT photocatalyst resulted in the formation of not only C–F bonds but also amide groups on the surface. This was evidenced by a clear N *1s* signal in XPS and an angle dependence at the N *K*-edge in NEXAFS (Figure 3A and S16). Though the identity of the nitrogen-containing group is not evident from the XPS or NEXAFS spectra, we were able to confirm the formation of amide groups by performing a series of control experiments with adamantane as a model substrate as well as with the diamond surface (See SI for more details). Furthermore, by replacing acetonitrile with 3,3,3-trifluoropropionitrile and trichloroacetonitrile, we also installed trifluoropropanamide and trichloroacetamide groups on the surface, respectively (Figure 3B). These amide groups present opportunities to subsequently attach molecules of interest through amide coupling reactions. Accordingly, amide-terminated surfaces were subjected to a hydrolysis reaction to form free amines. These amine groups are structurally similar to *tert*-butylamine and are thus expected to be considerably less reactive than typical primary amines due to their steric hindrance. Nevertheless, they still reacted with various acyl chlorides to install pentafluorobenzamide (Figure 3C), 3,5-bis(trifluoromethyl)benzamide, and heptafluorobutyramide moieties on the surface (Figure S24). Though amide and amine are indistinguishable in XPS, we observed distinct X-ray absorption features at the N *K*-edge and O *K*-edge in NEXAFS, consistent with prior literature reports[32,33] (Figure S25).

Collectively, this diverse set of functionalization reactions not only enables the construction of various structural motifs on the surface but also allows us to understand the mechanisms behind these reactions on single-crystal diamond surfaces. Density functional theory (DFT) calculations estimate the bond dissociation energy (BDE) of $C_{surface}$–H to be ≈ 98 kcal/mol (see SI, Figure S28), which is similar to that of nonactivated tertiary C–H bonds in small molecules (e.g., C–H BDE in adamantane ≈ 99 kcal/mol)[34]. In the fluorination reaction, surface C–H bonds could be abstracted by the excited state decatungstate anion ($^*[W_{10}O_{32}]^{4-}$),[35] the aminium radical cation (N–H BDFE ≈ 100 kcal/mol)[36] generated from Selectfluor, or the disulfonamidyl radical (N–H BDFE (≈ 105 to 110) kcal/mol)[37] produced from NFSI. The resulting carbon radicals could intercept either Selectfluor or NFSI to form new C–F bonds on the surface. Similarly, in the chlorination and xanthylation reactions, electrophilic amidyl radicals (N–H BDFE ≈ 110 kcal/mol)[38] generated from the light-driven homolysis of amides **1** and **2** could abstract surface C–H bonds to furnish the same carbon radical intermediates, which then react with **1** or **2** to install chlorine or xanthate groups, respectively (Figure S13)[30,31]. In the C–N bond formation reaction, the alkyl radicals undergo single electron transfer with Selectfluor (e.g., $E_{red}$ (Selectfluor) ≈ –0.04 V vs SCE in MeCN[39], $E_{ox}$ (*tert*-butyl radical) = 0.09 V vs SCE in MeCN)[40]. This process produces tertiary carbocations that can be trapped by a nitrile nucleophile to form C–N bonds via a Ritter mechanism (Figure S17)[36]. We note that such a photochemical Ritter amidation reaction involving only Selectfluor and nitriles has not been previously reported for small molecule substrates and that our photochemical protocol is the first example of direct amidation of diamond surfaces under ambient conditions. In addition to these successful functionalization reactions, other HAT protocols were investigated with the goal of C–C bond formation at the surface, although no reactivity was observed (SI, Figure S26). By contrast, similar approaches were shown to be effective for functionalization of diamonoids[41] and nanodiamonds[42–44]. These results further highlight a difference in the reactivity of high-quality, single-crystal diamond surfaces compared with other structurally analogous diamonds.

We next demonstrated that our surface functionalization strategy is compatible with shallow NV centers with long spin coherence. It has been widely established that coherence properties of single NV centers depend sensitively on surface termination and morphology[5,6,13,45]. Following previous work[5], we prepared smooth and low-defect oxygen-terminated surfaces, introduced nitrogen via ion implantation, and annealed to form NV centers. Next, the samples were hydrogen terminated and subsequently subjected to surface functionalization reactions. It has been established that hydrogen-terminated surfaces can lead to NV center passivation[46] or charge state instability[6]. To address these challenges, we explored three methods of hydrogen termination, including two plasma-based techniques and annealing in forming gas (SI). While all three methods produced appreciable hydrogen surface coverage (Figure S1), annealing under forming gas was the only method that was reliably compatible with shallow NV centers. Forming gas annealing does not produce significant quantities of atomic hydrogen that can diffuse readily in the diamond lattice and passivate NV centers or lead to Fermi level pinning[46]. Furthermore, it does not etch the surface and therefore does not etch away shallow NV layers or result in observable changes in surface morphology, which is a proxy for surface damage (Figure S4, S5). Additionally, it is well-known that hydrogen termination of diamond can lead to charge transfer to surface adsorbates, resulting in band bending that destabilizes the $NV^-$ charge state[6]. This band-bending depends on the surface chemistry, nature of surface adsorbates, and dopant concentration. In this work, we empirically observed that processing with either oxidizing acids or a low-temperature oxygen anneal was helpful to stabilize the $NV^-$ charge state after surface functionalization without removing functional groups (Figure S29).

We studied the impact of surface functionalization on NV center performance by measuring the spin coherence properties of shallow NV centers under the modified surfaces. We prepared three samples: Samples 1a and 1b were cut from the same larger diamond and processed together to produce 'triacid cleaned' surfaces with a disordered oxygen termination[5]. Sample 1a was interrogated under this oxygen termination, and Sample 1b was subjected to additional hydrogen termination followed by fluorination under our photochemical conditions prior to measurement. Sample 2 was amidated with 3,3,3-trifluoropropionitrile (more sample information in SI). For all samples, we measured the Hahn echo coherence time ($T_2$) as well as the depth[47] for a random selection of NV centers and observed coherent NV centers within 10 nm of the surface. To benchmark the impact of surface noise on NV properties, we compared the measured $T_2$ as a function of NV depth (Figure 4A). We found that the spin coherence times for NV centers in Sample 1b were slightly improved compared to Sample 1a, suggesting the functionalization procedure did not cause significant additional surface damage. The improvement in coherence properties between the 'triacid cleaned' and functionalized samples is comparable to the improvement after previously-reported high-purity oxygen annealing[5]. We also probed the spectral density of the noise bath using CPMG and XY-8 dynamical decoupling sequences[48]. In some cases, we were able to extend the coherence time to 79.7 $\mu$s ± 4.3 $\mu$s, demonstrating shallow NV centers with coherence properties comparable to the state of the art under functionalized surfaces (Figure 4B). We note that some properties of shallow NV centers are worse for functionalized samples than for state-of-the-art surfaces, including a shorter $T_1$ (Figure S32), reduction of $T_2$ after long exposure to green laser illumination, and changing optically detected electron spin resonance (OD-ESR) contrast under green laser illumination (Figure S33). Improving these parameters is an important subject for future work.

The shallow NV center properties we observed enable the detection of excess noise arising at the Larmor frequency of nuclear spins within the functional groups attached to the surface. We demonstrated this functionality using ensembles of shallow NV centers as a proof-of-principle experiment. The ensemble samples were prepared by implanting with a higher dose of $^{15}N$ (2.5 keV, $2 \times 10^{12}$ ions/cm$^2$) and annealing to form NV centers. We found that the dynamically decoupled coherence properties of the ensemble samples were slightly improved after functionalization (Figure 4C) and that our OD-ESR readout contrast was largely preserved (Figure S36). This slight improvement may arise from reduced disorder in the surface terminated with a mixture of hydrogen, fluorine, and oxygen relative to the fully oxygen-terminated surface, which has been established to comprise a number of different oxygen species[5,49]. However, we did observe a reduction in the double quantum lifetime ($T_1$) (Figure S37) and that the OD-ESR contrast and charge state properties of NV centers depend sensitively on the environment and time under green laser illumination (Figure S35).

We used these ensemble samples to detect the statistical polarization of $^{19}F$ nuclear spins attached to the surface. These nuclear spins precess at the Larmor frequency determined by their gyromagnetic ratio and the applied magnetic field, generating an AC magnetic signal, which can be detected using correlation spectroscopy (Figure 4D)[50,51]. In this experiment, the NV center is driven with an XY-8 pulse sequence that flips the NV center synchronously with the Larmor frequency of the $^{19}F$ signal. The NV center accumulates phase due to noise arising from $^{19}F$ nuclear spins during an initial XY-8 block. After this block, a $\pi/2$ pulse stores the phase information in the amplitude of the NV center. After a delay time ($\tau_p$), a second phase accumulation block occurs.

Scanning the time between the two phase accumulation blocks leads to a signal that oscillates at the Larmor frequency of the nuclear spins.

We performed this experiment on Sample 3, which was fluorinated using TBADT and Selectfluor, and detected the magnetic signal associated with $^{19}F$. To optimize photon collection efficiency while reducing the proton background, the samples were mounted in deuterated propylene carbonate (SI). We repeated the XY-8 sequence five times (k = 5) to optimize sensitivity while suppressing the proton signal. Taking the Fourier transform of this trace (Figure S38) showed a peak at the $^{19}F$ frequency (Figure 4E). The power spectrum of the $^{19}F$ NMR peak has a linewidth of 25.44 kHz ± 2.02 kHz, which is similar to previously measured linewidths of protons in polymers[50] and surface-bound fluorine groups[12] (Figure S42). To verify that the signal arose from $^{19}F$, we swept the magnetic field and observed a shift in peak frequency consistent with the gyromagnetic ratio of $^{19}F$ (Figure 4E, inset, also see Figure S39). Finally, we repeated the same experiment for a non-functionalized sample and did not detect a $^{19}F$ signal, confirming the absence of spurious signals from unintentional fluorine contamination sources (Figure S41).

**Outlook:**

Here, we have demonstrated a novel photochemical strategy to directly functionalize high-quality, single-crystal diamond surfaces that enables the covalent attachment of sensing targets for NV-based NMR sensing. The method relies on the HAT activation of inert C–H bonds to generate highly reactive free radical and carbocation intermediates, which engage in a range of bond-forming reactions to install diverse functional groups with surface coverage up to 10% (1.6 molecules/$nm^2$) under mild conditions. More broadly, we have developed a reaction discovery pipeline using multimodal spectroscopic characterization that is readily generalizable to other classes of chemistry. Notably, we have shown that our surface functionalization strategy is compatible with coherent shallow NV centers and we use NV centers to detect the Larmor precession of $^{19}F$ spins that are covalently attached to the diamond surface. Future work includes further diversifying the scope of functionalization reactions, in particular attaching biomolecules, such as proteins and nucleic acids, to enable nanoscale NMR studies. A further goal will be to establish that biomolecules and other sensing targets retain their functionality upon attachment to the surface and remain stable under optical and microwave excitation conditions during sensing experiments. Combining with advances in sensing protocols[52,53] or biophysical techniques[54], this new experimental capability has the potential to probe the magnetic signature of small ensembles of biomolecules, down to the single-molecule regime, enabling novel studies of dynamic structural changes in individual biomolecules.

Disclaimer:

Certain commercial equipment, instruments, or materials are identified in this paper in order to specify the experimental procedure adequately, and do not represent an endorsement by the National Institute of Standards and Technology.

Acknowledgements:

We thank Peter Maurer for helpful discussions and Zihuai Zhang for help with experimental challenges. This work was primarily supported by the Princeton Catalysis Initiative and the U.S. Department of Energy, Office of Science, Office of Basic Energy Sciences, under Award No. DE-SC0018978. The surface processing and spectroscopy pipeline was supported by the NSF CAREER program Grant No. DMR1752047, ensemble NMR and single center sensing were supported by the NSF Grant No. OMA-1936118, and some surface functionalization was supported by the Center for Molecular Quantum Transduction, an Energy Frontier Research Center funded by the U.S. Department of Energy, Office of Science, Office of Basic Energy Sciences, under Award No. DE-SC0021314CMQT. L.V.H.R. acknowledges support from the Department of Defense through the National Defense Science and Engineering Graduate Fellowship Program. J.H.C. acknowledges the National Science Foundation for a Graduate Research Fellowship (Grant No. #DGE-2039656). This research used resources of the Spectroscopy Soft and Tender Beamlines (SST-1 and SST-2) operated by the National Institute of Standards and Technology, located at the National Synchrotron Light Source II. NSLS-II is operated by the U.S. Department of Energy Office of Science Facilities at Brookhaven National Laboratory under contract no. DE-SC0012704. The authors acknowledge the use of Princeton's Imaging and Analysis Center (IAC), which is partially supported by the Princeton Center for Complex Materials (PCCM), a National Science Foundation (NSF) Materials Research Science and Engineering Center (MRSEC; DMR-2011750. This research was partly undertaken on the Soft X-Ray Spectroscopy beamline at the Australian Synchrotron, part of ANSTO. S.A.M., L.B.H., and A.B.J. gratefully acknowledge support from the US Department of Energy (BES grant No. DE-SC0019241), as well as the use of shared facilities at the UCSB Quantum Foundry through the Q-AMASE-i program (Grant No. NSF DMR-1906325). L.B.H. acknowledges the support from the NSF Graduate Research Fellowship Program (Grant No. DGE 2139319) and the UCSB Quantum Foundry. S.A.M. acknowledges the support from the UCSB Quantum Foundry. A. G. acknowledges the Hungarian NKFIH grant no. KKP129866 of the National Excellence Program of Quantum-coherent materials project, the support for the Quantum Information National Laboratory from the Ministry of Culture and Innovation of Hungary (NKFIH grant no. 2022-2.1.1-NL-2022-00004), the EU EIC Pathfinder project "QuMicro" (grant no. 101046911), and the EU QuantERA for the project MAESTRO. A.P. also acknowledges the support of Bolyai János Research Fellowship of the Hungarian Academy of Sciences and the computational resources provided by the Governmental Information Technology Development Agency (KIFÜ) of Hungary. S.S. acknowledged funding support from the NSRF via the Program Management Unit for Human Resources & Institutional Development, Research and Innovation [grant number B05F650024].

## References

1. Maze, J. R. *et al.* Nanoscale magnetic sensing with an individual electronic spin in diamond. *Nature* **455**, 644–647 (2008).
2. Lovchinsky, I. *et al.* Nuclear magnetic resonance detection and spectroscopy of single proteins using quantum logic. *Science* **351**, 836–841 (2016).
3. Abobeih, M. H. *et al.* Atomic-scale imaging of a 27-nuclear-spin cluster using a quantum sensor. *Nature* **576**, 411–415 (2019).
4. Szunerits, S., Nebel, C. E. & Hamers, R. J. Surface functionalization and biological applications of CVD diamond. *MRS Bull.* **39**, 517–524 (2014).
5. Sangtawesin, S. *et al.* Origins of Diamond Surface Noise Probed by Correlating Single-Spin Measurements with Surface Spectroscopy. *Phys. Rev. X* **9**, 031052 (2019).
6. Hauf, M. V. *et al.* Chemical control of the charge state of nitrogen-vacancy centers in diamond. *Phys. Rev. B* **83**, 081304 (2011).
7. Cui, S. & Hu, E. L. Increased negatively charged nitrogen-vacancy centers in fluorinated diamond. *Appl. Phys. Lett.* **103**, 051603 (2013).
8. Abendroth, J. M. *et al.* Single-Nitrogen–Vacancy NMR of Amine-Functionalized Diamond Surfaces. *Nano Lett.* **22**, 7294–7303 (2022).
9. Grinolds, M. S. *et al.* Subnanometre resolution in three-dimensional magnetic resonance imaging of individual dark spins. *Nat. Nanotechnol.* **9**, 279–284 (2014).
10. Glenn, D. R. *et al.* High-resolution magnetic resonance spectroscopy using a solid-state spin sensor. *Nature* **555**, 351–354 (2018).
11. Xie, M. *et al.* Biocompatible surface functionalization architecture for a diamond quantum sensor. *Proc. Natl. Acad. Sci.* **119**, e2114186119 (2022).

12. Liu, K. S. *et al.* Surface NMR using quantum sensors in diamond. *Proc. Natl. Acad. Sci.* **119**, e2111607119 (2022).

13. Stacey, A. *et al.* Evidence for Primal sp $^2$ Defects at the Diamond Surface: Candidates for Electron Trapping and Noise Sources. *Adv. Mater. Interfaces* **6**, 1801449 (2019).

14. Krueger, A. Current issues and challenges in surface chemistry of nanodiamonds. in *Nanodiamonds* 183–242 (Elsevier, 2017). doi:10.1016/B978-0-32-343029-6.00008-8.

15. Yeap, W. S. *et al.* Boron-Doped Diamond Functionalization by an Electrografting/Alkyne-Azide Click Chemistry Sequence. *ChemElectroChem* **1**, 1145–1154 (2014).

16. Agnès, C. *et al.* XPS study of ruthenium tris-bipyridine electrografted from diazonium salt derivative on microcrystalline boron doped diamond. *Phys. Chem. Chem. Phys.* **11**, 11647 (2009).

17. Strother, T. *et al.* Photochemical Functionalization of Diamond Films. *Langmuir* **18**, 968–971 (2002).

18. Zhang, G.-J. *et al.* DNA Micropatterning on Polycrystalline Diamond via One-Step Direct Amination. *Langmuir* **22**, 3728–3734 (2006).

19. Boukherroub, R. *et al.* Photochemical oxidation of hydrogenated boron-doped diamond surfaces. *Electrochem. Commun.* **7**, 937–940 (2005).

20. Nebel, C. E. Electronic properties of CVD diamond. *Semicond. Sci. Technol.* **18**, S1–S11 (2003).

21. Nichols, B. M., Butler, J. E., Russell, J. N. & Hamers, R. J. Photochemical Functionalization of Hydrogen-Terminated Diamond Surfaces: A Structural and Mechanistic Study. *J. Phys. Chem. B* **109**, 20938–20947 (2005).

22. Wang, X., Colavita, P. E., Streifer, J. A., Butler, J. E. & Hamers, R. J. Photochemical Grafting of Alkenes onto Carbon Surfaces: Identifying the Roles of Electrons and Holes. *J. Phys. Chem. C* **114**, 4067–4074 (2010).

23. Hamers, R. J. *et al.* Molecular and biomolecular monolayers on diamond as an interface to biology. *Diam. Relat. Mater.* **14**, 661–668 (2005).

24. Holmberg-Douglas, N. & Nicewicz, D. A. Photoredox-Catalyzed C–H Functionalization Reactions. *Chem. Rev.* **122**, 1925–2016 (2022).

25. Capaldo, L., Ravelli, D. & Fagnoni, M. Direct Photocatalyzed Hydrogen Atom Transfer (HAT) for Aliphatic C–H Bonds Elaboration. *Chem. Rev.* **122**, 1875–1924 (2022).

26. Stöhr, J. *NEXAFS Spectroscopy*. vol. 25 (Springer Berlin Heidelberg, 1992).

27. Bachman, B. F. *et al.* High-Density Covalent Grafting of Spin-Active Molecular Moieties to Diamond Surfaces. *Langmuir* **37**, 9222–9231 (2021).

28. Szpera, R., Moseley, D. F. J., Smith, L. B., Sterling, A. J. & Gouverneur, V. The Fluorination of C−H Bonds: Developments and Perspectives. *Angew. Chem. Int. Ed.* **58**, 14824–14848 (2019).

29. Lantaño, B. & Postigo, A. Radical fluorination reactions by thermal and photoinduced methods. *Org. Biomol. Chem.* **15**, 9954–9973 (2017).

30. Quinn, R. K. *et al.* Site-Selective Aliphatic C–H Chlorination Using *N* -Chloroamides Enables a Synthesis of Chlorolissoclimide. *J. Am. Chem. Soc.* **138**, 696–702 (2016).

31. Czaplyski, W. L., Na, C. G. & Alexanian, E. J. C–H Xanthylation: A Synthetic Platform for Alkane Functionalization. *J. Am. Chem. Soc.* **138**, 13854–13857 (2016).

32. Latham, K. G., Dose, W. M., Allen, J. A. & Donne, S. W. Nitrogen doped heat treated and activated hydrothermal carbon: NEXAFS examination of the carbon surface at different

temperatures. *Carbon* **128**, 179–190 (2018).

33. Zubavichus, Y., Shaporenko, A., Grunze, M. & Zharnikov, M. Innershell Absorption Spectroscopy of Amino Acids at All Relevant Absorption Edges. *J. Phys. Chem. A* **109**, 6998–7000 (2005).

34. Yang, H.-B., Feceu, A. & Martin, D. B. C. Catalyst-Controlled C–H Functionalization of Adamantanes Using Selective H-Atom Transfer. *ACS Catal.* **9**, 5708–5715 (2019).

35. Ravelli, D., Fagnoni, M., Fukuyama, T., Nishikawa, T. & Ryu, I. Site-Selective C–H Functionalization by Decatungstate Anion Photocatalysis: Synergistic Control by Polar and Steric Effects Expands the Reaction Scope. *ACS Catal.* **8**, 701–713 (2018).

36. Aguilar Troyano, F. J., Merkens, K. & Gómez-Suárez, A. Selectfluor® Radical Dication (TEDA $^{2+\cdot}$ ) – A Versatile Species in Modern Synthetic Organic Chemistry. *Asian J. Org. Chem.* **9**, 992–1007 (2020).

37. Bordwell, F. G., Harrelson Jr, J. A. & Lynch, T. Y. Homolytic bond dissociation energies for the cleavage of. alpha.-nitrogen-hydrogen bonds in carboxamides, sulfonamides, and their derivatives. The question of synergism in nitrogen-centered radicals. *J. Org. Chem.* **55**, 3337–3341 (1990).

38. Gentry, E. C. & Knowles, R. R. Synthetic Applications of Proton-Coupled Electron Transfer. *Acc. Chem. Res.* **49**, 1546–1556 (2016).

39. Timofeeva, D. S., Ofial, A. R. & Mayr, H. Kinetics of Electrophilic Fluorinations of Enamines and Carbanions: Comparison of the Fluorinating Power of N–F Reagents. *J. Am. Chem. Soc.* **140**, 11474–11486 (2018).

40. Wayner, D. D. M., McPhee, D. J. & Griller, D. Oxidation and reduction potentials of transient free radicals. *J. Am. Chem. Soc.* **110**, 132–137 (1988).

41. Weigel, W. K., Dang, H. T., Feceu, A. & Martin, D. B. C. Direct radical functionalization methods to access substituted adamantanes and diamondoids. *Org. Biomol. Chem.* **20**, 10–36 (2022).

42. Barzegar Amiri Olia, M., Donnelly, P. S., Hollenberg, L. C. L., Mulvaney, P. & Simpson, D. A. Advances in the Surface Functionalization of Nanodiamonds for Biological Applications: A Review. *ACS Appl. Nano Mater.* **4**, 9985–10005 (2021).

43. Zhong, Y. L. & Loh, K. P. The Chemistry of C–H Bond Activation on Diamond. *Chem. - Asian J.* **5**, 1532–1540 (2010).

44. Girard, H. A. *et al.* Surface properties of hydrogenated nanodiamonds: a chemical investigation. *Phys. Chem. Chem. Phys.* **13**, 11517 (2011).

45. Yuan, Z. *et al.* Charge state dynamics and optically detected electron spin resonance contrast of shallow nitrogen-vacancy centers in diamond. *Phys. Rev. Res.* **2**, 033263 (2020).

46. Stacey, A. *et al.* Depletion of nitrogen-vacancy color centers in diamond via hydrogen passivation. *Appl. Phys. Lett.* **100**, 071902 (2012).

47. Pham, L. M. *et al.* NMR technique for determining the depth of shallow nitrogen-vacancy centers in diamond. *Phys. Rev. B* **93**, 045425 (2016).

48. De Lange, G., Wang, Z. H., Ristè, D., Dobrovitski, V. V. & Hanson, R. Universal Dynamical Decoupling of a Single Solid-State Spin from a Spin Bath. *Science* **330**, 60–63 (2010).

49. Kaviani, M. *et al.* Proper Surface Termination for Luminescent Near-Surface NV Centers in Diamond. *Nano Lett.* **14**, 4772–4777 (2014).

50. Aslam, N. *et al.* Nanoscale nuclear magnetic resonance with chemical resolution. *Science* **357**, 67–71 (2017).

51. Laraoui, A. *et al.* High-resolution correlation spectroscopy of 13C spins near a nitrogen-vacancy centre in diamond. *Nat. Commun.* **4**, 1651 (2013).

52. Rovny, J. *et al.* Nanoscale covariance magnetometry with diamond quantum sensors. *Science* **378**, 1301–1305 (2022).

53. Boss, J. M., Cujia, K. S., Zopes, J. & Degen, C. L. Quantum sensing with arbitrary frequency resolution. *Science* **356**, 837–840 (2017).

54. Orrit, M., Ha, T. & Sandoghdar, V. Single-molecule optical spectroscopy. *Chem. Soc. Rev.* **43**, 973 (2014).

55. Lichtman, D., Craig, J. H., Sailer, V. & Drinkwine, M. AES and XPS spectra of sulfur in sulfur compounds. *Appl. Surf. Sci.* **7**, 325–331 (1981).

56. Smentkowski, V. S. & Yates, J. T. Fluorination of Diamond Surfaces by Irradiation of Perfluorinated Alkyl Iodides. *Science* **271**, 193–195 (1996).

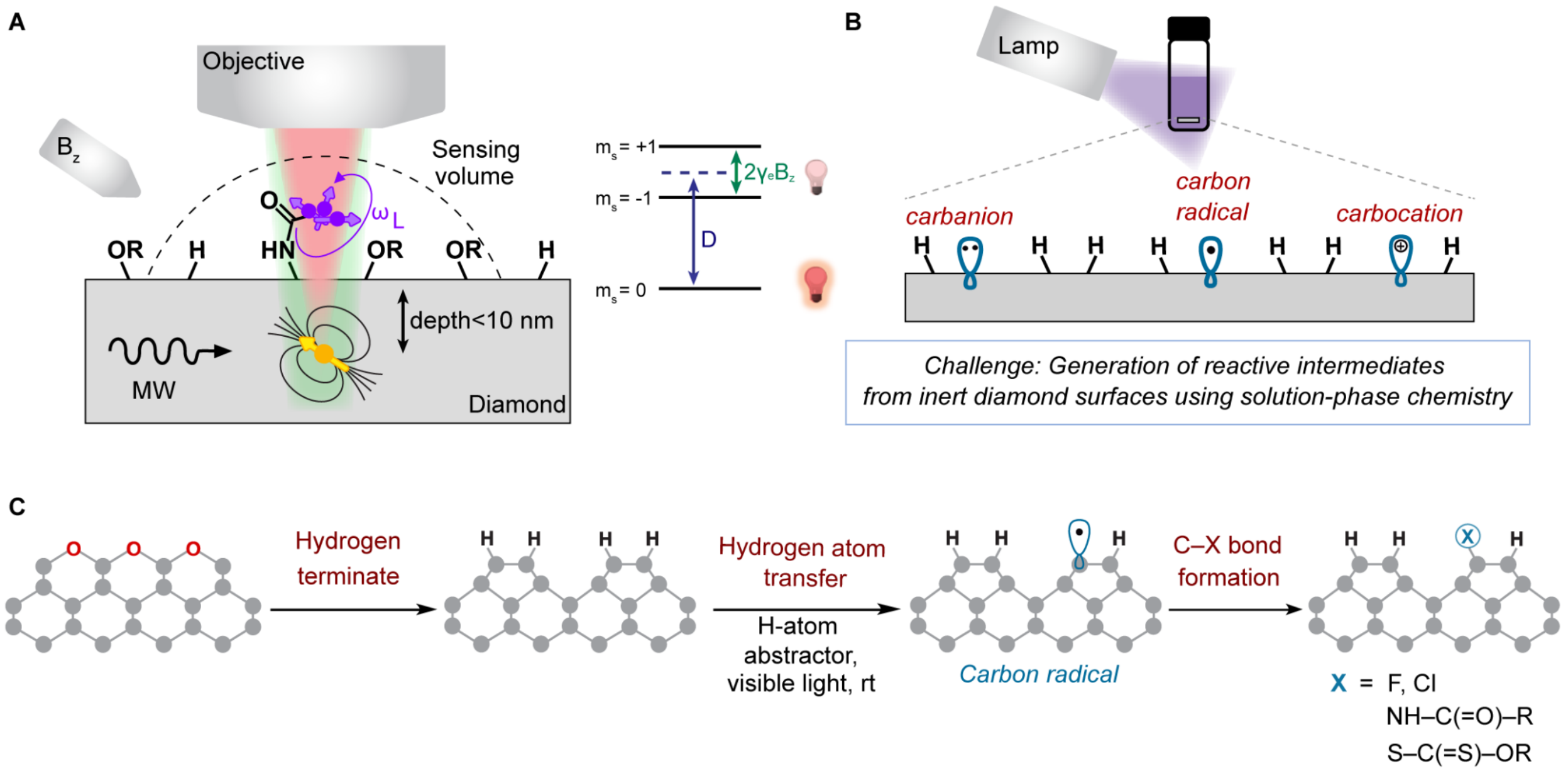

**Figure 1.** (**A**) NV-based sensing schematic showing a shallow NV center (yellow) in a host diamond crystal (gray). The NV center can probe the magnetic field originating from functional groups containing nuclear spins (purple) that are covalently bonded to the surface and that precess at the Larmour frequency ($\omega_L$) determined by their gyromagnetic ratio and the applied magnetic field ($B_z$). A green laser is used to optically initialize and read out the spin state of the NV center at room temperature and the spin-dependent red fluorescence is collected by a microscope objective. Top right: NV ground state energy level diagram with spin-dependent optical fluorescence showing zero field (D) and Zeeman splittings ($2\gamma_e B_z$). (**B**) The generation of reactive intermediates from inert surfaces using solution-phase chemistry is a long-standing challenge in diamond surface functionalization. (**C**) C–H bond activation via photochemical hydrogen atom transfer enables the formation of a wide range of carbon–heteroatom bonds on diamond surfaces.

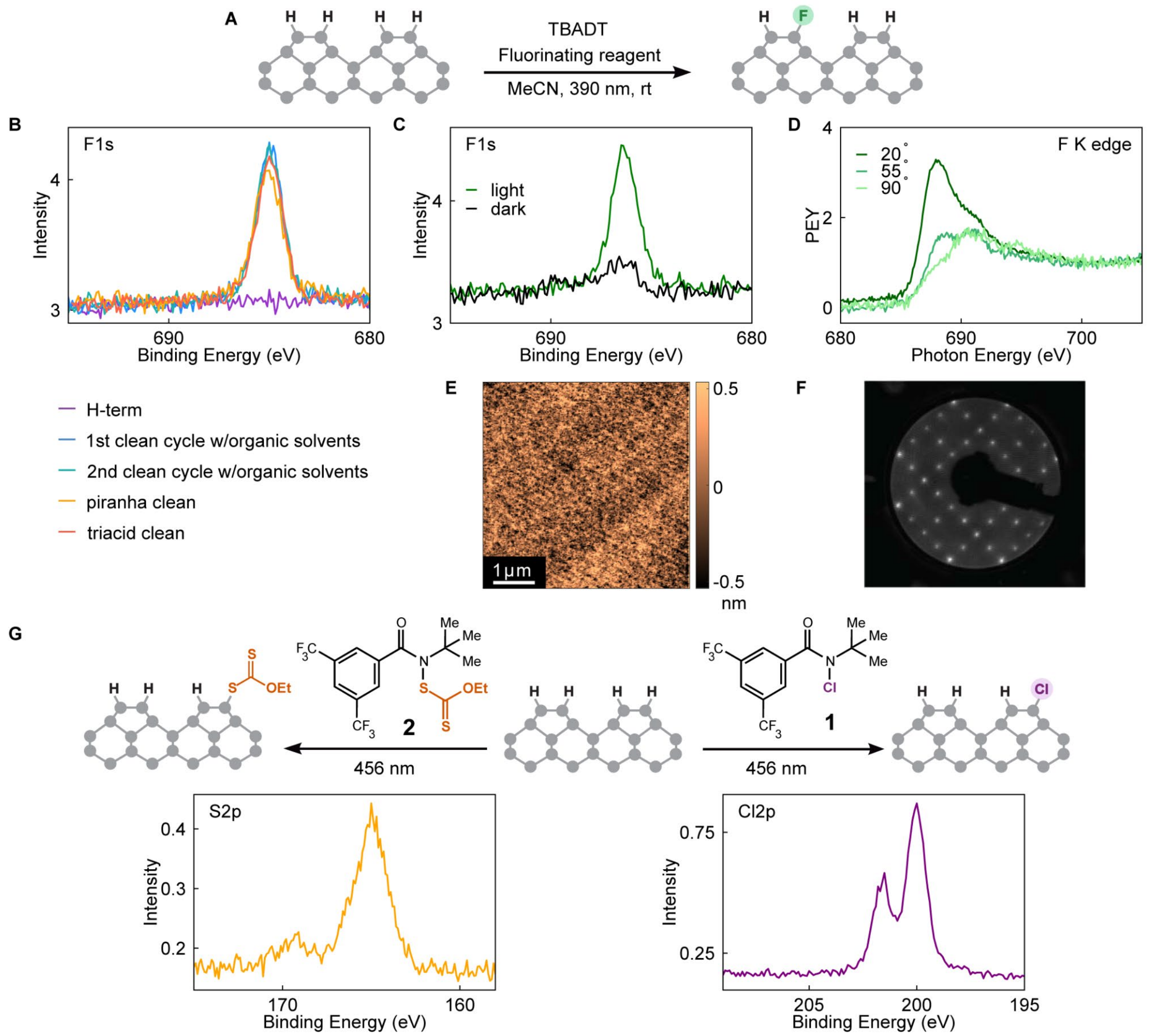

**Figure 2.** Activation of C–H bonds via HAT enables fluorination, chlorination, and xanthylation reactions on diamond surfaces. (**A**) Reaction scheme for photochemical fluorination reaction. (**B**) XPS F *1s* spectra of a fluorinated surface after iterative cleaning procedures. (**C**) F *1s* spectra for control experiments performed with and without light, showing that light is necessary to achieve fluorination. (**D**) NEXAFS F *K*-edge spectra acquired at different angles of incidence. The observed angle dependence indicates the fluorine atoms are well-oriented on the surface, consistent with covalent bond formation. (**E**) AFM image of a functionalized diamond surface. The resulting surface is smooth ($R_a$ = 201 pm ± 6 pm), and there is no evidence of physisorbed contamination. Samples in (**B**–**D**) were fluorinated with NFSI, and samples in (**E**), and (**F**) were fluorinated with Selectfluor. XPS and NEXAFS spectra for samples fluorinated with Selectfluor are included in the SI (Figure S10, S11). (**F**) LEED image of a fluorinated surface shows a 2×1 pattern, which is consistent with a mixed F/H termination and indicates that the surface is highly ordered. (**G**) Top: reaction schemes for xanthylation (left) and chlorination (right) reactions. Bottom left: XPS S *2p* spectrum of a xanthylated surface (binding energy ≈ 165 eV). The peak around 170 eV could correspond to a group containing sulfur in higher oxidation states,[55] possibly resulting from the oxidation of the xanthate ester group. Bottom right: XPS Cl *2p* spectrum of a chlorinated surface.

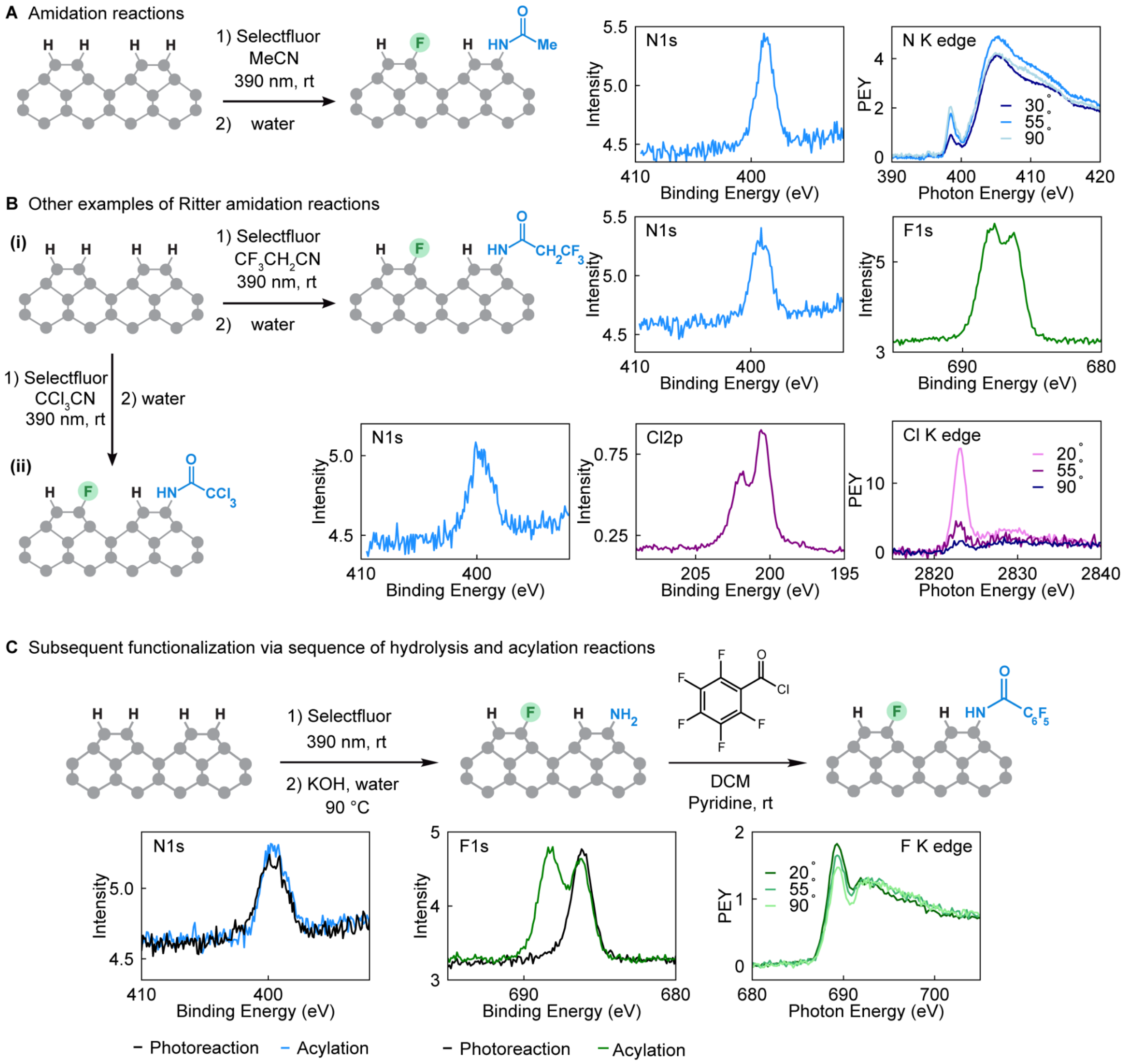

**Figure 3.** Amidation of diamond surfaces via a photochemical Ritter reaction. (**A**) Left: Control experiment scheme for the fluorination reaction in the absence of the TBADT photocatalyst leading to the discovery of C–N bond formation at the surface. Middle: XPS N *1s* spectrum of post-irradiation surface. Right: NEXAFS N *K*-edge spectra acquired at different incident X-ray angles relative to the surface after functionalization. The observed angle dependence indicates that the nitrogen groups are well-oriented on the surface, consistent with covalent C–N bond formation. (**B**) Examples of amidation reactions with halogenated nitrile nucleophiles. (**i**) Left: Reaction scheme for surface amidation with 3,3,3-trifluoropropionitrile. Middle: XPS N *1s* spectrum. Right: XPS F *1s* spectrum indicating the presence of both $C_{(diamond)}$–F bonds (686 eV) and the trifluoromethyl group ($CF_3$) in the amide (689 eV)[56]. (**ii**) Left: Reaction scheme for surface amidation with trichloroacetonitrile. Middle left: XPS N *1s* spectrum. Middle right: XPS Cl *2p* spectrum indicating the presence of trichloromethyl groups on the surface. Right: NEXAFS Cl *K*-

edge spectra acquired at different incident X-ray angles relative to the surface bearing trichloroacetamide groups. (**C**) Top: Subsequent surface functionalization via a sequence of hydrolysis and acylation reactions. Bottom left and middle: XPS N *1s* and F *1s* spectra after amidation and acylation reactions. Bottom right: NEXAFS F *K*-edge spectra acquired at different incident X-ray angles relative to a surface bearing pentafluorobenzamide groups.

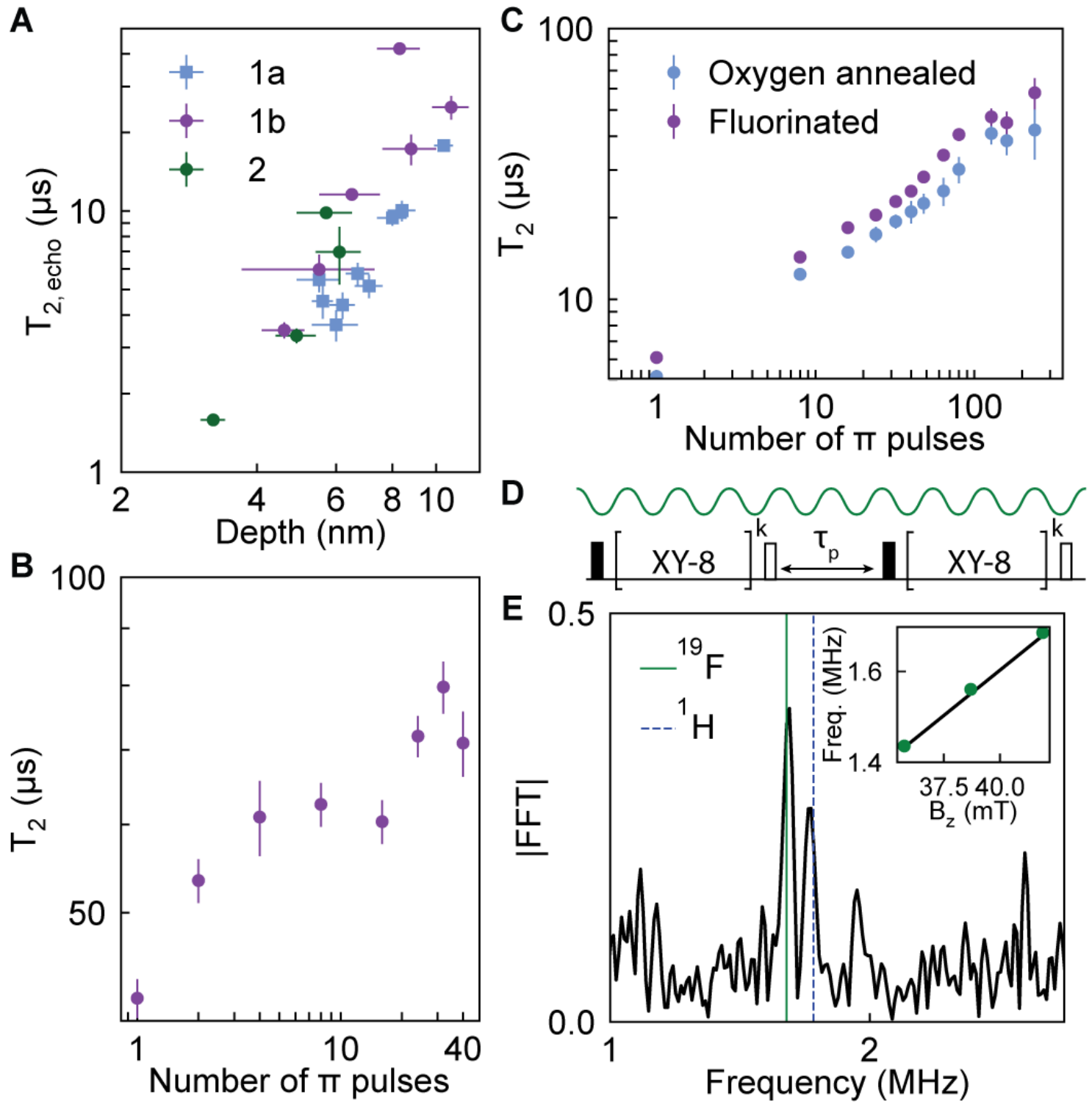

**Figure 4.** NV sensors under functionalized surfaces. (**A**) Hahn echo coherence time ($T_{2,\,echo}$) as a function of NV center depth under functionalized surfaces. Samples 1a and 1b are pieces of the same diamond that was implanted and processed before dicing. Sample 1a was characterized under an 'acid cleaned' oxygen terminated surface[5], while Sample 1b was hydrogen terminated followed by fluorination and acid cleaning. Sample 2 was functionalized using the Ritter reaction with 3,3,3-trifluoropropionitrile and subjected to a low-temperature oxygen anneal. (**B**) Coherence times with dynamical decoupling for a single 8.3 nm ± 0.9 nm deep NV center in Sample 1b. Increasing the number of π pulses extends the coherence time to 79.7 $\mu$s ± 4.3 $\mu$s. (**C**) Dynamically decoupled coherence times for a shallow NV ensemble in Sample 3 as a function of the number of π pulses. The same sample is interrogated after oxygen annealing and subsequent fluorination. (**D**) Diagram of the NV-NMR sensing sequence. The green wave depicts an oscillating magnetic field from nuclear spins. The black pulse diagram shows the correlation sequence used to detect nuclear spins. Two XY-8 blocks allow the NV center to accumulate phase from $^{19}$F nuclear spins. Each block is repeated 'k' times. The time between the blocks, $\tau_p$, is scanned during the experiment. (**E**) Fourier transform of correlation sequence data (Figure S38) reveals a peak at the expected frequency for $^{19}$F at a magnetic field of 42 mT, arising from covalently attached fluorine. This dataset was acquired by averaging the time domain signal for 10 spots on the sample over 5 hours and has a signal to noise ratio of around 10. The magnetic field dependence of the peak position is consistent with the gyromagnetic ratio of $^{19}F$ (inset).

**Diamond Surface Functionalization via Visible Light-Driven C–H Activation for Nanoscale Quantum Sensing**

Lila V. H. Rodgers[1,*], Suong T. Nguyen[2,*], James H. Cox[2], Kalliope Zervas[1], Zhiyang Yuan[1], Sorawis Sangtawesin[3], Alastair Stacey[4,5], Cherno Jaye[6], Conan Weiland[6], Anton Pershin[7,8], Adam Gali[7,8], Lars Thomsen[9], Simon A. Meynell[10], Lillian B. Hughes[11], Ania C. Bleszynski Jayich[10], Xin Gui[2], Robert J. Cava[2], Robert R. Knowles[2,†], Nathalie P. de Leon[1,†]

*[1]Princeton University, Department of Electrical and Computer Engineering, Princeton, New Jersey 08540, USA*
*[2]Princeton University, Department of Chemistry, Princeton, New Jersey 08540, USA*
*[3]School of Physics and Center of Excellence in Advanced Functional Materials, Suranaree University of Technology, Nakhon Ratchasima 30000, Thailand*
*[4]School of Physics, University of Melbourne, Parkville VIC 3010, Australia*
*[5]School of Science, RMIT University, Melbourne, VIC 3000, Australia*
*[6]Material Measurement Laboratory, National Institute of Standards and Technology, Gaithersburg, Maryland 20899, USA*
*[7]Wigner Research Centre for Physics, POB 49, Budapest, H-1525, Hungary*
*[8]Department of Atomic Physics, Budapest University of Technology and Economics, Műegyetem rakpart 3., Budapest, H-1111, Hungary*
*[9]Australian Synchrotron, Australian Nuclear Science and Technology Organisation, 800 Blackburn Road, Clayton, VIC 3168, Australia*
*[10]Physics Department, University of California Santa Barbara, Santa Barbara, California 93106, USA*
*[11]Materials Department, University of California, Santa Barbara, Santa Barbara, California 93106, USA*

**these authors contributed equally to the work*
*† corresponding authors: rknowles@princeton.edu, npdeleon@princeton.edu*

## Supplementary Information

## I. General information

***Sample preparation for reaction discovery***

All diamond samples in this work were purchased from Element Six and are single crystal samples grown by chemical vapor deposition. Standard grade samples with (100) surfaces were used for reaction discovery and testing different hydrogen termination procedures. These samples were prepared with smooth surfaces following previous work[1] prior to being subjected to various hydrogen termination and functionalization procedures. For the purposes of reaction discovery, we treat the hydrogen-terminated surface as the starting condition before subjecting samples to reactions.

***Surface spectroscopy measurement details***

XPS data were acquired at normal incidence using a Thermo Fisher K-Alpha XPS and X-Ray Spectrometer tool with an aluminum anode (1486.8 eV) and a 250 µm spot size. A flood gun was used to mitigate issues with sample charging. All XPS data were processed by shifting the binding energy based on C *1s* $sp^3$ reference (284.8 eV) and normalized to the sum of the total signal collected in the survey scan to account for focusing variation.

All AFM images were taken with a Bruker Icon3 tool operating in tapping mode (AFM tip from Oxford Instruments Asylum Research, part number AC160TS-R3, resonance frequency 300 kHz). Error values for average roughness ($R_a$) were computed by dividing the image into quadrants, computing the $R_a$ for each quadrant, and then calculating the standard deviation of those values.

LEED images were acquired using a OCI Vacuum BDL-450 w/ Z-translation (190 mm + 10 mm WD) installed in a custom ultra-high vacuum chamber. The spectra in this work were collected with approximately 252 eV beam energy.

NEXAFS spectra were collected using the polarized soft X-ray sources at both the Australia Synchrotron as well as the SST-1 and SST-2 beamlines at NSLS-II, Brookhaven National Lab. In all cases, the data were collected in partial electron yield (PEY) mode. A flood gun was used to mitigate sample charging.

The N *K*-edge measurements shown in Figure 3A and Figure S16 and the F *K*-edge measurements shown in Figure 2B, Figure S10, and Figure S16 were collected at the Australia Synchrotron in the high-throughput system. This system had a base pressure of $10^{-7}$ mbar and used a retarding grid detector. The retarding grid bias was set to –100 V and –550 V for N and F *K*-edge measurements, respectively. The C *K*-edge measurements shown in Figure S2 were collected at the Australia Synchrotron in the Prevac endstation with a base pressure of $10^{-10}$ mbar.

For SST-1 and SST-2 measurements, data were collected using channeltron electron multipliers. Energy selection was accomplished using a variable line spacing plane grating monochromator at SST-1 and a double Si (111) crystal monochromator at SST-2. The N *K*-edge and O *K*-edge measurements shown in Figure S25 were collected at SST-1 with the entrance grid bias set to –280 V and –400 V for F and N *K*-edges, respectively. The F *K*-edge and C K-edge measurements shown in Fig 3C in the main text and Figure S29 were collected at SST-1 with an entrance grid bias of –550 V and –150 V, respectively. All Cl *K*-edge measurements shown in this work were collected at SST-2 and the entrance grid bias set between –1800 V and –2000 V.

The NEXAFS data were processed by first normalizing to the total intensity of the incident beam which was measured by the drain current off a gold grid ($I_0$). Data acquired at the Australia Synchrotron were 'double normalized' to a signal from a photodiode in the end chamber after dividing the photodiode signal by the corresponding $I_0$. Next, the average value of the PEY signal from the pre-edge region was subtracted. For some datasets, the pre-edge PEY signal was fit to a linear function and this was subtracted instead of the mean value. Finally, the data were normalized to the average PEY signal in the post-edge region. Photon energy for datasets acquired at SST-1 were shifted based on a reference energy calibration. Photon energy for datasets acquired at the Australia synchrotron were shifted based on C *K*-edge features ($sp^2$ peak, diamond exciton peak) such that they aligned with SST-1 datasets that were shifted based on the measured energy reference. Cl *K*-edge datasets acquired at SST-2 were not shifted.

## II. Hydrogen-terminated surface preparation and characterization

In this work, we employed three methods of hydrogen terminating (100) diamond surfaces. Here, we detail the experimental procedures to create these surfaces and surface spectroscopy measurements of these surfaces.

***'As grown' plasma***: The samples were processed in a CVD diamond growth chamber. Notably, this procedure etches and regrows the top few microns of material. The procedure has been described previously[2], but we provide a brief description for completeness. Diamond growth and termination was conducted in a 'clamshell' type Seki reactor (Seki SDS 6500 with a palladium-based hydrogen purifier) under an hydrogen plasma at 85 Torr ($1.1 \times 10^4$ Pa) and 4500 W. The growth step consisted of exposure to 4% methane for 5 min followed by a hydrogen termination step (no methane addition) for about 3 min with microwave power decreased to 3200 W. The sample temperature is approximately 700 °C at plasma extinction.

***'Gentle' plasma***: The goal was to use a plasma to gently terminate the surface without etching or removing material. The termination was done in a SEKI AX6300 chemical vapor deposition diamond growth chamber with a base pressure of $2 \times 10^{-6}$ Pa. After loading the sample, the hydrogen plasma was initiated with an RF power of 750 W and a flow rate of 400 sccm and a corresponding pressure of 25 Torr ($3.3 \times 10^3$ Pa). Over a duration of 5 minutes, the sample temperature was gradually elevated to 800 °C. Subsequently, the sample was held at 800 °C for (15 to 50) minutes before undergoing a controlled temperature reduction to 200 °C over a period of 20 minutes. To mitigate hydrogen dissociation during the cooling process, the plasma settings were adjusted to a reduced RF power of 400 W, a flow rate of 30 sccm, and a pressure of 3.5 Torr ($0.5 \times 10^3$ Pa). The plasma was deactivated upon reaching the target temperature of 200 °C following a 20 minute cooldown period.

***Hydrogen annealing***: Adapted from literature reports of producing hydrogen-terminated single crystal diamond by annealing under pure hydrogen gas[3], we produced hydrogen terminated diamond by annealing under an atmosphere of forming gas (5% $H_2$, 95% Ar, Airgas

X02AR95C3000993) in a tube furnace (Thermo Scientific Lindberg/Mini Mite). This same procedure was used in another recent work[4]. The sample was heated to 100 °C over 1 h, held at 100 °C for 2 h for degassing, ramped to 800 °C over 2.5 h, and held at 800 °C for 72 h. In the course of developing this procedure, we tried annealing durations ranging from 24 h to 2 weeks with similar results. The majority of the samples in this work were annealed for 72 h for consistency. The setup consisted of a tube furnace with a quartz tube, a mass flow controller to control the flow rate, and a bubbler with mineral oil at the output of the furnace to isolate the chamber from atmosphere. The forming gas was filtered before entering the furnace (SAES MC1-203F). The samples were loaded in a high-purity alumina boat purchased from CoorsTek.

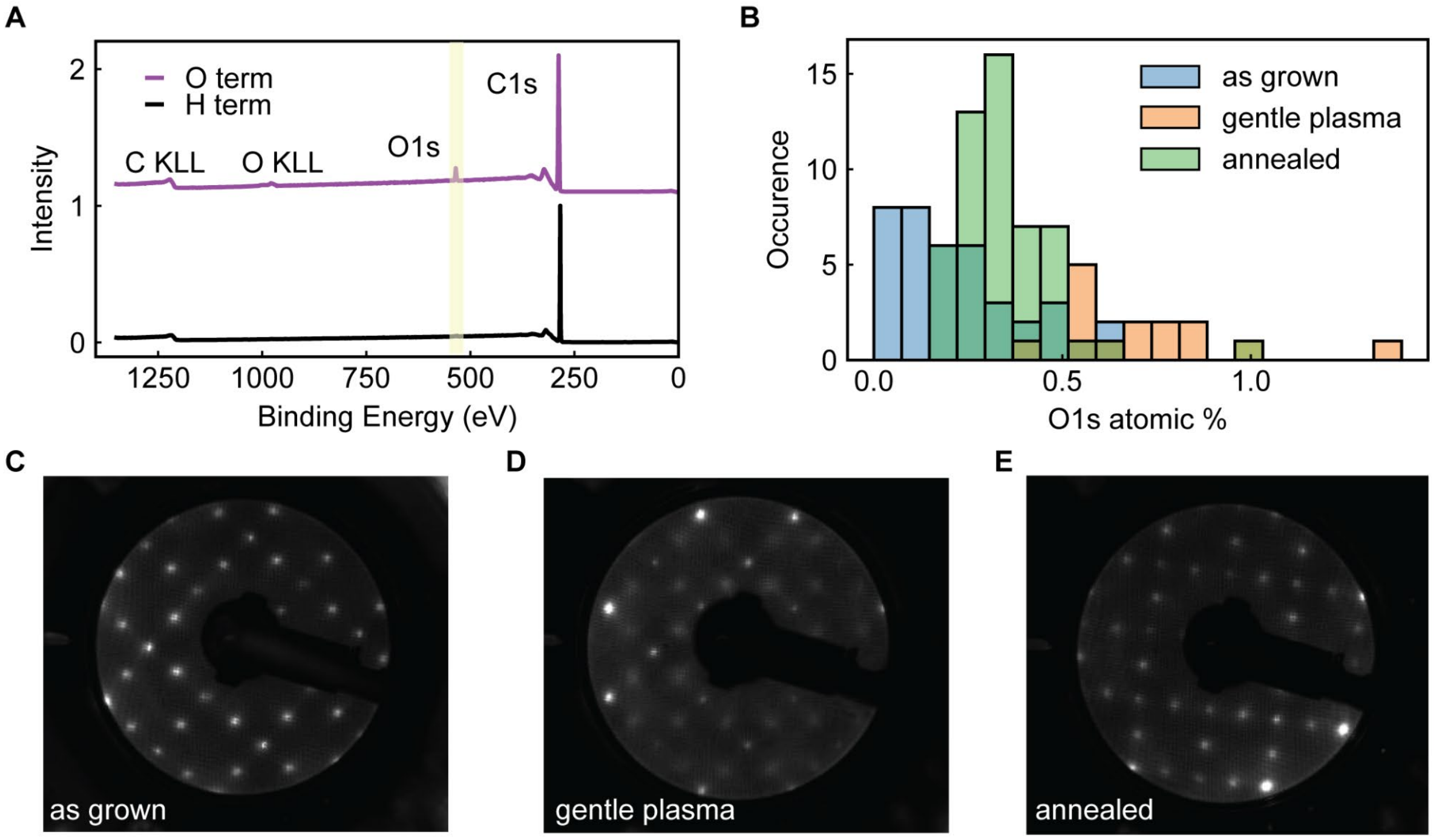

**Figure S1.** Characterization of hydrogen-terminated surfaces. (**A**) XPS characterization of diamond before (purple) and after (black) hydrogen termination. The survey spectra showed a reduction in O *1s* signal after hydrogen termination. (**B**) Histogram of O *1s* atomic percentage measured by XPS after three different methods of hydrogen termination. All methods resulted in a reduced oxygen signal (monolayer coverage corresponds to approximately 7.6 atomic %[1]). The 'as grown' surfaces on average showed the lowest oxygen signal. We note that adventitious carbon contamination can also contribute to the O *1s* signal. (**C–E**) LEED images of hydrogen-terminated surfaces prepared with the three different methods all showed 2×1 patterns characteristic of hydrogen-terminated diamond.

A key signature of hydrogen termination is a reduction in the O *1s* peak observed in XPS, since XPS is unable to directly detect hydrogen (Figure S1A). We routinely performed XPS on all of our samples after hydrogen termination to check O *1s* coverage and checked for absence of common contaminants such as silicon, sodium, or chlorine. As shown in Figure S1B, we observed differences in O *1s* levels across samples terminated with various procedures. Surfaces terminated with the ‘as grown’ plasma procedure consistently showed the lowest O *1s* coverage. We note that ubiquitous adventitious carbon contamination can also contribute to the O *1s* signal.

Next, we used low energy electron diffraction (LEED) to study the surface reconstruction (Figure S1C-E). Here, we looked for a 2×1 LEED pattern typical of a hydrogen-terminated diamond[5]. In all cases, we observed 2×1 LEED patterns. Some variation between quality of LEED images could result from sample-to-sample variation, surface miscut, or the presence of residual oxygen on the surface.

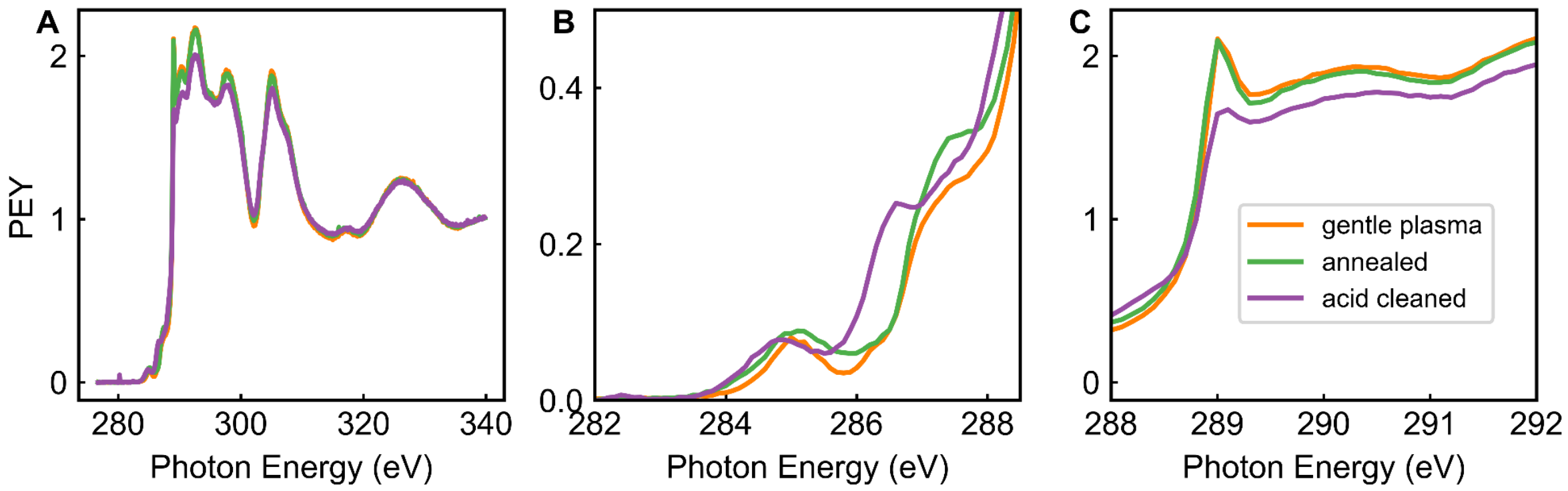

**Figure S2.** NEXAFS studies of hydrogen-terminated surfaces. (**A**) NEXAFS C *K*-edge full spectra of three diamond samples: one prepared with the gentle plasma a few months prior (orange), one that was freshly annealed (green), and one that was acid cleaned and had never been hydrogen terminated (purple). (**B**) Zoom-in on the pre-edge region of the same data. The feature at 285 eV is associated with $sp^2$ carbon, the feature at 286.5 eV is associated with carbon–oxygen bonds, and the shoulder at 287 eV is associated with carbon–hydrogen bonds[2]. (**C**) Zoom-in on the exciton feature of the C *K*-edge spectra at 289 eV. The hydrogen-terminated samples showed a sharper exciton feature compared to the acid cleaned sample. Since the ‘gentle plasma’ sample was not freshly prepared for this synchrotron experiment, it is likely not instructive to read into differences in the amplitude and sharpness of the hydrogen termination feature at 287 eV.

Additionally, we obtained NEXAFS spectra of hydrogen-terminated samples prepared by the ‘gentle’ plasma and annealing methods to prove that they were hydrogen terminated (Figure S2). C *K*-edge NEXAFS spectra of samples terminated by the ‘as grown’ procedure has been previously reported[2]. These spectra were collected after annealing *in situ* to remove adventitious carbon. The C *K*-edge spectra were consistent with typical diamond NEXAFS spectra (Figure S2). The pre-edge region of all samples showed a $sp^2$ peak at 285 eV. All of the samples that were hydrogen terminated show varying degrees of a shoulder peak at 287 eV (Figure S2). This feature has been attributed to C–H bonds at the diamond surface[2]. In contrast, a sample that was acid cleaned and not hydrogen terminated showed a dominant peak closer to 286.5 eV, which is associated with carbon–oxygen bonds[2]. Since the ‘gentle’ plasma hydrogen-terminated sample was not prepared freshly for these synchrotron measurements, it is likely not informative to read into differences in the height and sharpness of the 287 eV feature across samples. Finally, we observed that the hydrogen-terminated samples displayed a sharper exciton feature at 289 eV compared to the acid cleaned sample (Figure S2).

Finally, we used AFM to survey the surface morphology. AFM scans of the ‘as grown’ plasma surfaces showed rough morphological features on the micron scale but were smooth on the tens to hundreds of nanometers scale (Figure S3). We found that the morphology after the ‘gentle’ plasma varied dramatically between different termination batches. In some cases, we observed no changes in morphology, but other times, we observed mild to severe pitting (Figure S4). We attribute these differences to slight variations in plasma or chamber conditions. So far, we have not observed any changes in morphology from annealing (Figure S5). Since maintaining a smooth surface is critical for coherent shallow NV centers, we found the hydrogen annealing approach to be most promising.

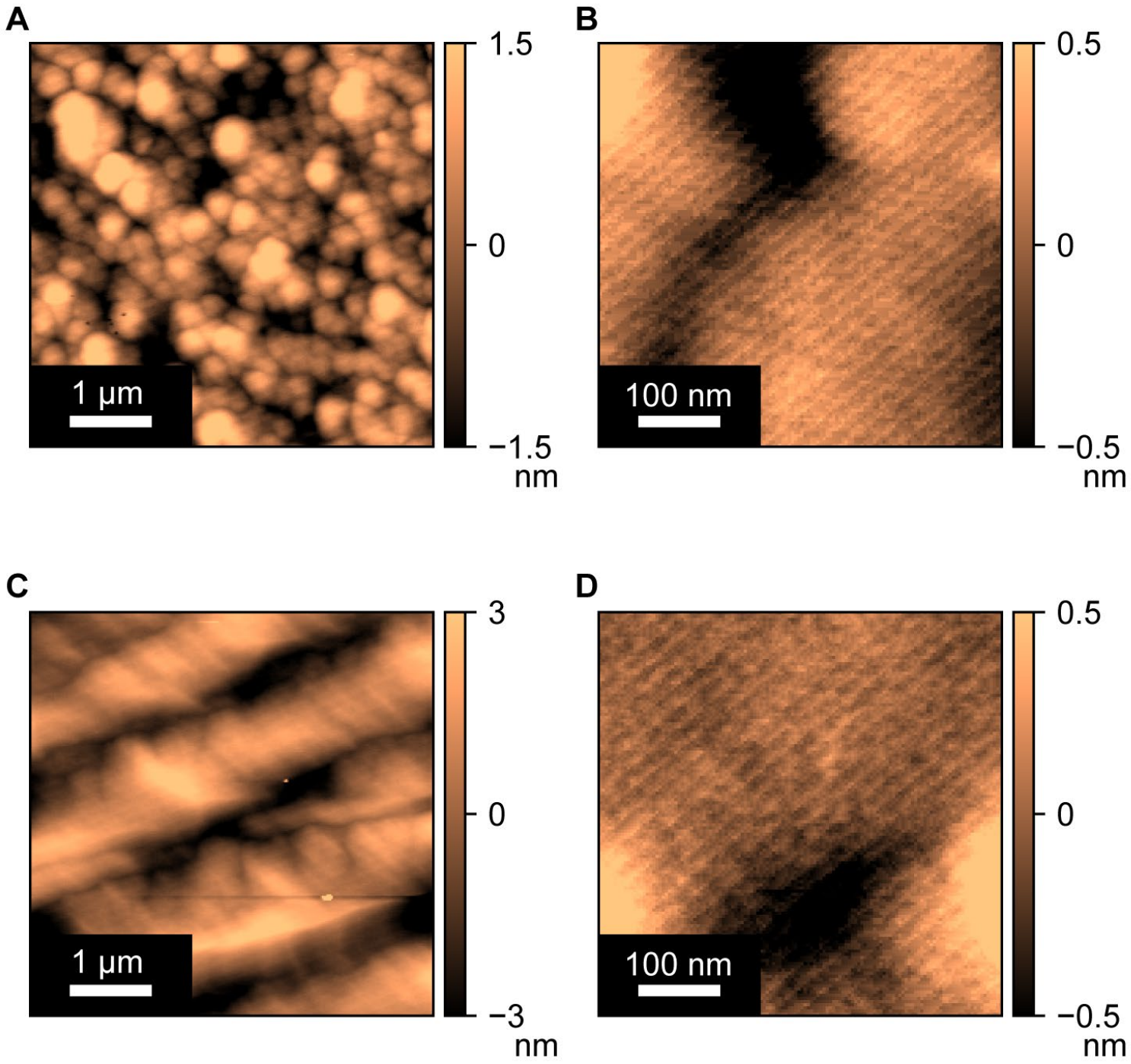

**Figure S3.** AFM images showing representative surface morphology for samples terminated with the ‘as grown’ plasma approach. (**A**) 5 µm x 5 µm scan showed large scale features, $R_a$ = 0.672 nm ± 0.905 nm. (**B**) 500 nm x 500 nm scan of the same region in (**A**) showed that the surface was smooth on the tens to hundreds of nm scale, $R_a$ = 0.220 nm ± 0.092 nm. (**C**) A 5 µm x 5 µm scan for a second sample, $R_a$ = 1.256 nm ± 0.096 nm. (**D**) 500 nm x 500 nm scan for the same region shown in (**C**), $R_a$ = 0.177 nm ± 0.109 nm.

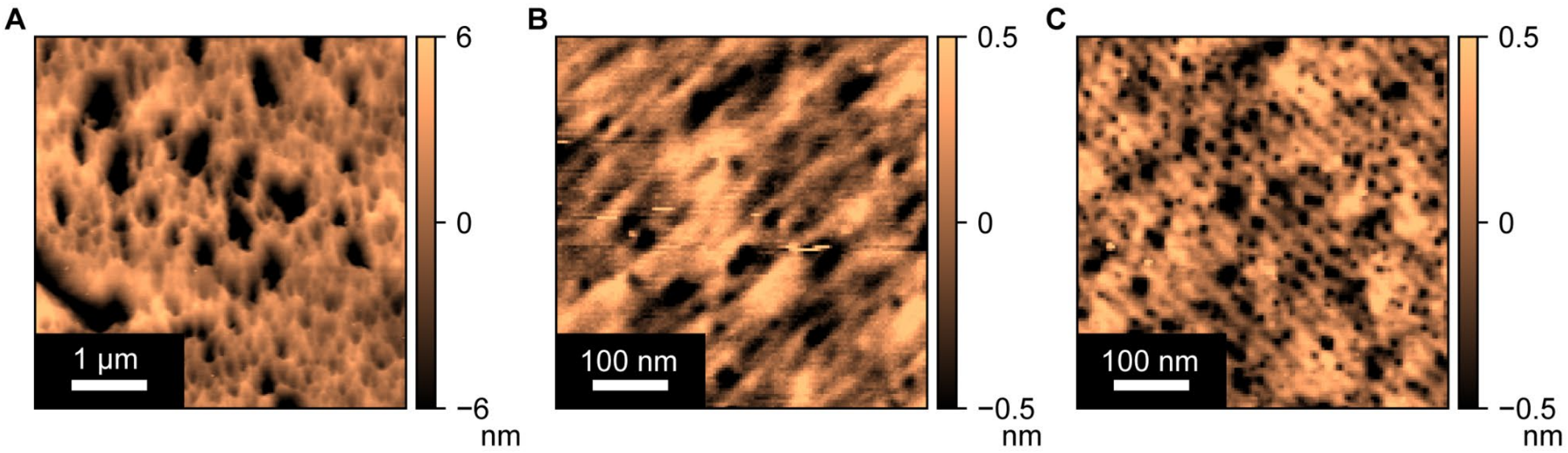

**Figure S4.** AFM images showing surface roughening after hydrogen termination with plasma. Each AFM image is from a different sample that was subjected to separate rounds of 'gentle' plasma termination. (**A**) Sample showed severe surface roughening after hydrogen termination. $R_a$ = 2.124 nm ± 0.305 nm. (**B**) Sample showed mild roughening after hydrogen termination. $R_a$ = 0.217 nm ± 0.014 nm. (**C**) NV sample that showed mild pitting after hydrogen termination. We were unable to observe OD-ESR contrast on this sample after functionalization. $R_a$ = 0.214 nm ± 0.005 nm.

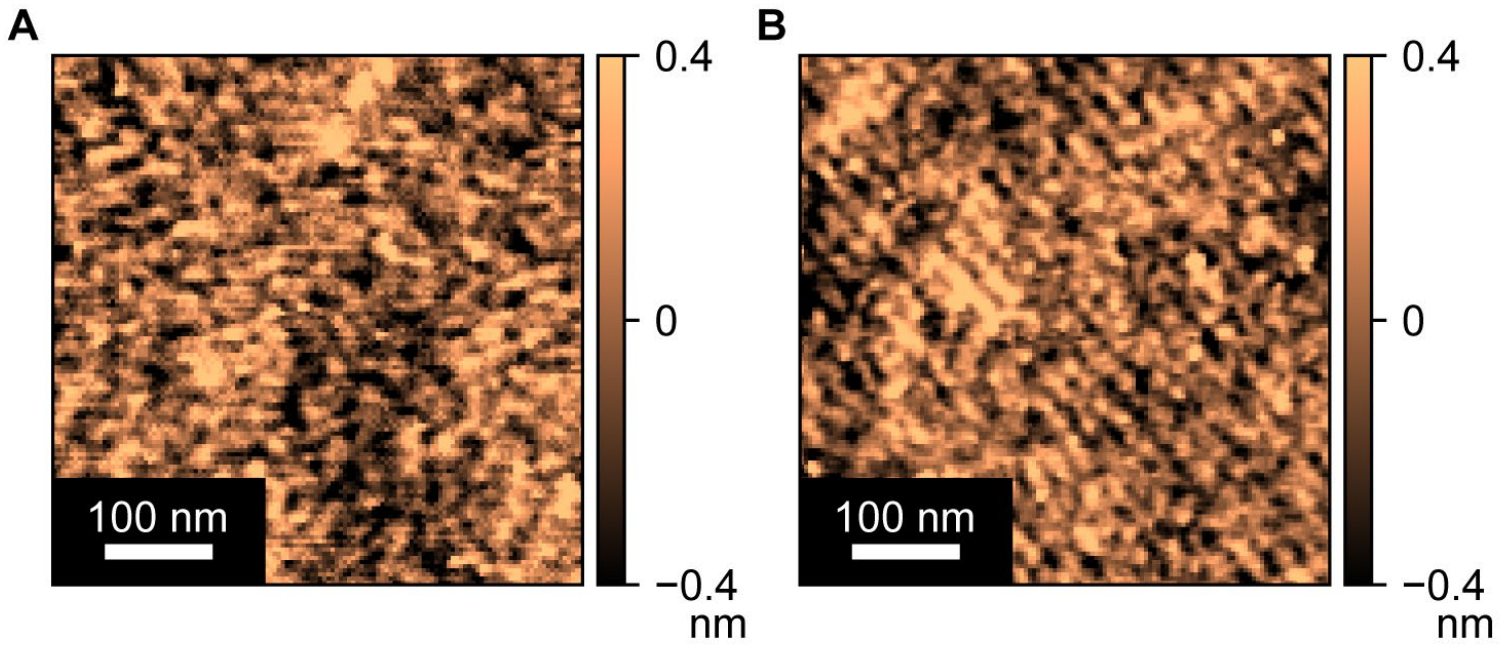

**Figure S5.** AFM images of a diamond surface before (**A**) and after (**B**) hydrogen terminating by annealing in forming gas. We observe no additional roughening or changes in morphology. $R_a$ values for left and right images are 0.182 nm ± 0.005 nm and 0.170 nm ± 0.015 nm, respectively. Diagonal lines visible in the right image are imaging artifacts from the instrument which change orientation when scan direction changes.

***Application to shallow NV experiments***

While the effectiveness of the functionalization reactions did not depend on the surface termination procedure, these techniques have dramatic implications for shallow NV centers. The 'as grown' termination procedure etches and regrows several microns of surface material. In this work, we created shallow NV centers prior to hydrogen termination, and these shallow NV centers would be etched away by this 'as grown' termination procedure. We did some preliminary tests with samples that were hydrogen terminated by the 'gentle' plasma approach. On one sample with deeper NV centers (50 keV implant), we were able to recover coherent NV centers with observable OD-ESR contrast and reasonable spin coherence times after functionalization. On shallower samples terminated in a different plasma run, we were able to observe bright fluorescent spots in a confocal microscope image, consistent with $NV^-$ centers, but were unable to observe OD-ESR contrast. We observed sporadic uncontrolled roughening from the 'gentle' plasma procedure (Figure S4), which indicates surface damage. We suspect that the observed surface damage combined with known issues with hydrogen diffusion[6] makes this procedure less amenable to shallow NV applications. By contrast, we found that hydrogen termination via the annealing procedure in forming gas was most promising for shallow NV applications, likely because it did not lead to any noticeable changes in morphology (Figure S5). An additional advantage of this annealing procedure is that it is the most technologically accessible, since it does not require plasmas generated in CVD growth chambers.

## III. Cleaning procedures for post-functionalization diamond samples

A key challenge with using XPS for reaction discovery is that insufficient cleaning after the reaction can lead to spurious XPS results arising from nonspecific binding (Figure S6). We addressed this issue by developing a thorough cleaning procedure to remove contamination while leaving surface-bonded functional groups intact. Typically, we first sonicated samples in the solvent of the reaction, heated them to 50 °C in dimethyl sulfoxide (DMSO) for 1 h, then sonicated them for 20 min in each of the following: *N*-methylpyrrolidone (NMP) or DMSO, chloroform, deionized (DI) water, acetone, and isopropyl alcohol (IPA). This solvent sequence was adjusted based on the stability of various functional groups. We then used XPS to check for the presence of the heteroatom associated with a particular functional group and the absence of other contaminants from the reaction (Figure S7, Figure S8). Additionally, since physisorbed contaminants rarely result in an evenly distributed monolayer, we used AFM to survey dried contamination on the surface (Figure S9). We routinely repeated the cleaning procedure and surface analysis to check for consistency in the XPS results through several cleans.

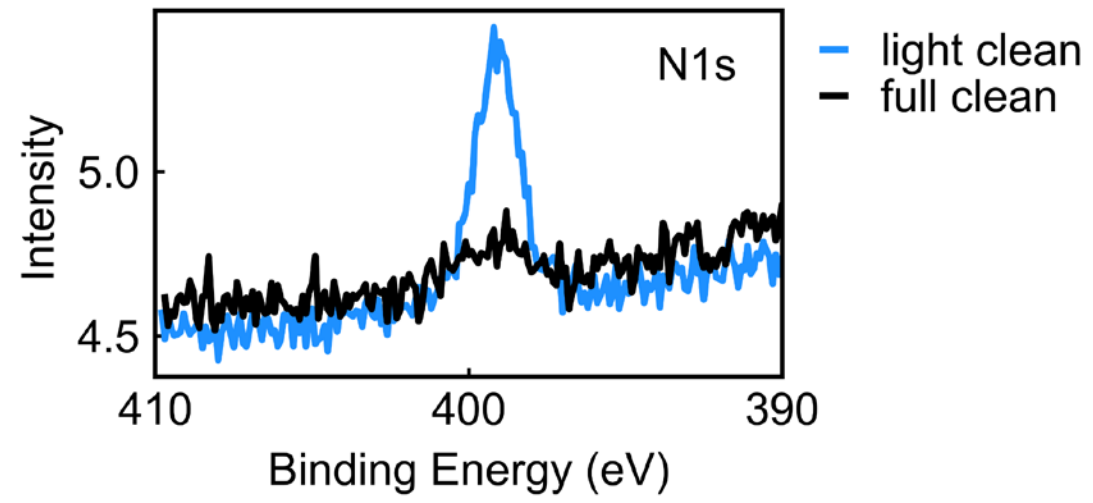

**Figure S6.** Example of a spurious XPS peak resulting from insufficient cleaning after the reaction. The post-reaction sample was cleaned by sonication for 10 minutes in DI water, acetone, and IPA, after which a clear N *1s* peak in XPS was observed (light clean). However, after additional cleaning by sonication for 20 min in additional solvents (NMP, chloroform, DI water, acetone, IPA), the peak disappeared (full clean).

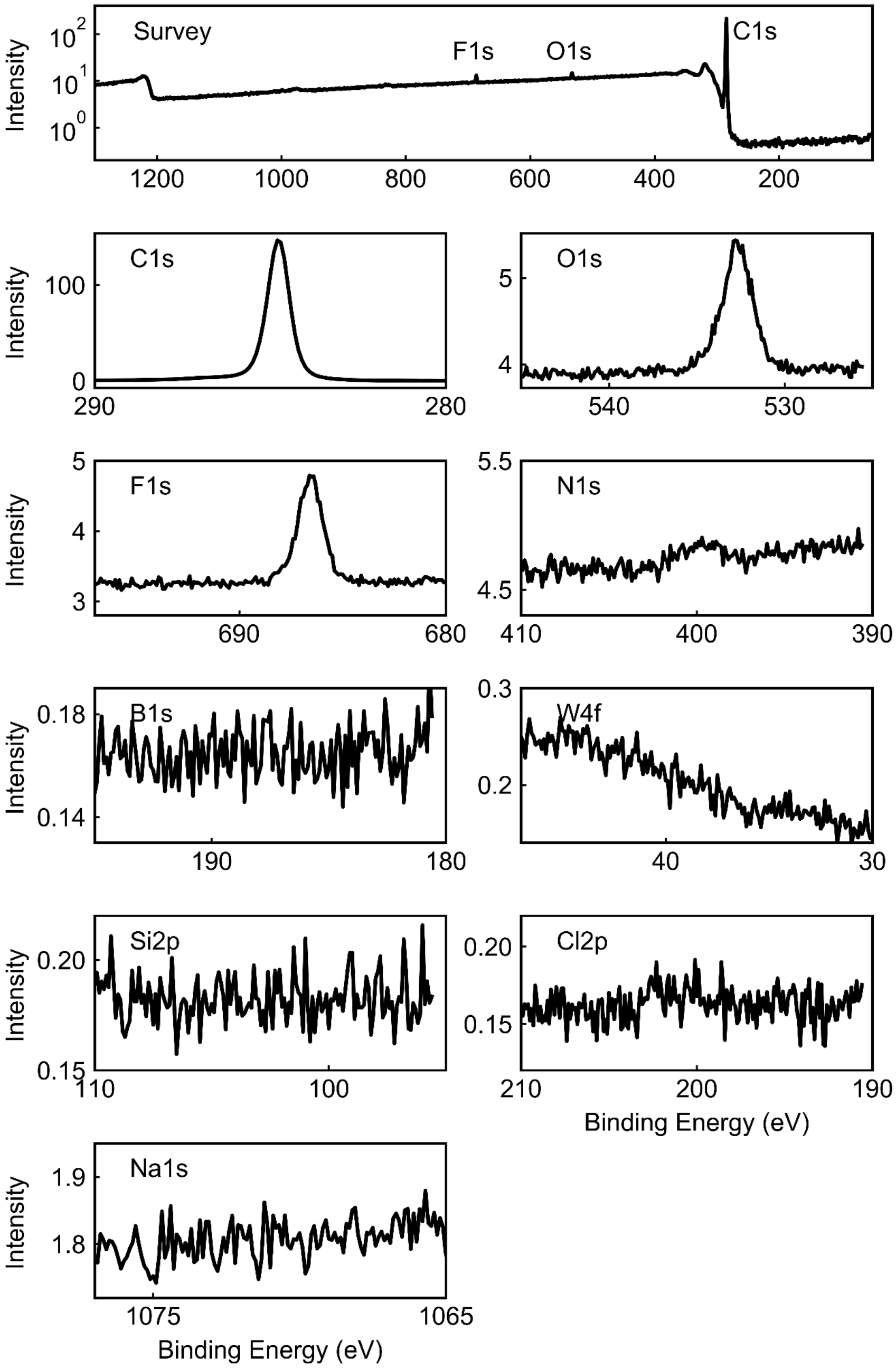

**Figure S7.** Full XPS spectra of a diamond surface after the photochemical fluorination reaction with Selectfluor. These spectra showed that neither impurities from reactions (B *1s*, W *4f*, N *1s*) nor common impurities from handling and packaging (Si *2p*, Cl *2p*, Na *1s*) were present on the surface, confirming the effectiveness of our cleaning procedure to completely remove any physisorbed contaminants remaining on the surface at the end of the reaction.

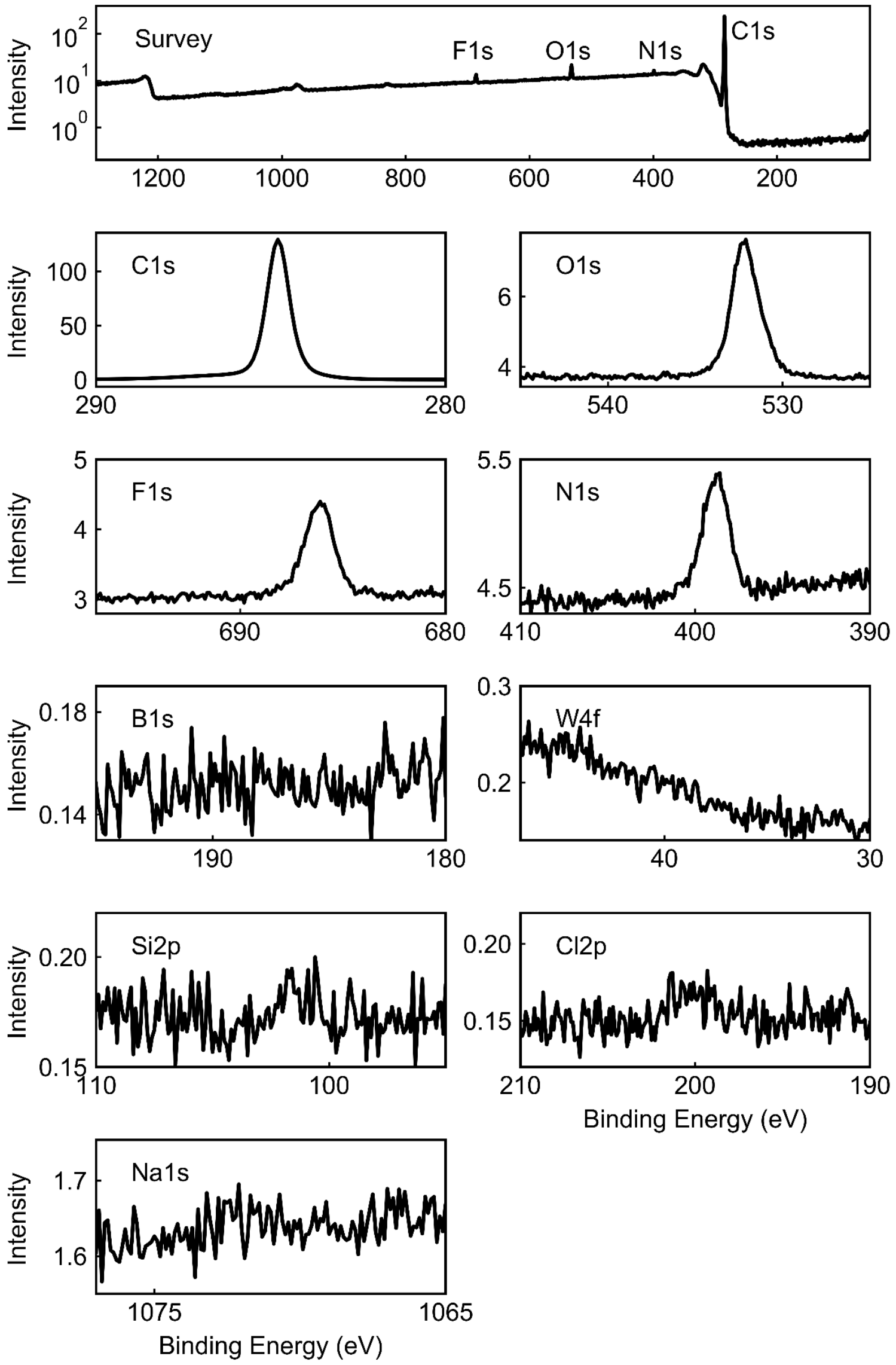

**Figure S8.** Full XPS spectra of a diamond surface after the photochemical fluorination and amidation reactions with Selectfluor. Neither impurities from reactions (B *1s*, W *4f*) nor common impurities from the handling process (Si *2p*, Cl *2p*, Na *1s*) were present on the surface.

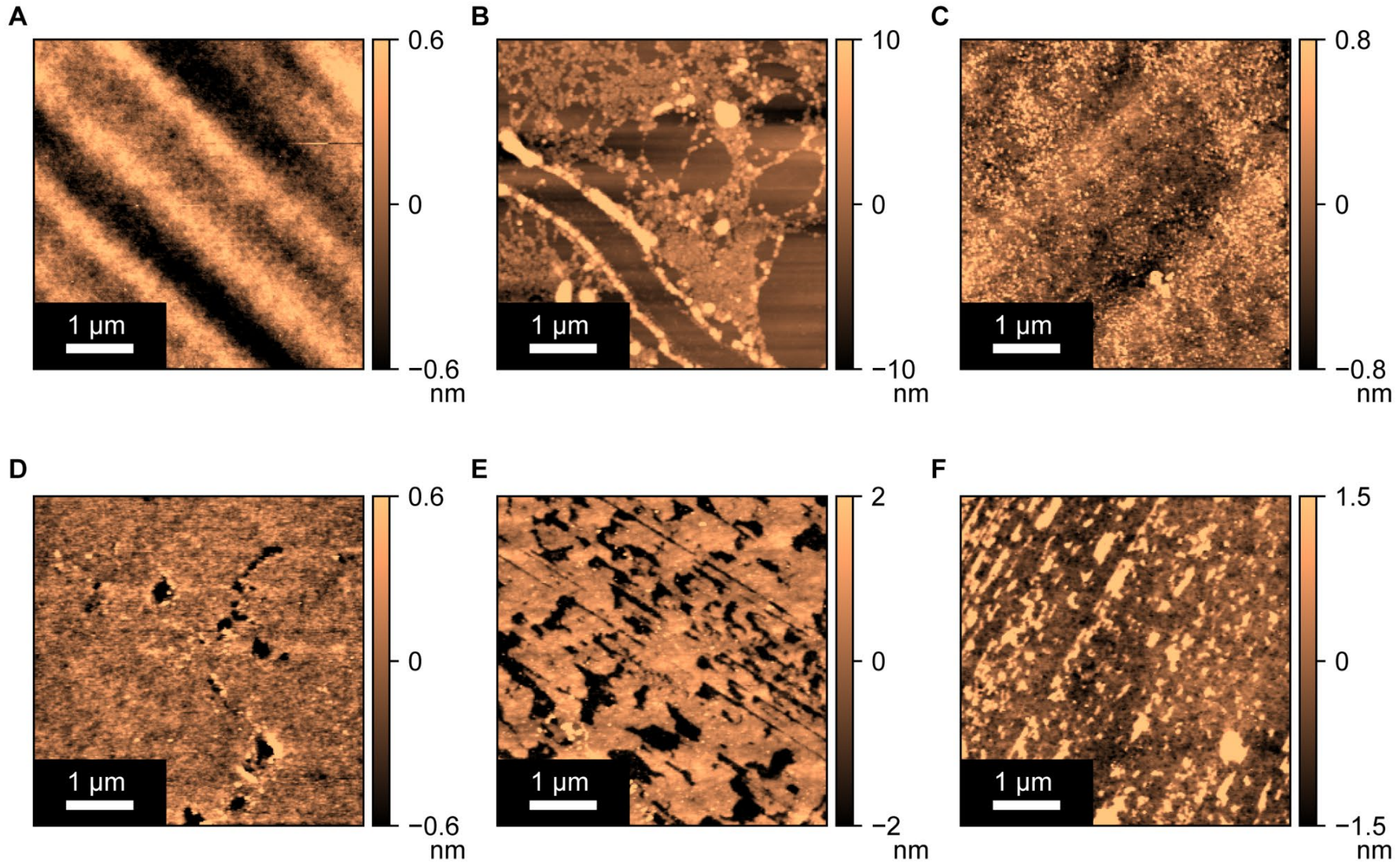

**Figure S9.** AFM images showing physisorbed contamination on diamond surfaces. (**A**) An example of a clean surface. (**B–F**) Examples of physisorbed contamination on insufficiently cleaned post-functionalization samples.

This rapid feedback approach using lab-based tools allowed us to screen for promising reactions. However, the spectral resolution of XPS is insufficient to provide chemical shift information to prove bond formation. To confirm our results, we performed near edge X-ray absorption fine structure (NEXAFS) spectroscopy at a synchrotron. With NEXAFS, we looked for an angle dependence at the edge associated with the relevant heteroatom (see 'Surface functionalization reactions' section for more detailed results). We interpret angle dependence to indicate that functional groups are well-oriented on the surface. This is a signature of covalent attachment and is inconsistent with physisorbed contaminants.

## IV. Surface functionalization reactions

### *General information*

Purple Kessil lamps used in this work are Kessil H150B LED Grow Light (390 and 456 nm) purchased from Kessil. All reactions were carried out in well-ventilated fume hoods. Gas chromatography (GC) was performed on an Agilent Technologies 7890A GC system equipped with a split-mode capillary injection system and flame ionization detectors. *N*-(*tert*-butyl)-*N*-chloro-3,5-bis(trifluoromethyl)benzamide and *N*-(*tert*-butyl)-*N*- ((ethoxycarbonothioyl)thio)-3,5-bis(trifluoromethyl)benzamide reagents that were used in chlorination and xanthylation reactions were synthesized according to previous reports[7,8]. Other reagents and solvents are commercially available.

### *Photochemical fluorination reaction*

*Proposed mechanism*: Under 390 nm light irradiation, the decatungstate anion ($[W_{10}O_{32}]^{4-}$) is photoexcited to a triplet excited state that possesses partial alkoxy radical character[9]. These electrophilic radical species can efficiently abstract hydrogen atoms from C–H bonds on the diamond surface to form alkyl radicals. The resulting alkyl radical can then intercept Selectfluor or NFSI and form a new C–F bond. After HAT, the reduced decatungstate anion ($[W_{10}O_{32}]^{-5}H^{+}$) can undergo disproportionation to afford the original oxidation state ($[W_{10}O_{32}]^{4-}$ and doubly-reduced $[W_{10}O_{32}]^{6-}2H^{+}$, which can undergo two sequential oxidations (possibly mediated by Selectfluor) to return to the $[W_{10}O_{32}]^{4-}$ state. Under these conditions, it is also possible that the aminium radical cation generated from Selectfluor or the disulfonamidyl radical generated from NFSI can mediate the abstraction of hydrogen atoms from the surface in a manner analogous to photoexcited decatungstate (Figure S17 below).

*General procedure*: A hydrogen-terminated, single-crystal diamond sample (2 mm x 2 mm) was added to a 1-dram vial along with a magnetic stir bar, tetra-*n*-butylammonium decatungstate (10 mg), and the fluorinating reagent (e.g., *N*-fluorobenzenesulfonimide (NFSI) or Selectfluor) (100 mg). The reaction mixture was degassed and backfilled with argon three times, after which 0.5 mL of dry acetonitrile was added. The reaction was then stirred at room temperature for 48 h under the irradiation of 390 nm purple Kessil lamps and kept near room temperature using fans. At the end of the reaction, the diamond sample was thoroughly washed using the cleaning procedure described above and dried using a nitrogen spray gun before characterization.

*Note on the fluorination reaction optimization study*

We found that both NFSI and Selectfluor are equally effective fluorinating reagents for diamond surfaces. We observed clear F *1s* peaks with XPS for samples fluorinated with either reagent (Figure S11A). Similar to the F *K*-edge NEXAFS angle dependence shown in the main text (Fig. 2D) for a sample fluorinated with NFSI, we observed a comparable angle dependence for a sample fluorinated with Selectfluor (Figure S10). The energy of the main $\sigma^*$ feature in the F *K*-edge NEXAFS data around 689 eV agrees with reported spectra for aliphatic fluorides[10]. In addition to Selectfluor and NFSI, we also tested other fluorinating reagents, including 1-fluoro-pyridinium tetrafluoroborate and 1-fluoro-3,3-dimethylbenziodoxole (fluoroiodane). We found that these reagents were also effective, albeit in slightly lower efficiencies than Selectfluor and NFSI (Figure S11A). We also tested the analogous thermal fluorination reaction involving potassium persulfate and Selectfluor II in a solution of aqueous acetonitrile at 50 °C, which was previously shown to be effective for the fluorination reaction of adamantane.[11] Interestingly, this thermal condition was completely ineffective at fluorinating our diamond surfaces, as shown by the absence of the fluorine peak F *1s* by XPS (Figure S11B).

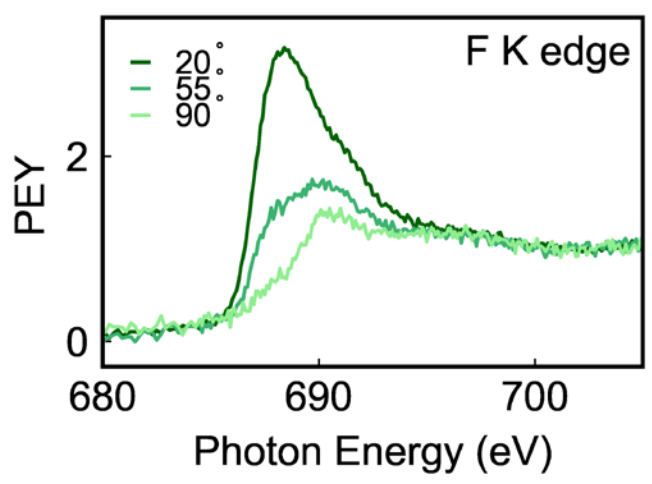

**Figure S10.** F *K*-edge NEXAFS angle dependence for a sample fluorinated with Selectfluor. F *K*-edge angle dependence for a sample fluorinated with NFSI is shown in the main text (Fig. 2D).

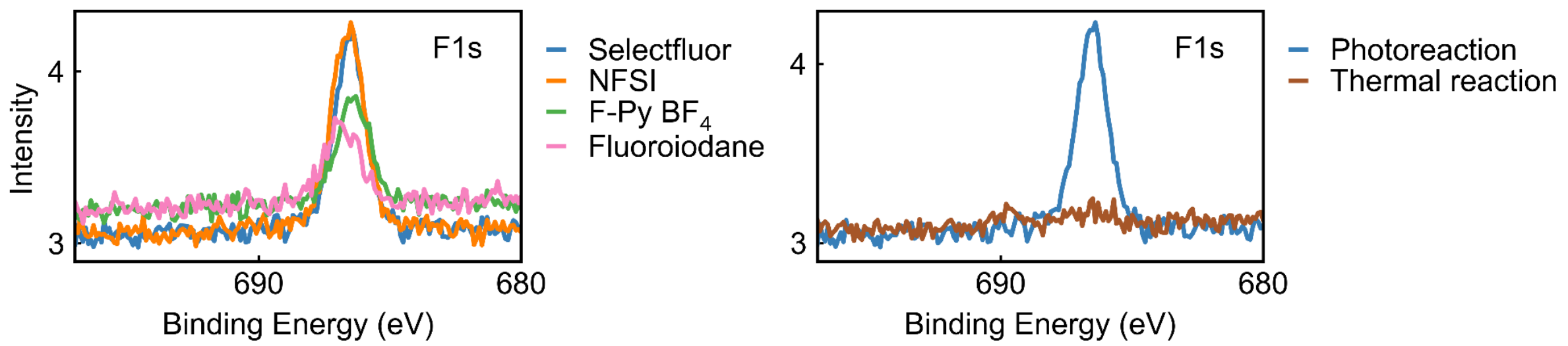

**Figure S11.** F *1s* XPS spectra of diamond surfaces after the fluorination reaction using different conditions. (**A**) Photochemical reactions with different fluorinating reagents: Selectfluor (blue), NFSI (orange), 1-fluoro-pyridinium tetrafluoroborate (F-Py $BF_4$) (green), and 1-fluoro-3,3-dimethylbenziodoxole (fluoroiodane, pink). (**B**) Comparison between (1) photochemical fluorination using TBADT and Selectfluor (blue) and (2) thermal fluorination (brown) using potassium persulfate (27 mg) and Selectfluor II (80 mg) in a solution of acetonitrile and water (0.5 mL, 3/2 v/v) at 50 °C[11].

*Evaluating the reactivity of surfaces prepared from different hydrogen termination procedures*

We used the fluorination reaction to benchmark the reactivity of samples prepared using different methods of hydrogen termination. We observed clear and consistent F *1s* signals in XPS for all samples regardless of hydrogen termination procedure (Figure S12). This suggests that the chemistry described in this work is robust to different hydrogen termination surface preparation procedures.

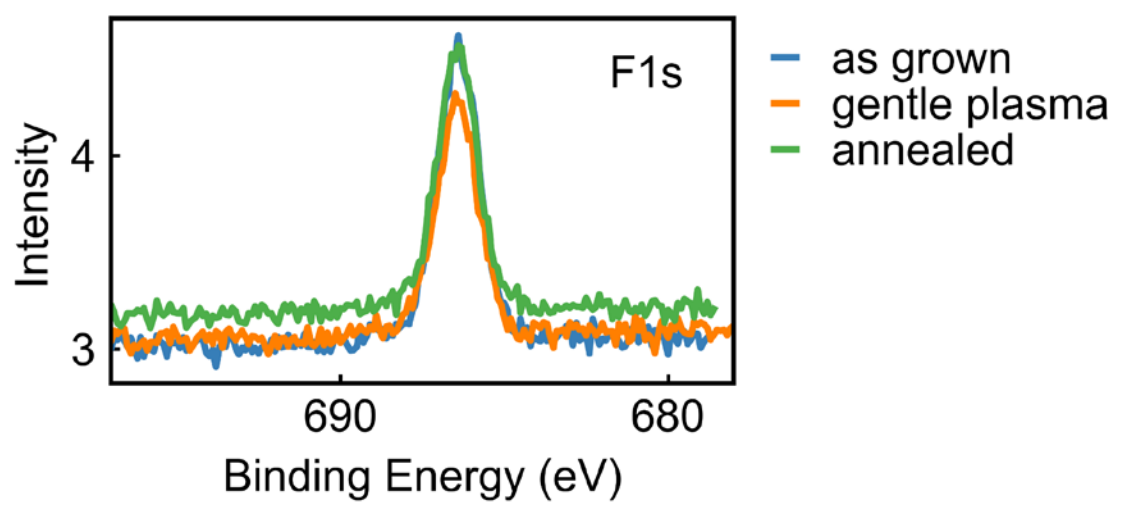

**Figure S12.** XPS F *1s* spectra after the fluorination reaction on samples prepared from three different methods of hydrogen termination. The resulting fluorine signals were consistent across the three samples, indicating that the functionalization protocol is robust to different methods of hydrogen termination.

***Chlorination and xanthylation reaction***

*Proposed mechanism*: In analogy to reports on small molecules[7,8,12], we propose that the N–Cl bond in *N*-(*tert*-butyl)-*N*-chloro-3,5-bis(trifluoromethyl)benzamide **1** and the N–S bond in *N*-(*tert*-butyl)-*N*-((ethoxycarbonothioyl)thio)-3,5-bis(trifluoromethyl)benzamide **2** can be homolyzed under 456 nm irradiation, forming the same electrophilic amidyl radical intermediate in both cases. This species can abstract a hydrogen atom from the diamond surface to afford an alkyl radical, which can react with another equivalent of **1** or **2** to install, respectively, a chlorine or a xanthate group on the surface and generate another reactive amidyl radical (Figure S13). This chain process then propagates independently of light.

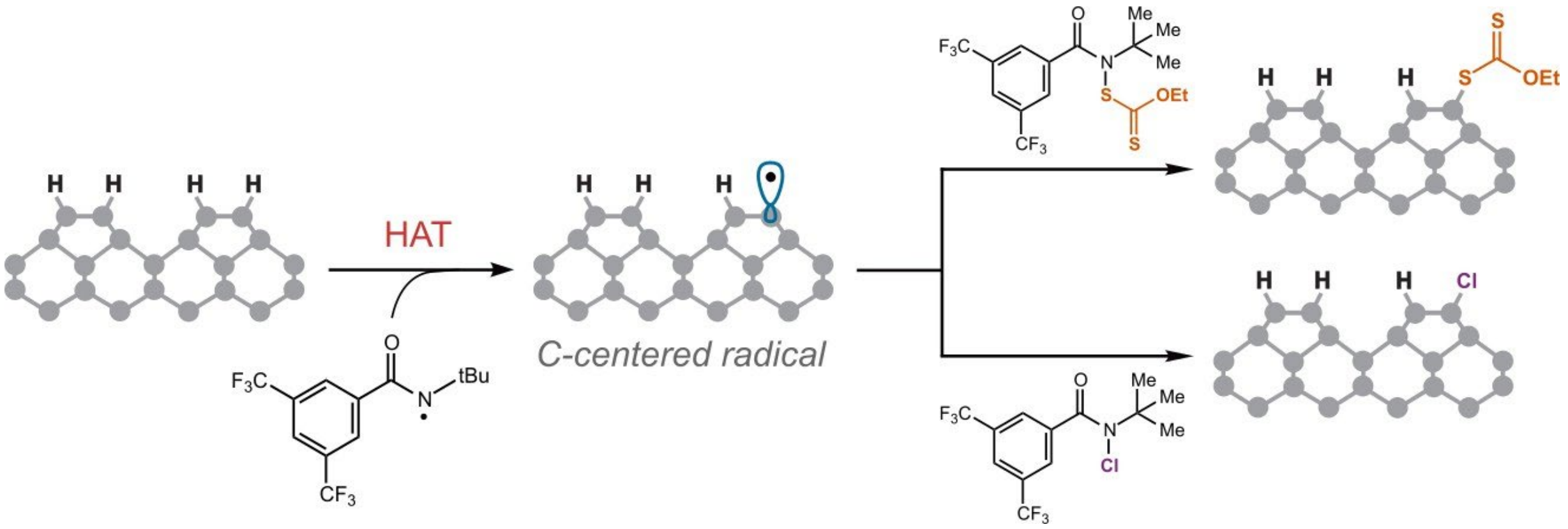

**Figure S13.** Proposed mechanism of chlorination and xanthylation reactions.

*Procedure for chlorination reaction*: A hydrogen-terminated, single-crystal diamond was added to a 2-dram vial along with a magnetic stir bar, *N*-(*tert*-butyl)-*N*-chloro-3,5-bis(trifluoromethyl)benzamide (50 mg), and cesium carbonate (47 mg). The reaction mixture was degassed and backfilled with argon three times, after which 0.4 mL of anhydrous benzene was added. The reaction was then stirred for 24 h under the irradiation of 456 nm blue Kessil lamps. The reaction temperature was maintained at approximately 65 °C over the course of the reaction by the heat produced by the lamps; no fans were used. At the end of the reaction, the diamond sample was thoroughly washed using the cleaning procedure described above and dried using a nitrogen spray gun before characterization.

*NEXAFS Cl K-edge studies:* We observed a possible angle dependence at the Cl *K*-edge for these samples, although issues with X-ray beam induced damage (Figure S14) and sample contamination were confounding factors for this measurement (Figure S15).

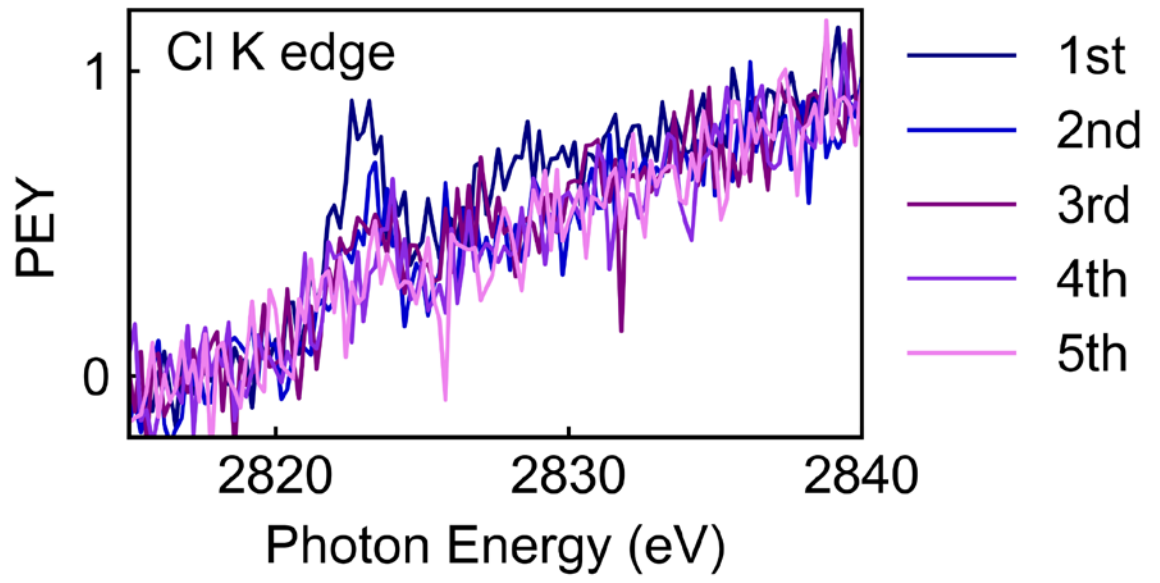

**Figure S14.** Observation of beam damage on sample bearing trichloroacetamide groups. Cl *K*-edge signal was reduced upon repeated measurements of the same spot.

We attempted to use NEXAFS to study samples with chlorine-containing functional groups. The main challenge we encountered was beam damage, since the Cl *K*-edge requires substantially higher beam energy than the F, N, C, or O *K*-edges. Evidence of beam damage for a sample bearing trichloroacetamide groups is shown in Figure S14. When the same spot is repeatedly scanned, the signal decreases. This required us to move to a 'fresh' region on the samples for each angle measurement. Since the beam spot size is on the order of 1 mm, we could only measure a few spots per sample.

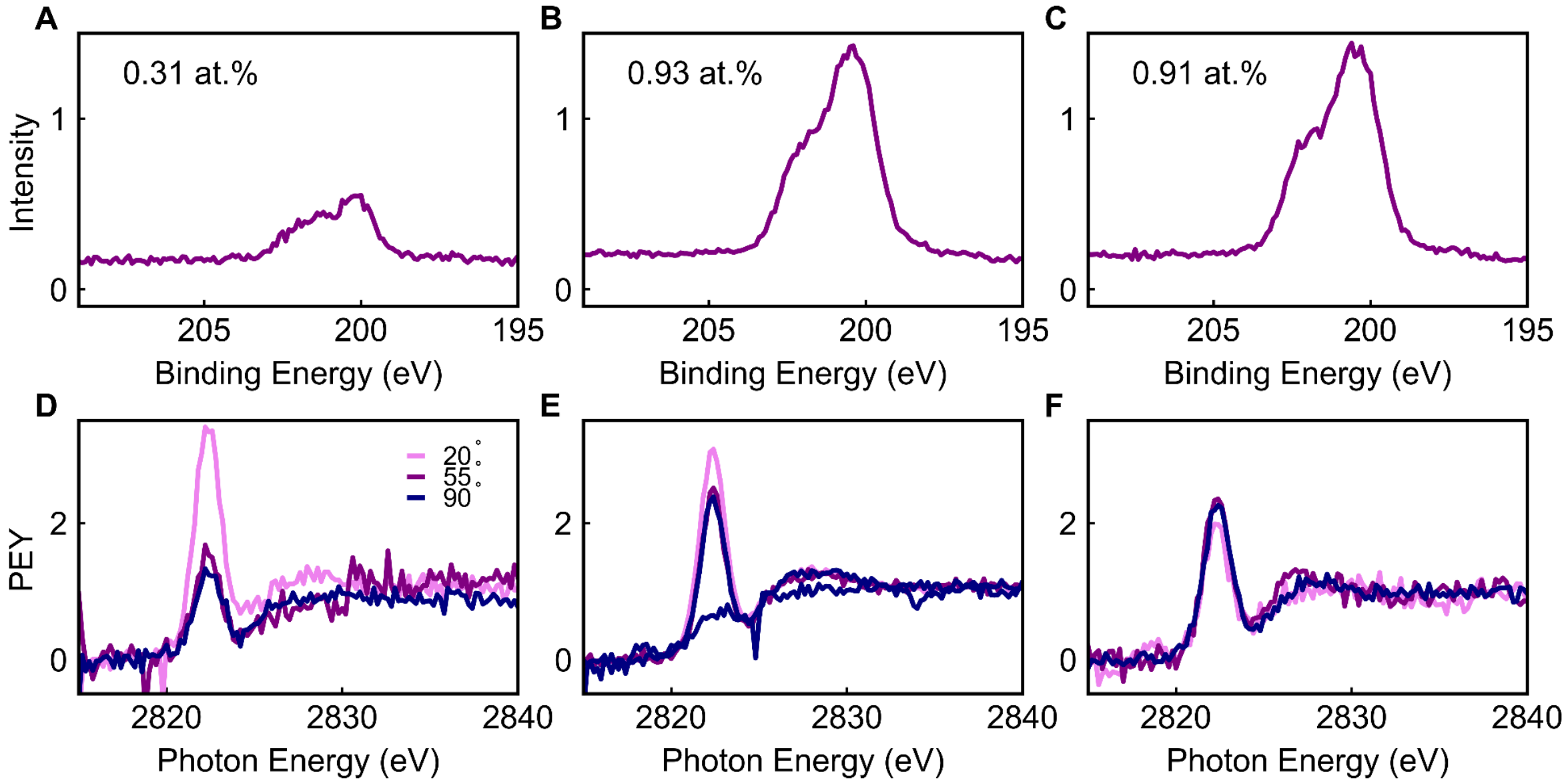

**Figure S15.** Further XPS and NEXAFS studies for direct chlorination reaction. (**A**–**C**) Cl *2p* XPS spectra for three samples. (**D**–**F**) NEXAFS Cl *K*-edge spectra for different angles for the same three samples. (**A**) and (**D**) are from the same sample, as are (**B**) and (**E**), and (**C**) and (**F**). The two dark blue lines shown in (**E**) show the same measurement on two distinct spots on the sample. Overall, these data present a possible angle dependence for the direct chlorination reaction and supports the hypothesis that an angle dependence can be used to distinguish between a well-ordered covalent bond and contamination from the reaction. Negative PEY points after normalization could possibly be due to a brief detector instability event.

With this in mind, we studied samples prepared with the direct chlorination reaction. We prepared three samples that showed dramatically different Cl *2p* signals in XPS (Figure S15A–C). The sample shown in (**A**) had the most 'typical' Cl *2p* signal, while the samples in (**B**) and (**C**) showed a Cl *2p* signal that was significantly higher than average. We performed Cl *K*-edge measurements on these same samples. In Figure S15, (**A**) and (**D**) correspond to the same sample, (**B**) and (**E**) are from the same sample, and (**C**) and (**F**) are from the same sample. For the sample in (**D**), we observed a possible angle dependence, however the difference between the 55° and 90° traces was not significant. For the sample shown in (**E**), we repeated the 90° measurement on two spots with dramatically different results. The 90° trace with the lowest amplitude for the π* feature around 2822 eV suggests an angle dependence. By contrast, the 90° trace with high amplitude was not consistent with an angle dependence and could be explained by contaminants from the reaction. The sample shown in (**F**) did not show an angle dependence. Taken together, these data are consistent with the picture that these samples had residual Cl contamination due to insufficient cleaning after the reaction and that contamination can be inhomogeneously distributed across the surface, resulting in no angle dependence. For example, the third sample in (**C**) and (**F**) showed a higher than average Cl *2p* signal and no angle dependence, both consistent with contamination. For the second sample, the dramatic change in 90° signal for two spots is consistent with contamination unevenly distributed across the surface.

Since we observed indications of an angle dependence, we present this as further evidence that the direct chlorination reaction was successful. Furthermore, these results support our statement that an angle dependence results from a well-ordered covalent bond, whereas contamination on the surface should not result in an angle dependence.

*Procedure for xanthylation reaction:* A hydrogen-terminated, single-crystal diamond was added to a 2-dram vial along with a magnetic stir bar and *N*-(*tert*-butyl)-*N*-((ethoxycarbonothioyl)thio)-3,5-bis(trifluoromethyl)benzamide (170 mg). The reaction mixture was degassed and backfilled with argon three times, after which 0.4 mL of anhydrous trifluorotoluene was added. The reaction was then stirred at room temperature for 24 h under the irradiation of 456 nm blue Kessil lamps. The reaction temperature was maintained near room temperature over the course of the reaction using fans. At the end of the reaction, the diamond sample was thoroughly washed using the cleaning procedure described above and dried using a nitrogen spray gun before characterization.

***Photochemical C–N bond formation reaction***

*General procedure*: A hydrogen-terminated, single-crystal diamond sample (2 mm x 2 mm) was added to a 1-dram vial along with a magnetic stir bar and Selectfluor (100 mg). The reaction mixture was degassed and backfilled with argon three times, after which 0.4 mL of degassed nitrile reagent* was added. The reaction was then stirred at room temperature for 48 h under the irradiation of 390 nm purple Kessil lamps and kept near room temperature using fans over the course of the reaction. At the end of the reaction, the diamond sample was thoroughly washed using the cleaning procedure described above and dried using a nitrogen spray gun before characterization.

**Note*: examples of nitrile reagents include acetonitrile, trichloroacetonitrile, and 3,3,3-trifluoropropanetrile. Full XPS results for a representative sample are shown in Figure S8.

*Note on using Selectfluor and NFSI in C–N bond formation reactions:* We found that both Selectfluor and NFSI were similarly effective reagents in the fluorination and amidation reaction. This was evidenced by the appearance of comparable F *1s* and N *1s* peaks for diamond samples subjected to amidation reactions with either reagent (Figure S16A and B). Additionally, we observed an angle dependence at the N *K*-edge for a sample subjected to the amidation reaction with NFSI (Figure S16C). The spectra shown in Figure S16C appears overall similar to the spectra for a sample subjected to the amidation reaction with Selectfluor shown in the main text (Figure 3A). We also observed an angle dependence at the F *K*-edge for samples subjected to amidation reactions with Selectfluor and NFSI (Figure S16D and E).

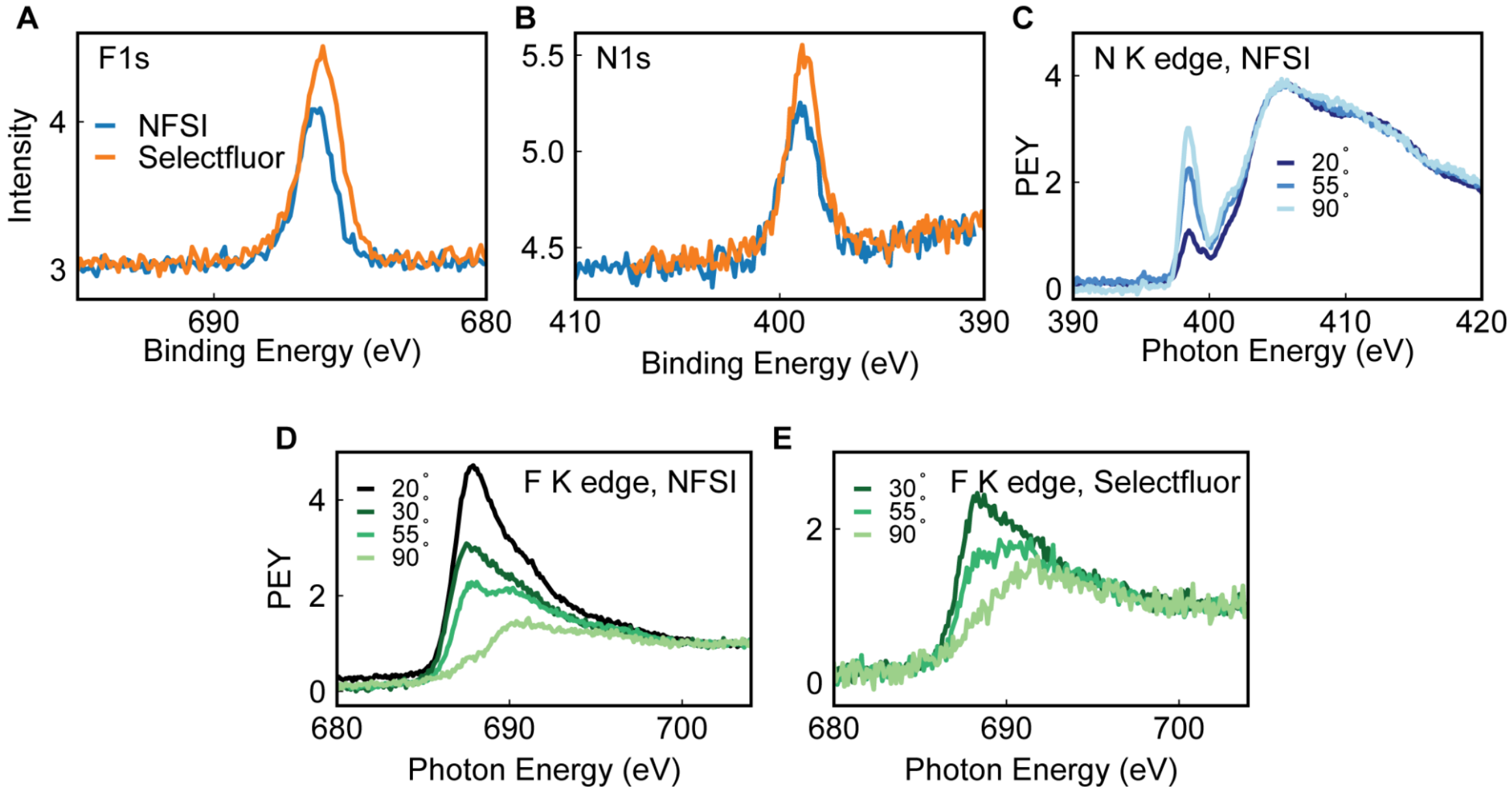

**Figure S16.** Comparison between Selectfluor and NFSI as fluorinating and amidating reagents. (**A**) and (**B**) show F *1s* and N *1s* spectra for samples subjected to the amidation reaction with NFSI and Selectfluor. (**C**) N *K*-edge NEXAFS angle dependance for a sample subjected to the amidation reaction with NFSI. (**D**) and (**E**) F *K*-edge NEXAFS angle dependence for samples subjected to amidation reactions with NFSI and Selectfluor, respectively.

***Control experiments to determine the identity of the nitrogen-containing groups formed on the surface***

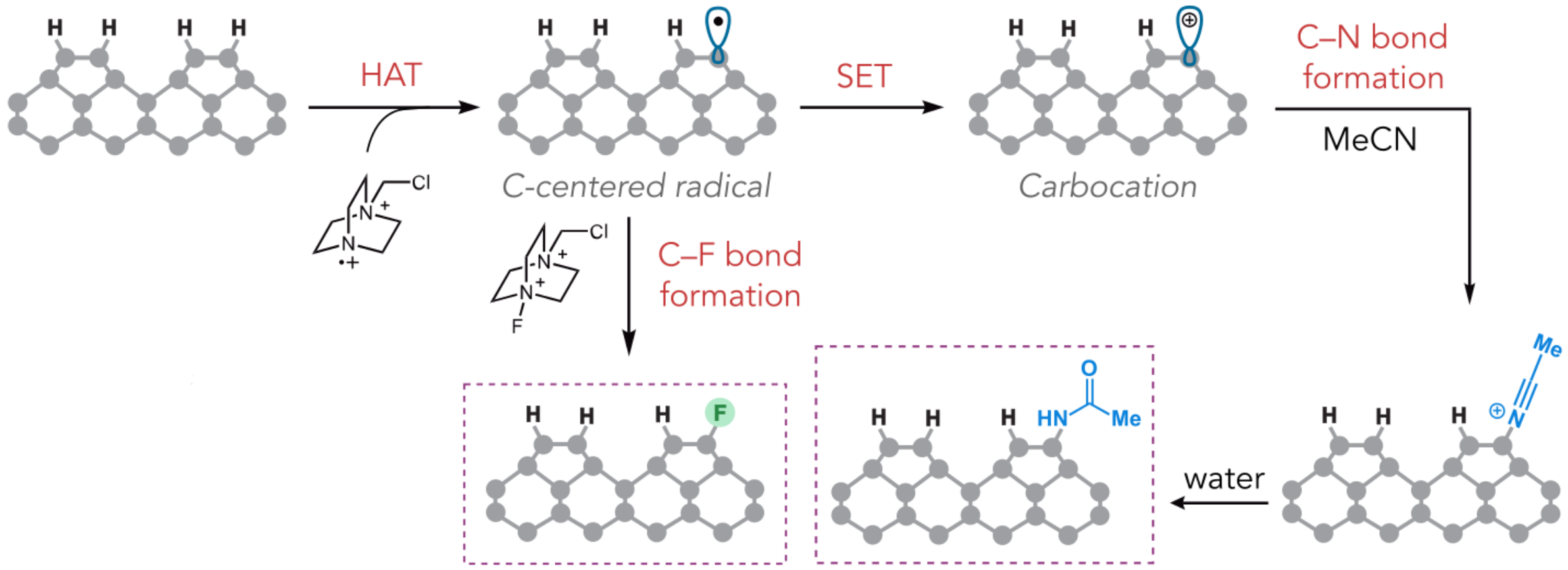

**Figure S17.** Proposed mechanism for C–N bond formation reaction through a Ritter-type amidation.

The proposed mechanism for the C–N bond formation reaction is described in Figure S17. We hypothesize that a reactive aminium radical cation (N–H BDFE approximately 100 kcal/mol) generated from Selectfluor through a chain mechanism is capable of abstracting C–H bonds on the surface to furnish alkyl radicals. These open-shell species can engage in two distinct processes: (1) intercepting with Selectfluor to form C–F bonds and (2) undergoing single electron transfer with Selectfluor to produce tertiary carbocations that are trapped by acetonitrile solvent to form C–N bonds via a Ritter mechanism.

To probe this proposed mechanism, we performed a series of control experiments. First, we posited that if this mechanism is operative, amide formation would not be observed in reactions where the solvent does not contain a nitrile group. Accordingly, we ran the same reaction using dichloroethane (DCE) solvent and indeed did not observe any N *1s* signal in XPS (Figure S18). In addition, we ran this photochemical reaction with adamantane as a model substrate. Similar to the reaction of the diamond surface, both 1-fluoroadamantane and 1-acetamidoadamantane were formed in 26% and 19% yield, respectively (quantified via gas chromatography) (Figure S20). The analogous reaction of adamantane with Selectfluor in acetonitrile at 80 °C was previously reported,[13] although these thermal conditions were not effective for the functionalization of diamond surface (Figure S19). By contrast, the photochemical Ritter amidation reaction using only Selectfluor and nitriles has not been reported for small molecule substrates. We attributed this unusual reactivity to the high stability of the carbocation intermediate on the surface, which has a sufficiently long lifetime to react with the solvent nucleophile but is less likely to engage in competing reactions commonly seen in small molecules (e.g., elimination reaction).

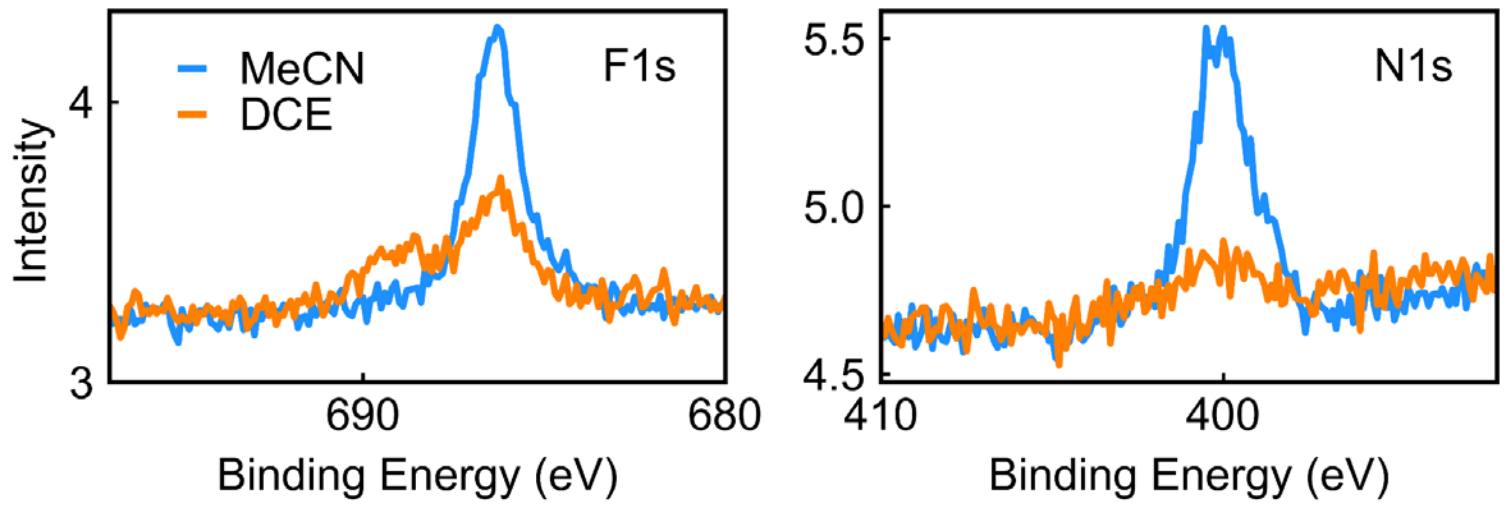

**Figure S18.** XPS spectra of diamond surfaces after fluorination and amidation reactions in acetonitrile (MeCN) (blue) and 1,2-dichloroethane (DCE) solvent (orange).

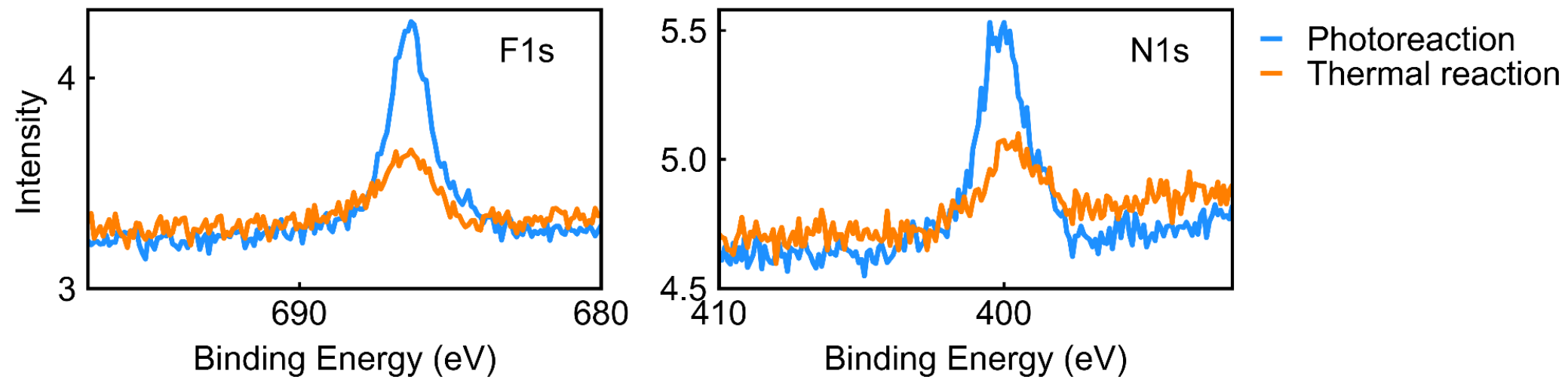

**Figure S19.** XPS spectra of diamond surfaces after fluorination and amidation photochemistry reactions (blue) and corresponding thermal reactions (orange).

*Control experiments with adamantane model substrate*

Photochemical reaction. To an oven-dried 2-dram vial was added adamantane (1.0 equiv, 0.05 mmol, 6.8 mg), Selectfluor (2.0 equiv, 0.1 mmol, 35.4 mg), and sodium bicarbonate (1.0 equiv, 0.05 mmol, 4.2 mg) (Figure S20). The mixture was then degassed and backfilled with Ar three times. Anhydrous acetonitrile (1 mL) was added to the mixture. The reaction was then stirred at room temperature for 24 h under the irradiation of 390 nm purple Kessil lamps and kept near room temperature using fans. At the end of the reaction, an internal standard (1-bromonaphthalene) was added to the reaction mixture. An aliquot of the reaction mixture was passed through a short silica plug and this solution was subjected to GC analysis to determine the yield of 1-fluoroadamantane (26%) and 1-acetyladamantane (19%).

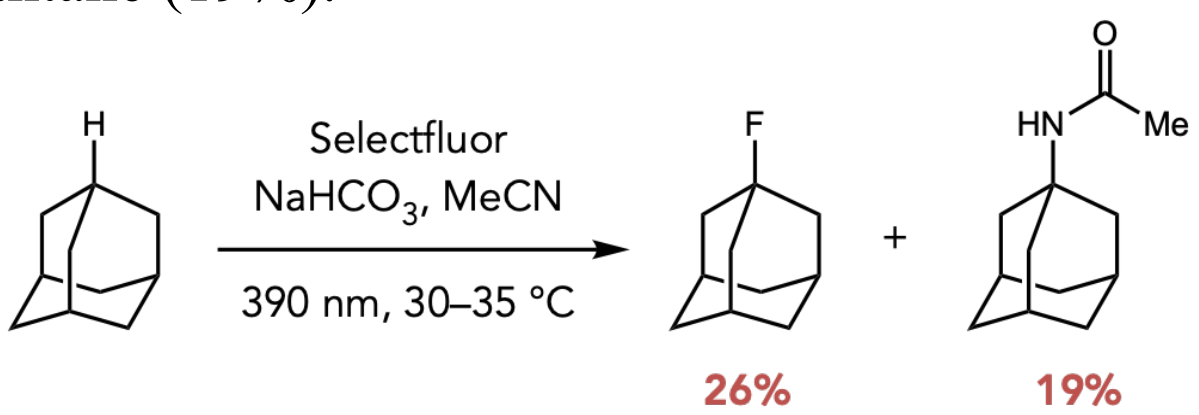

**Figure S20.** C–H functionalization reaction of adamantane via photochemistry.

Thermal reaction. To an oven-dried 2-dram vial was added adamantane (1.0 equiv, 0.05 mmol, 6.8 mg) and Selectfluor (2.0 equiv, 0.1 mmol, 35.4 mg) (Figure S21). The mixture was then degassed and backfilled with Ar three times. Anhydrous acetonitrile (1.0 mL) was added to the mixture. The reaction was then stirred at 80 °C for 24 h. At the end of the reaction, an internal standard (1-bromonaphthalene) was added to the reaction mixture. An aliquot of the reaction mixture was passed through a short silica plug and this solution was subjected to GC analysis to determine the yield of 1-fluoroadamantane (5%) and 1-acetyladamantane (6%). Note that these conditions were used according to a previously reported patent[13].

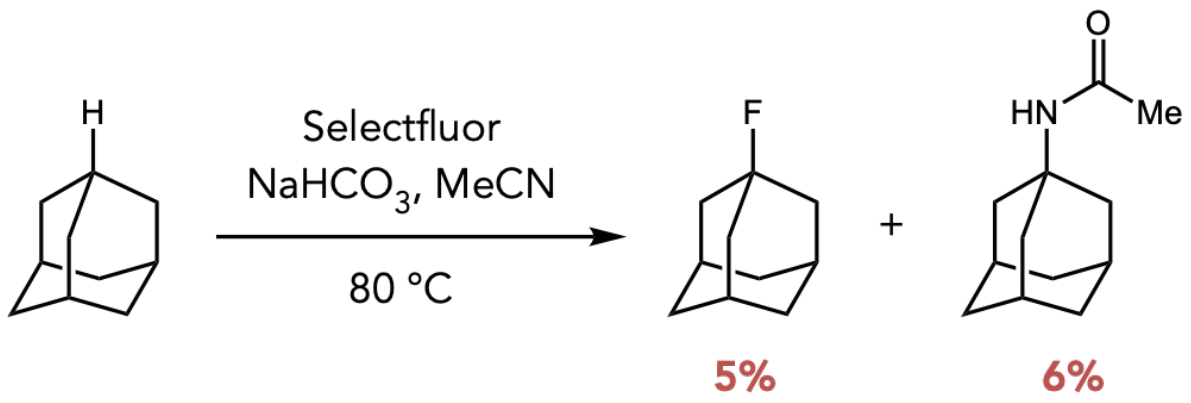

**Figure S21.** C–H functionalization reaction of adamantane under thermal conditions.

These results suggested that photochemical conditions were more efficient for the fluorination and amidation reaction of adamantane than previously reported thermal conditions (Figure S21). Similar results were also observed for these reactions on the diamond surface (Figure S19).

*Additional control experiments with the Ritter reaction using trichloroacetonitrile nucleophile*

In the Ritter reaction with trichloroacetonitrile, this reagent can potentially serve as a halogen atom transfer (XAT) reagent. More specifically, the tertiary carbon radical on the surface generated upon HAT with Selectfluor could potentially abstract the Cl atom from this solvent to form C–Cl bonds, where Cl atoms are directly bound to the surface. Thus, the appearance of a Cl peak in the XPS spectrum after the photoreaction (Figure 3B, main text) could result from two possible reactions: Ritter-type amidation and direct chlorination. To probe these mechanisms, we hydrolyzed the trichloroacetamide-terminated surface using an aqueous solution of potassium hydroxide. XPS measurements of the resulting surface showed a much smaller Cl *2p* signal while the N *1s* remained unchanged (Figure S22). This remaining Cl *2p* signal, which was attributed to the $C_{(diamond)}$–Cl bonds, indicated that the direct surface chlorination reaction might operate via XAT, but only to a limited extent. In addition, we performed another control experiment using a similar XAT reagent, ethyl trichloroacetate, which is unable to participate in the Ritter reaction and observed the formation of C–Cl bonds with surface coverage comparable to that measured after the hydrolysis reaction (Figure S22).

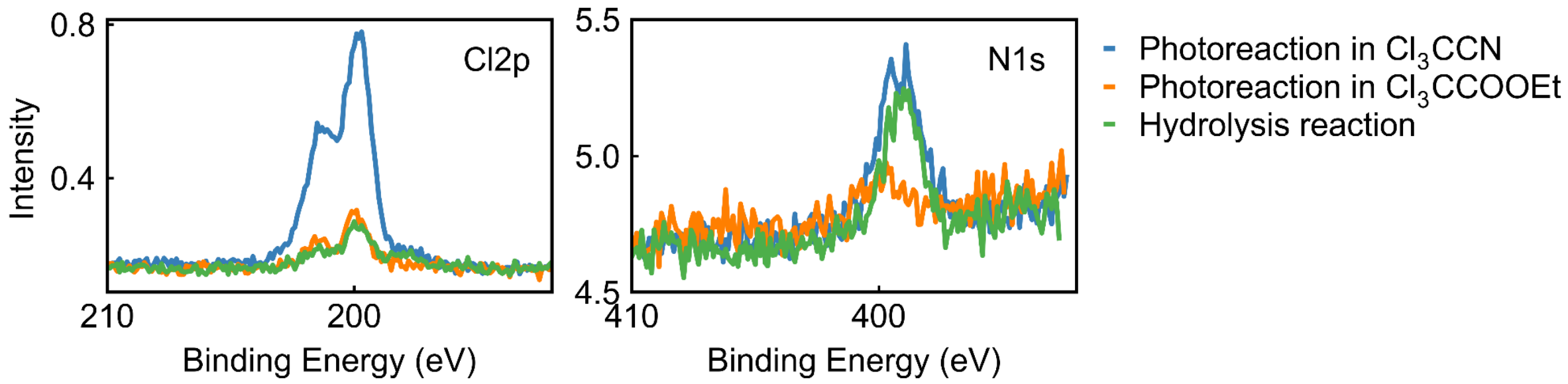

**Figure S22.** XPS spectra of diamond surfaces after (1) photoreaction with Selectluor and trichloroacetonitrile solvent (blue), (2) photoreaction with Selectfluor and ethyl trichloroacetate solvent (orange), and (3) 2-step sequence involving amidation reaction with Selectfluor and trichloroacetonitrile followed by hydrolysis under an aqueous solution of potassium hydroxide (green).

***General procedure for hydrolysis reaction***

A diamond sample with an amide-terminated surface, which was obtained after the Ritter reaction, was added to a 2-dram vial along with a magnetic stir bar and potassium hydroxide (56 mg). DI water (0.4 mL) was added and the reaction mixture was stirred under an inert atmosphere at 90 °C for 24 h. At the end of the reaction, the diamond sample was thoroughly washed using the cleaning procedure described above and dried using a nitrogen spray gun before characterization.

We used XPS to characterize our surfaces before and after hydrolysis. In a typical sample, as shown in Figure S23, we observed that the F *1s* and N *1s* peaks were unchanged before and after the hydrolysis.

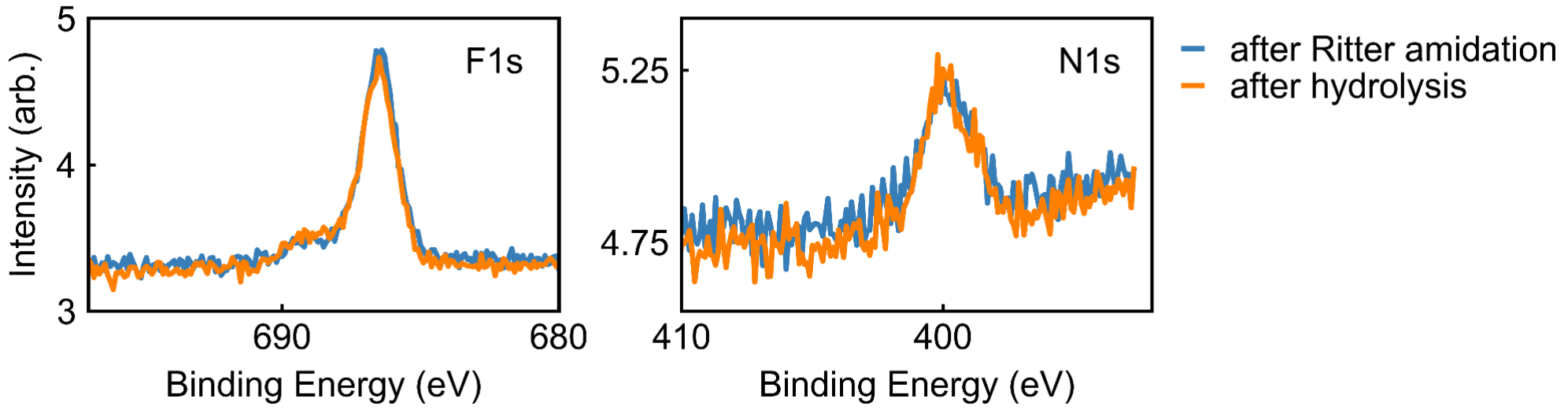

**Figure S23.** XPS spectra of the diamond surface after Ritter amidation to form amide (blue) and after hydrolysis to form amine (orange). Data show F *1s* and N *1s* peaks are unchanged.

***General procedure for acylation reaction***

A diamond sample with an amine-terminated surface, which was obtained after the hydrolysis reaction, was added to a 2-dram vial along with a magnetic stir bar. The vial was then degassed and backfilled with Ar three times. Next, anhydrous dichloromethane (0.3 mL), pyridine (30 μL), and acylation reagent (0.1 mL) were added. The vial was then sealed, placed under an inert atmosphere, and stirred at room temperature for 24 h. At the end of the reaction, the diamond sample was thoroughly washed using the cleaning procedure described above and dried using a nitrogen spray gun before characterization.

*Note*: examples of acylation reagents include pentafluorobenzoyl chloride, 3,5-bis(trifluoromethyl)benzoyl chloride, and heptafluorobutyric anhydride (Figure S24).

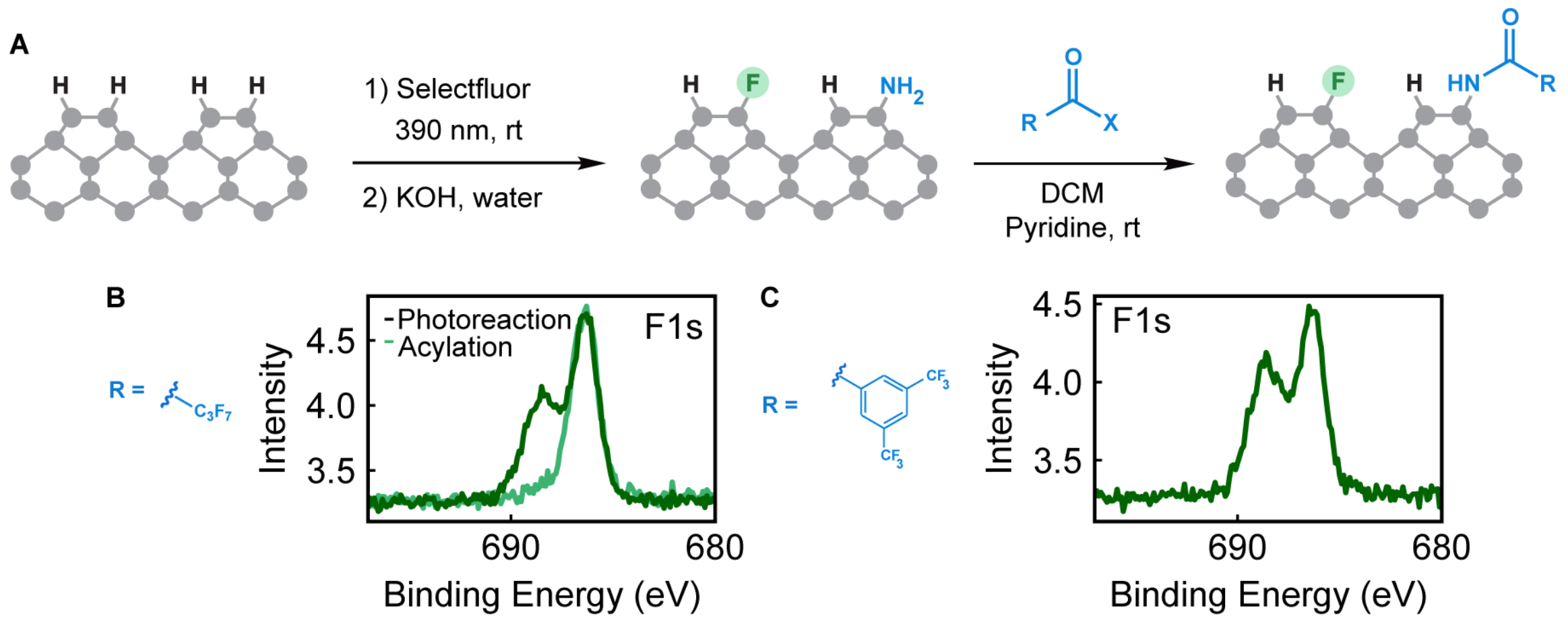

**Figure S24.** XPS spectra of the diamond surface after acylation reactions with 3,5-bis(trifluoromethyl)benzoyl chloride and with heptafluorobutyric anhydride. (**A**) Reaction scheme. (**B**) F *1s* XPS results after reaction with heptafluorobutyric anhydride. (**C**) F *1s* XPS results after reaction with 3,5-bis(trifluoromethyl)benzoyl chloride.

*Note on coupling efficiency:* To estimate the coupling efficiency for these reactions, we compare the atomic percentage of the N *1s* signal to the atomic percentage of the F *1s* peak around 689 eV. We also take into account the number of atoms in a particular functional group. For the examples shown in the main text Figure 3C and Figure S24, we find the coupling efficiency to be less than 30%.

### *NEXAFS studies on amide- and amine-terminated surfaces*

Since amide and amine groups are indistinguishable via XPS, we looked for differences using NEXAFS (Figure S25). At the N *K*-edge, we found that amide-terminated samples, which were subjected to a sequence of either Ritter amidation (red), or Ritter amidation followed by hydrolysis and acylation reactions (orange), showed a shifted feature around (398 to 399) eV, an enhanced feature around 401 eV, and a higher $\sigma^*$ shoulder compared with those of amine-terminated samples that were amidated and hydrolyzed (Figure S25A, blue). Previous reports have attributed the 401 eV feature to the N *1s* → π* transition in amides, supporting our structural assignments[14,15]. Additionally, at the O *K*-edge, we found a shifted π* feature for amide-terminated samples compared with the amine-terminated samples (Figure S25B). The O *K*-edge signal on these surfaces can arise from oxygen groups bound to the surface, adventitious carbon, or the oxygen contained in the functional group in the case of amide-terminated samples. This shift could be due to the presence of the carbonyl group on the amide or from electronic interactions between the amide and surrounding oxygen groups on the diamond surface.

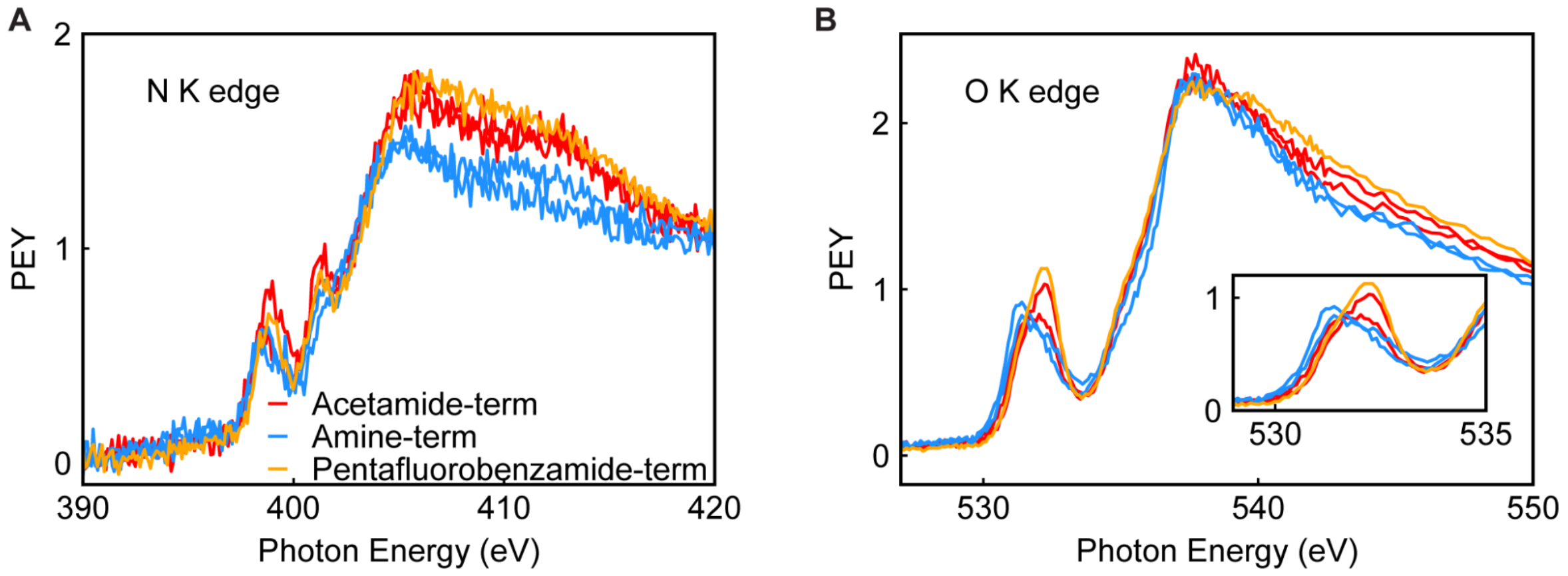

**Figure S25.** Comparison of NEXAFS spectra (*i.e.*, N *K*-edge (left) and O *K*-edge (right)) of amide-terminated versus amine-terminated surfaces. The amide-terminated surfaces were obtained after Ritter amidation reaction in acetonitrile (red) and after a 3-step reaction sequence involving Ritter amidation, hydrolysis, and acylation with pentafluorobenzoyl chloride (yellow). The amine-terminated surface was obtained after a 2-step reaction sequence involving Ritter amidation and hydrolysis (blue).

***Attempts for C–C bond formation reactions on the surface***

We screened various thermal and photochemical conditions for the alkylation reaction of diamond surfaces. Several HAT reagents with BDFE values of at least 100 kcal/mol were employed, such as benzoyl radical, amidyl radical, chlorine radical, aminium radical cation, and the excited-state TBADT (Figure S26). In addition, a number of electron-deficient olefins were also employed in these transformations. Unfortunately, surface alkylation was not observed in any of these tests. We posit that while the HAT process might be effective, the addition of the carbon radical to the alkene acceptor might be reversible and the quenching of the resulting electrophilic carbon radical is not fast enough to drive the transformation.

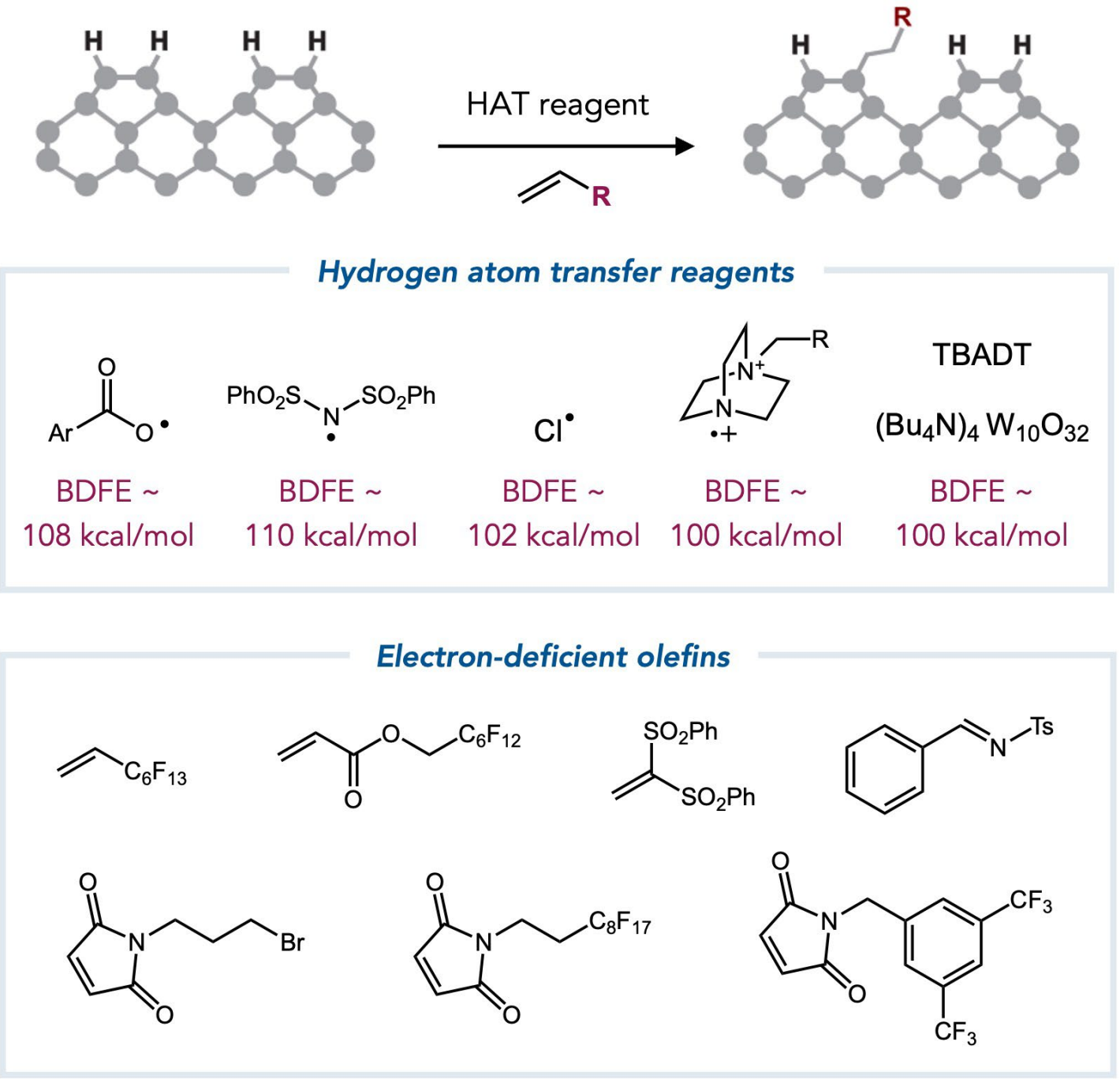

**Figure S26.** Examples of HAT reagents (with their corresponding BDFE values) and electron-deficient olefins that were investigated for the surface alkylation.

## V. Density functional theory calculations

In our study, density functional theory calculations were performed using the VASP software[16]. A symmetric slab of (001) diamond was constructed, including the 2×1 surface reconstruction pattern, and the dangling bonds were passivated by either hydrogen or fluorine atoms. The slab consisted of 768 C-atoms along with 64 atoms representing different coverage. The lateral size of 14.178 Å corresponded to the standard 512 C-atom diamond supercell, and sufficient vacuum was added along the c-direction to prevent interaction with periodic images. All calculations were performed using the Heyd, Scuseria, and Ernzerhof (HSE06) functional[17]. The plane wave cutoff energy was set to 370 eV, and the Gamma point scheme was used for sampling the Brillouin zone. To calculate electron affinity ($\chi$), we determined the energy difference between the conduction band minimum and the vacuum level in agreement with our previous study[18].

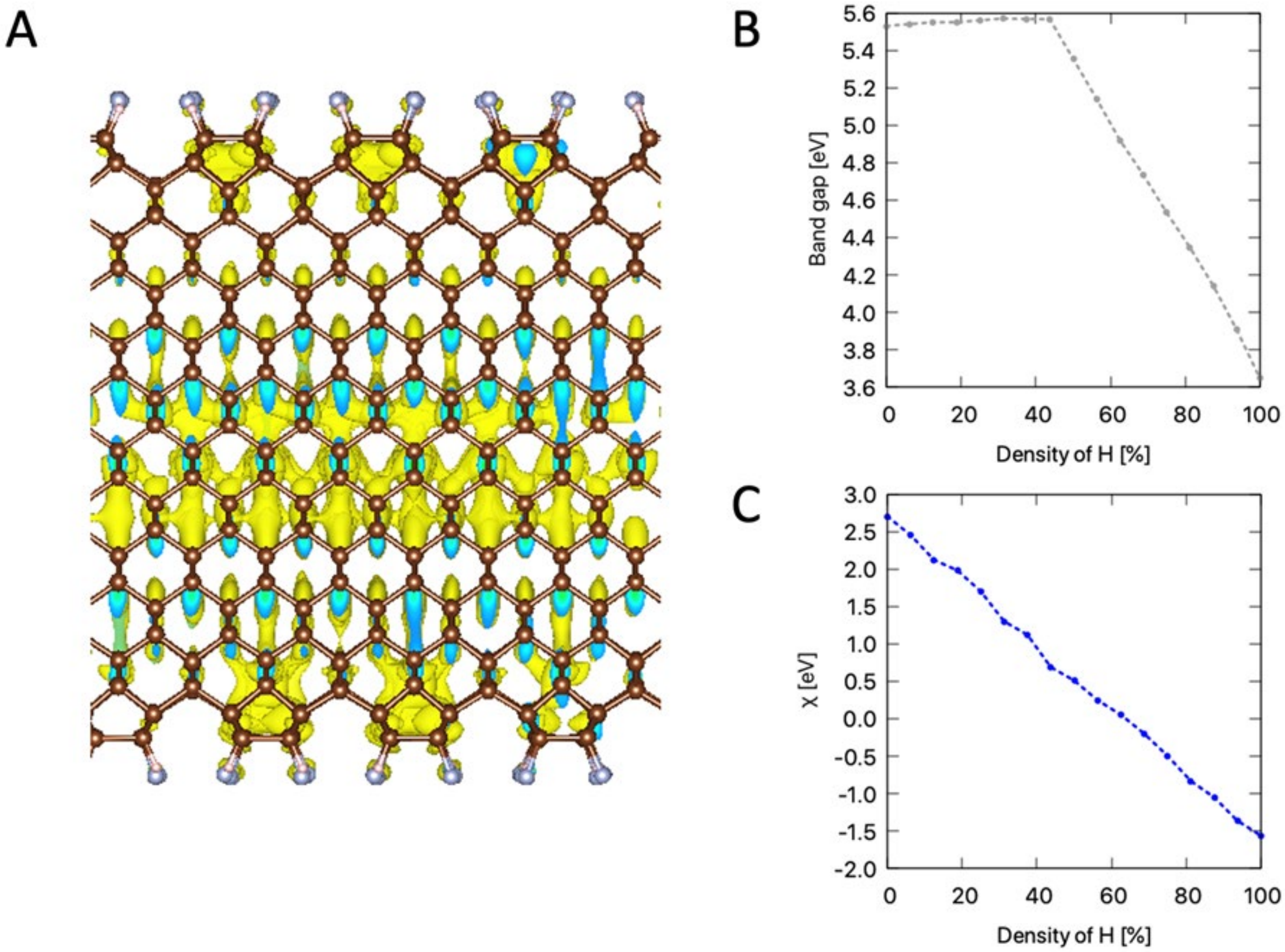

**Figure S27.** (**A**) The structure of the symmetric (001) diamond slab used in this study. The iso-surface reflects the surface state in the case of 70% fluorine and 30% hydrogen coverage. (**B**) The energy gap (eV) between the valence band maximum and the lowest unoccupied orbital as a function of hydrogen density. (**C**) The electron affinity (eV) as a function of hydrogen density.

Our theoretical investigation focused on studying the changes in the electronic structure when transitioning from hydrogen to fluorine coverage on the diamond surface. In the case of hydrogen termination, we observed the appearance of a surface level inside the band gap, which destabilizes the negative charge state of NV centers that are close to the surface. Our density functional theory calculations indicated that this level is located around 3.7 eV from the valence band maximum (Figure S27B). As we increased the density of fluorine atoms, we observed a gradual shift of this level towards the conduction band. This trend is also reflected in the changes in calculated electron affinity (Figure S27C) where the value varies between –1.5 and 2.7 eV with respect to hydrogen density. We note that these values are in close agreement with the reported experimental data (–1.3 eV[19] and 2.6 eV[20], respectively). Interestingly, we also observed a surface state for fully fluorinated diamond located about 0.1 eV below the conduction band minimum. However, we found that the presence of this state can be largely reduced by using a mixed fluorine-hydrogen termination. For instance, in the case of 70% fluorine and 30% hydrogen coverage (Figure S27A), the surface states merge with the conduction band. This behavior is consistent with the positive values of electron affinity in Figure S27. Based on these findings, we anticipate that fluorination densities ranging from 50% to 80% would provide the optimal conditions for hosting stable $NV^-$ centers. An interesting future experimental direction is to find conditions to create this surface using solution-phase chemistry as this could remove the need for additional processing to stabilize shallow $NV^-$ centers under fluorinated surfaces. We note that our experimentally prepared surfaces also contain oxygen groups which are not included in this model.

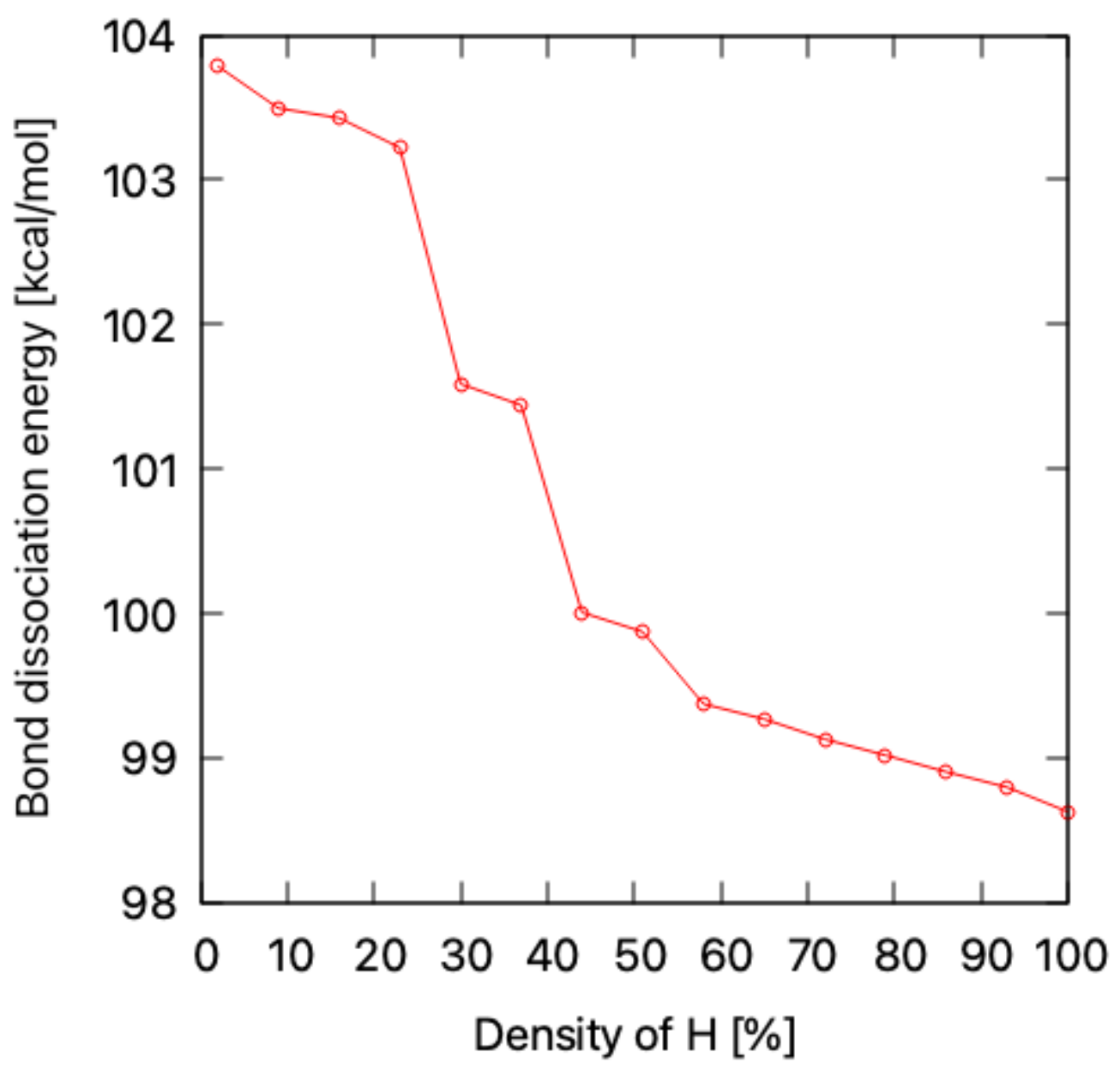

**Figure S28.** The C-H bond dissociation energy (kcal/mol) as a function of hydrogen density calculated using the slab model from Figure S27.

It is important to note that, in the calculations we considered different mutual arrangements of fluorine and hydrogen atoms at the surface. However, no significant differences were observed in the energies of the empty band inside the band gap nor in the electron affinity for a given density of hydrogen atoms. Nevertheless, we found that the total energy of a configuration decreases substantially (up to 1 eV) when the surface atoms are evenly distributed. This is attributed to the stabilizing effect of the Coulombic interaction between hydrogen atoms bearing a partial positive charge and fluorine atoms bearing a partial negative charge, favoring a sparse distribution of atoms. Specifically, as the density of hydrogen atoms decreases, a decrease in C–H bond length from 1.10 to 1.09 Å was observed. Figure S28 illustrates this effect for the bond dissociation energy, showing an increase in values from ~99 to 104 kcal/mol. These results qualitatively align with our experimental measurements, suggesting that fluorination becomes more challenging as the density of fluorine atoms at the surface increases.

## VI. Additional processing to stabilize shallow NV centers

In our experiments, we observe a coherent and charge-stable spin signal from single $NV^-$ centers with some additional processing after surface functionalization. First, we studied single NV centers under a fluorinated surface (Sample 1b in main text Figure 4A) and were able to readily observe ODMR only after refluxing the samples in a mixture of sulfuric, nitric, and perchloric acids ("triacid clean"). Figure S29G shows a histogram of oxygen content measured with XPS across many samples after two iterative rounds of cleaning with organic solvents, cleaning in piranha solution (hydrogen peroxide, sulfuric acid), and triacid cleaning. While there was a large spread in oxygen content, in part because adventitious carbon contamination could contribute to the signal, the triacid cleaned samples exhibited a higher atomic percentage of oxygen compared to the other cleaning procedures. We hypothesized that the triacid cleaning procedure could oxidize the residual C–H bonds to form C–O bonds on the surface, helping to stabilize the $NV^-$ charge state. Alternatively, the triacid cleaning procedure could change the nature of surface adsorbates which could impact band bending.

However, the triacid cleaning procedure is harsh, and may remove functional groups of interest. An example for a Ritter reaction installing $C_{diamond}$–F bonds and 3,3,3-trifluoropropanamide groups is shown in Figure S29A–C) . The data show that after the triacid clean, the O *1s* signal increased, however the F *1s* signal associated with the $CF_3$ group decreased. This indicates that the amide is hydrolyzed, causing the removal of $CF_3$ groups after the triacid cleaning process.

To combat this challenge, we developed a low-temperature oxygen annealing procedure, which increased the amount of oxygen, while leaving other functional groups largely intact (Figure S29 (D–F)). In this procedure, we placed the samples in a tube furnace and flowed high-purity oxygen gas. We first heated the sample to 100 °C over 1 h, held at 100 °C for 2 h for degassing, ramped to 300 °C over 2 h, and held at 300 °C for 3.5 h before shutting off the furnace.

Although we suspect these procedures are helpful to stabilize the $NV^-$ charge state because they replace residual hydrogen with oxygen, an alternative hypothesis is that we are instead changing the nature of the surface adsorbates. Since the proposed band bending mechanism[21] on diamond involves charge transfer from the diamond to surface adsorbates, it is reasonable to suspect they could impact NV charge state behavior. Although surface adsorbates are ubiquitous and hard to study, we can use NEXAFS to look at the carbon *K*-edge (Figure S29). We found that samples that were subjected to a low-temperature oxygen anneal showed a slightly reduced $sp^2$ signal as well as an overall reduction of signal in the pre-edge region, which has previously been associated with adventitious carbon[1].

This additional processing could be avoided in the future by (1) directly preparing mixed surfaces with different relative coverages of functional groups, oxygen-containing groups, and hydrogen, (2) controlling the nature of surface adsorbates, or (3) working with samples with higher dopant density to screen the electric field from the surface.

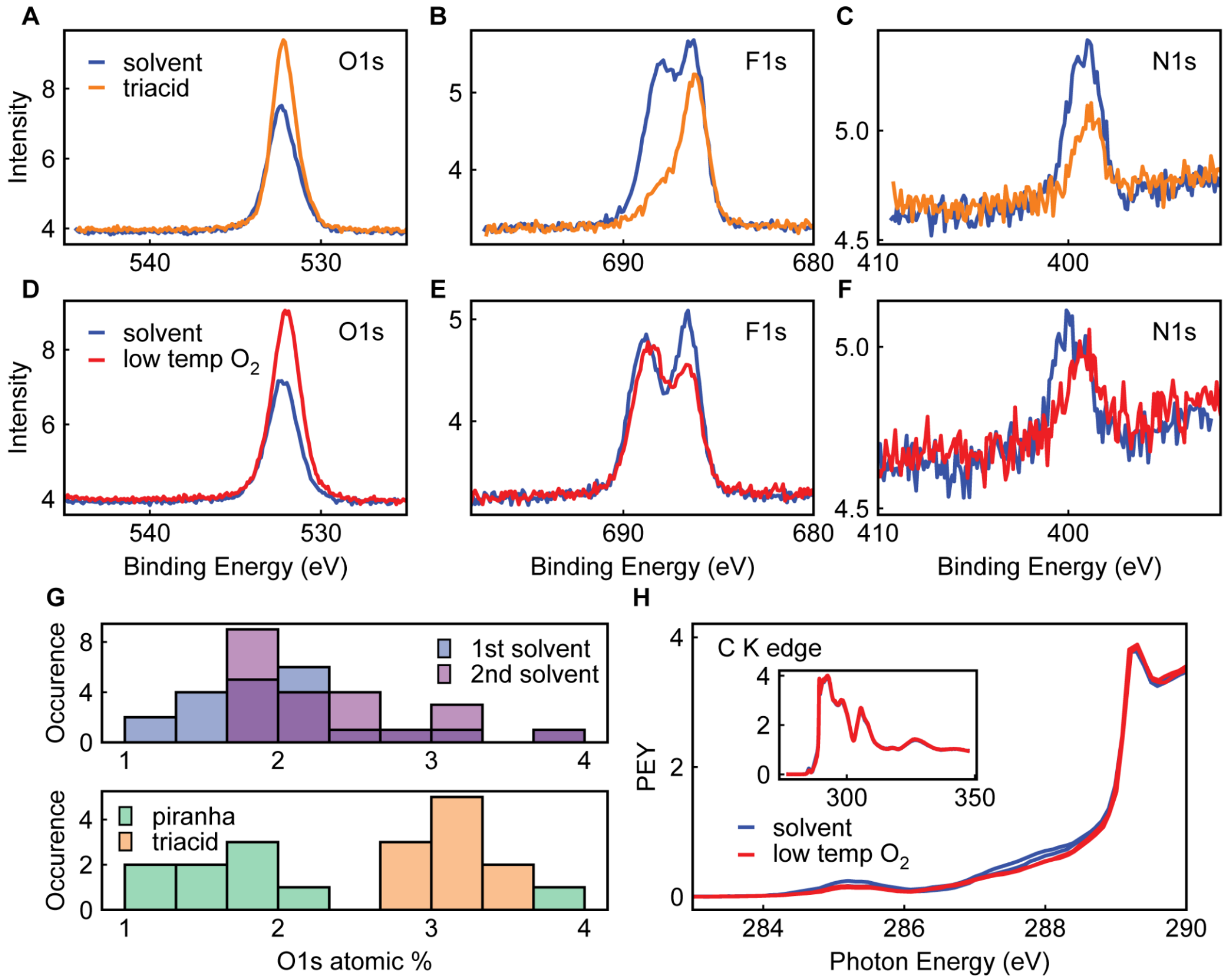

**Figure S29.** Investigation into the empirical observation that triacid cleaning or low-temperature oxygen annealing helps to recover an $NV^-$ spin signal. (**A**–**C**) XPS data for a sample that was subjected to Ritter reaction with 3,3,3-trifluoropropionitrile and underwent cleaning processes with organic solvents and triacid mixture. After the triacid clean, the O *1s* signal increased, but the peak at 689 eV associated with the $CF_3$ group was reduced. (**D**–**F**) XPS data from another sample that was subjected to Ritter reaction with 3,3,3-trifluoropropionitrile and underwent a cleaning process with organic solvents and low-temperature annealing. After the anneal, we observed a higher O *1s* signal and the F *1s* and N *1s* peaks remain unchanged, suggesting the functional group remains largely intact. (**G**) Histograms of O *1s* signal atomic percent after functionalized samples were treated with various different cleaning procedures. While there was a large spread, the triacid cleaning procedure led to the highest O *1s* coverage. We note that adventitious carbon could also contribute to the O *1s* signal. (**H**) C *K*-edge NEXAFS data for four samples. The blue lines correspond to two samples that were functionalized and solvent cleaned. The two red lines correspond to samples that went through the same procedure but were subjected to an additional low-temperature oxygen anneal.

We verified that the low-temperature oxygen annealing procedure was compatible with subsequent hydrolysis and acylation reactions. We checked this by performing identical procedures on two samples, one of which was subjected to an additional low-temperature oxygen anneal prior to hydrolysis and acylation. Both samples gave similar results (Figure S30).

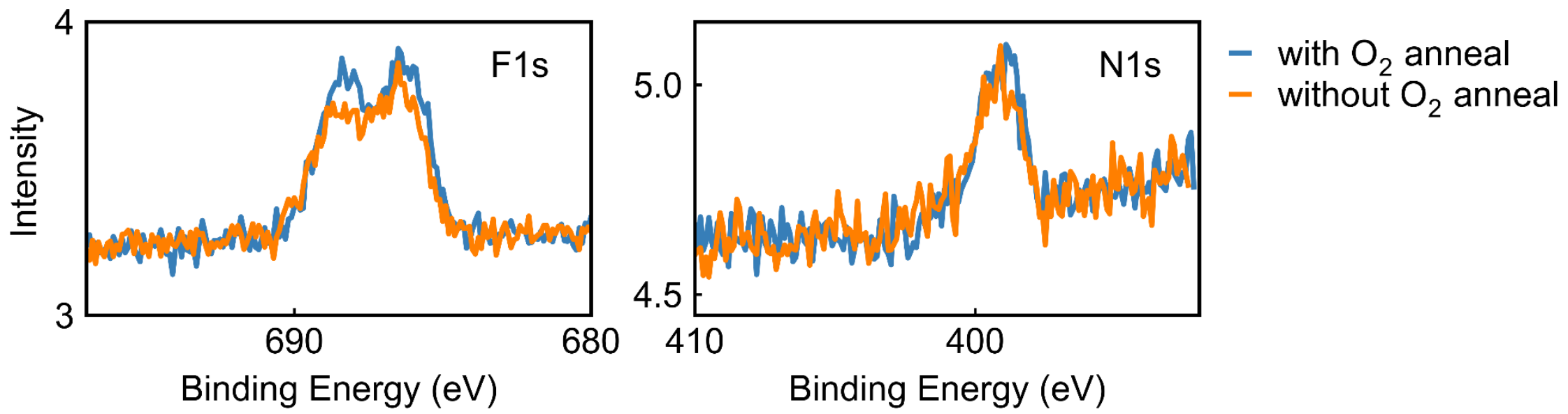

**Figure S30.** XPS results of two samples after acylation reaction installing $C_7F_{15}$ groups on the surface. One sample was treated with an additional low-temperature oxygen anneal prior to the hydrolysis and acylation reactions. F *1s* and N *1s* results were consistent across two samples, demonstrating the annealing procedure did not adversely affect the reaction efficiency.

## VII. Single NV center measurements

The details of our single NV measurement setup have been published elsewhere[1,22]. Single NV center measurements were performed on a home-built confocal microscope (Figure S31). NV centers are excited by a 532 nm optically pumped solid state laser (OPSL, Coherent Sapphire) which is modulated with an acousto-optic modulator (AOM). The beam is scanned using galvo mirrors and projected into an oil immersion objective (Nikon, Plan Fluor 100X, NA = 1.30) with a telescope in a 4f configuration. Laser power at the back of the objective was kept between (60 to 100) μW, approximately 25% of the saturation power of a single NV center, in order to avoid irreversible photobleaching. A dichroic beamsplitter separates the excitation and collection pathways, and fluorescence is measured using a fiber-coupled avalanche photodiode (Excelitas SPCM-AQRH-44-FC). A neodymium permanent magnet was used to introduce a DC magnetic field for Zeeman splitting and the orientation of the magnetic field was aligned to within 1° of the NV center axis using a goniometer.

Spin manipulation on the NV center was accomplished using microwaves. The output of a signal generator was gated with fast SPDT switches (Mini-Circuits ZASWA-2-50DR+) before being amplified by Mini-Circuits ZHL-16W-43+. The signal was then delivered to the sample via a coplanar stripline. The stripline was fabricated by depositing 10 nm Ti, 1000 nm Cu, and 200 nm Au on a microscope coverslip at a commercial facility. Following metallization, the stripline was photolithographically defined and etched with gold etchant and hydrofluoric acid. Finally, a 100 nm layer of $Al_2O_3$ was deposited on top of the fabricated stripline via atomic layer deposition (ALD) to protect the metal layer. Pulse timing was controlled with a Spincore PulseBlaster ESR-PRO500 with 2 ns timing resolution.

Coherence time measurements for samples 1a and 1b were collected at 190 mT and NV depth measurements were performed around 27 mT. Coherence time and depth measurements for Sample 2 ($^{12}C$ enriched) were collected at 46 mT. For all samples, depth measurements were estimated using the NMR proton signal from microscope immersion oil, following the protocol outlined in previous work[23]. The dataset shown in Figure 4B used the CPMG sequence for (1 to 4) π pulses and the XY-8 sequence for (8 to 40) π pulse measurements. Lifetime ($T_1$) measurements were conducted at low field (3 mT).

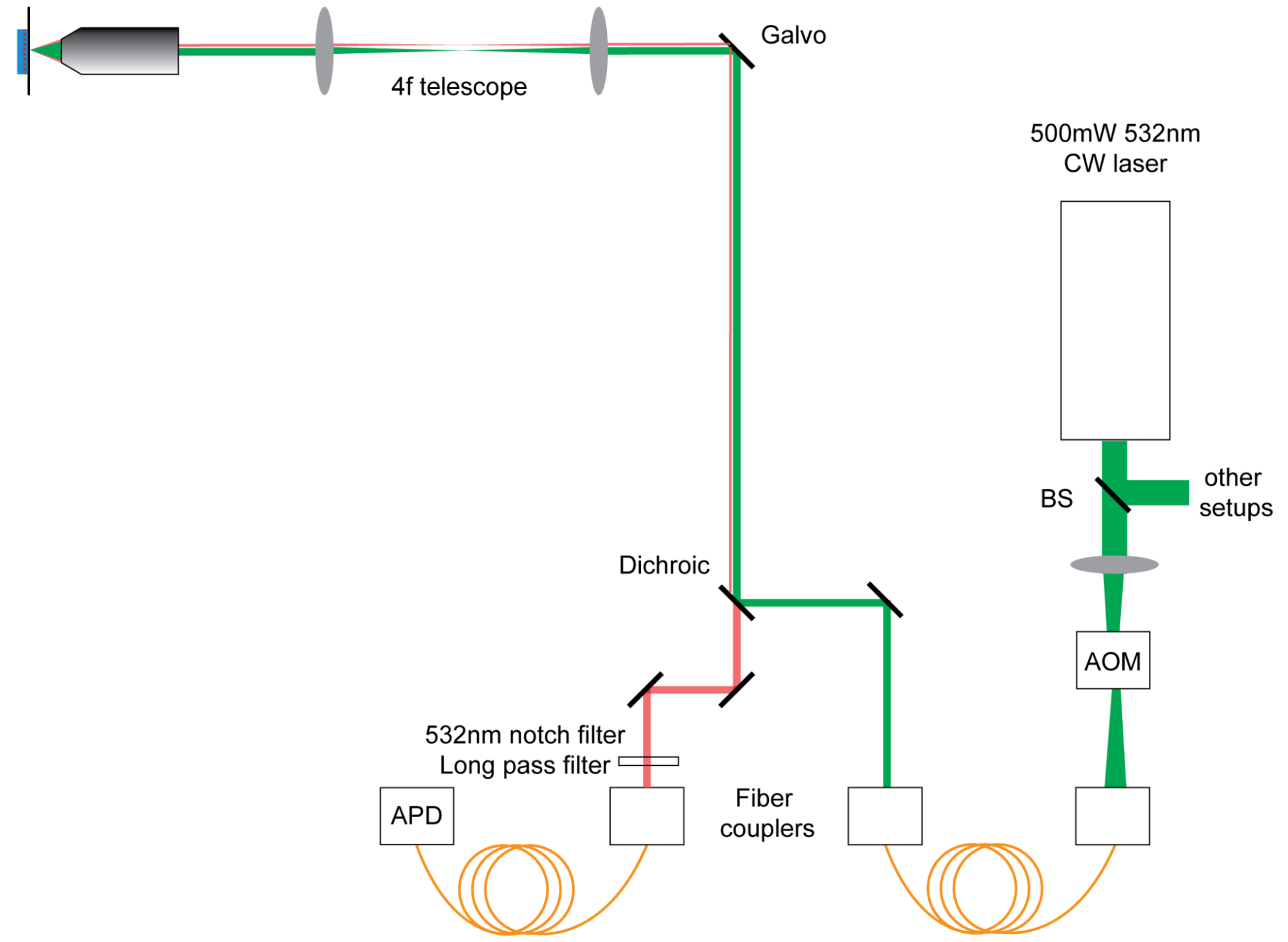

**Figure S31.** Single NV scanning confocal setup diagram

***Single NV center sample preparation***

All samples were purchased or obtained from Element Six.

**Sample 1a**. An electronic grade single crystal (denoted as ELSC by Element Six) grade sample purchased from Element Six with a natural abundance of $^{13}C$ nuclear spins. The sample was implanted with a low density of $^{15}N$ (3 keV, $10^9/cm^2$, 0° tilt) and annealed to form NV centers. Prior to measurement, the sample was oxygen annealed and then 'reset' with another 800 °C anneal in a vacuum tube furnace. The activation, oxygen, and reset anneals followed procedures outlined in previous work[1]. The activation and reset annealing procedures produce NV centers with comparable coherence properties[1].

**Sample 1b**. An ELSC grade sample cut from the same larger diamond as Sample 1a. The two samples were implanted with the same ion implantation parameters and processed together (subjected to activation anneal, oxygen anneal, reset anneal). The sample was then subjected to anneal in forming gas (24 h anneal) followed by the direct fluorination reaction and an acid clean.

**Sample 2**. An ELSC grade sample with an isotopically-purified $^{12}C$ layer. The surface was 'as grown' (i.e., was not subjected to polishing or etching). This sample was implanted with low dose $^{15}N$ ions to form single NV centers (1.5 keV, $10^9/cm^2$, 0° tilt) and annealed in a vacuum tube furnace to 800 °C to form NV centers. The sample was then subjected to a 72 h anneal in forming gas, functionalized using the Ritter reaction with 3,3,3-trifluoropropionitrile, and then subjected to a low-temperature oxygen anneal prior to NV measurements.

***Other single NV center behaviors***

We observed some decreased performance of NV centers after functionalization that are an important area for future improvement. First, we measured a short double-quantum (DQ) and single-quantum (SQ) $T_1$ lifetimes of some NV centers, down to 30 microseconds, perhaps from charge state instability (Figure S32). For NV centers under state-of-the-art surfaces, we typically measure SQ and DQ $T_1$ lifetimes in the range of several hundred microseconds to a few milliseconds[1]. Next, even after acid cleaning or low-temperature oxygen annealing, we observed that the ESR contrast can change as a function of green illumination time, as shown in Figure S33. In Figure S33A, we observed an increase in Rabi contrast as a function of measurement time for a single NV for a sample that was functionalized and then acid cleaned. In Figure S33A, we observed the ESR contrast increase for a NV center on a sample that was functionalized and then subjected to a low-temperature oxygen anneal. Finally, we also observed that the apparent coherence times of NV centers decreased over the timescale of several hours of measurement on the same NV center. This posed significant challenges for using single NV centers for sensing experiments that require long averaging times. These observations are likely due to changes in charge state properties or surface adsorbates, and may be mitigated by judicious selection of immersion oil or aqueous solution, additional processing, screening the electric field leading to band bending with dopants, or developing functionalization strategies for oxygen-terminated diamond surfaces.

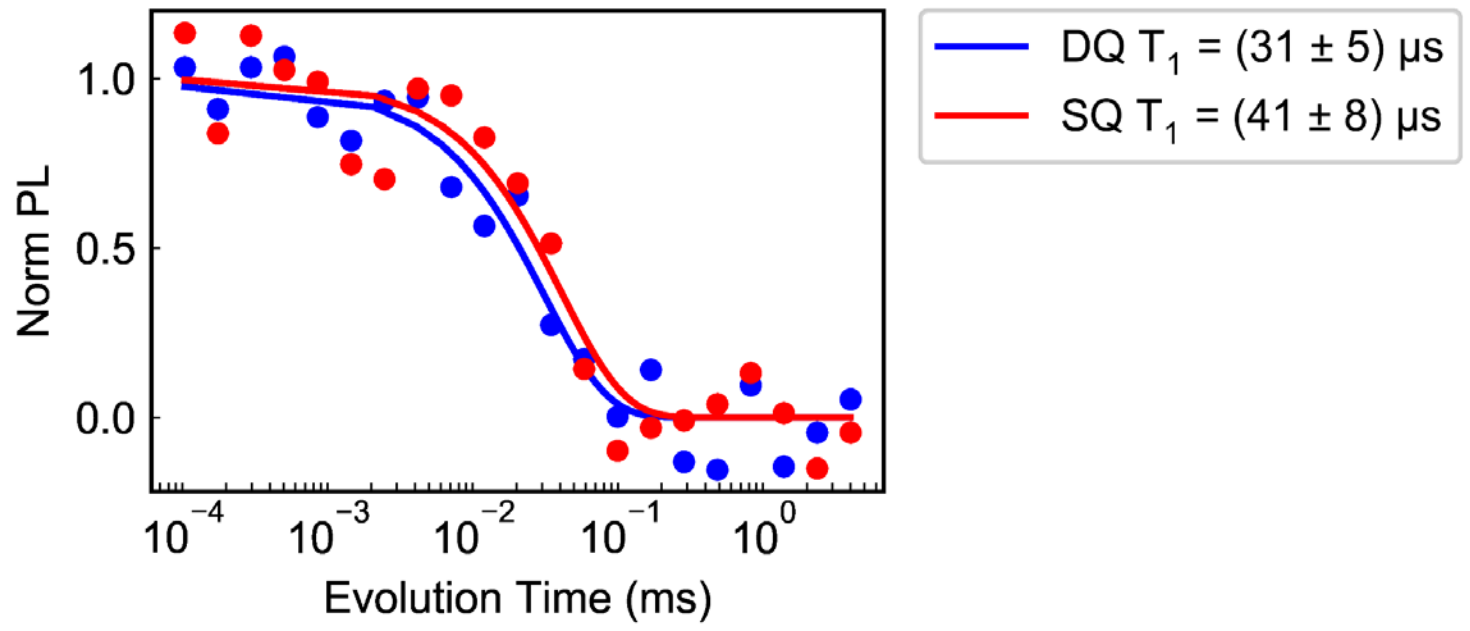

**Figure S32.** Representative double quantum (DQ) and single quantum (SQ) lifetimes for a single NV center in Sample 2 after functionalization.

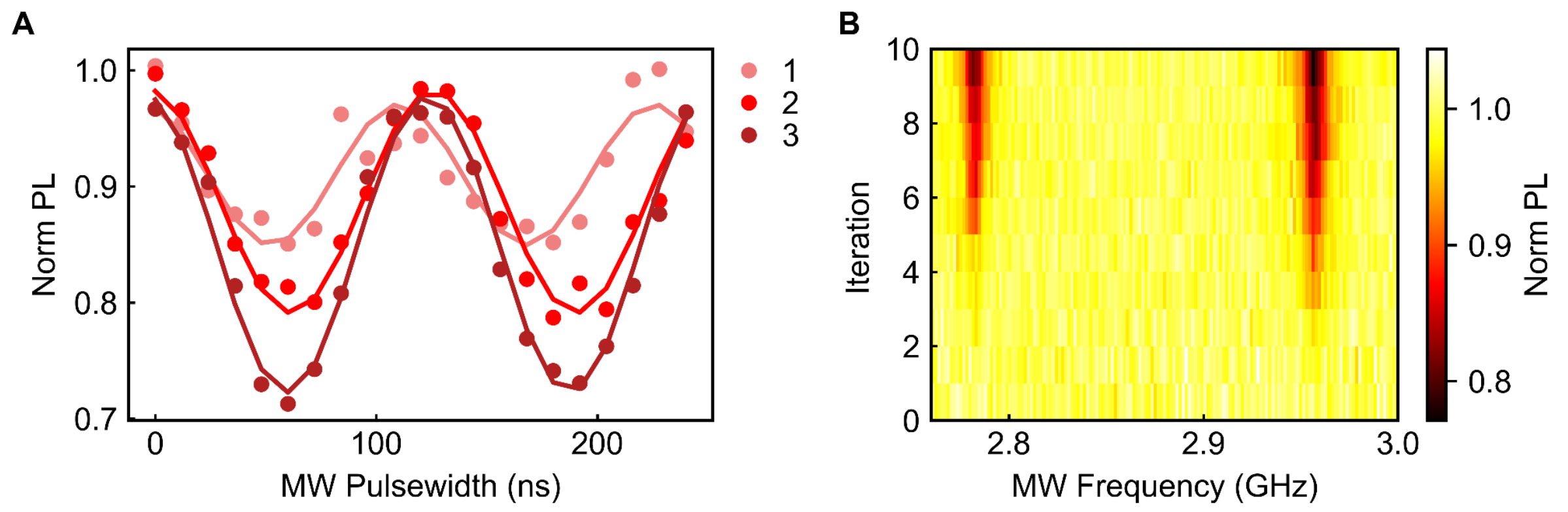

**Figure S33.** (**A**) Rabi contrast of a single NV center after a sample was functionalized and acid cleaned. The Rabi contrast increased over the course of several hours of measurement. (**B**) The ESR contrast improved when a single NV center measured in Sample 2 was subjected to many averages of an ODMR experiment over the course of approximately 1 hour. Sample 2 was functionalized and subjected to a low-temperature oxygen anneal before measurement.

## VIII. NV ensemble measurements

### *Ensemble measurement setup*

We modified our home-built single-center scanning confocal microscope for wide-field ensemble measurements (Figure S34). This allowed the excitation of a large number of NV centers at once, boosting the sensitivity for proof-of-principle sensing experiments. Most notably, ensemble experiments require higher laser power (new laser, no fibers), a change in optics to excite a large spot size, and a different type of photodetector that can detect higher amounts of fluorescence.

NV centers were excited using a 532 nm green laser (Lighthouse Photonics, Sprout-H-5W). The laser output was set to 1 W, which corresponded to approximately 300 mW before the dichroic. The diameter of the laser beam was reduced with a telescope before entering the AOM. After the AOM, a polarizer was used to rotate polarization to maximize the NV signal from the NV centers that are aligned along the magnet axis. Next, a telescope expanded the beam and the beam width was adjusted based on the desired spot size on the diamond. An iris was placed at the focal plane of the first telescope lens to spatially reject the other modes of the AOM. The aperture of this iris was adjusted to allow for an appropriate extinction ratio, following previous widefield experiments[24]. Next, a lens focused the green laser on the back focal plane of the objective. The distance between this lens and the objective was adjusted such that the light coming out of the objective was approximately collimated. The spot size on the sample is determined by the beam diameter going into the focusing lens, the length of the focusing lens, and the effective focal length of the objective. In our setup, the spot size can be most easily tuned by changing the telescope lenses before the focusing lens. NV fluorescence was collected by an oil immersion objective with 100× magnification and 1.3 NA (Nikon N100x-PFO) and separated from green excitation by a short pass dichroic (Thorlabs DMSP550). This dichroic was chosen to reflect both $NV^0$ and $NV^-$ fluorescence. Microscope immersion oil (Nikon, Type N) was used between the objective and coverglass and the front aperture of the objective was routinely cleaned. In the collection path, a removable mirror mount allowed emission from both charge states to be studied on a spectrometer by passing the emission through a 532 nm notch filter (Semrock NF03-532E-25) onto a reflective collimator with a fiber coupler. This proved useful for troubleshooting charge state changes under green laser illumination. With the mirror mount removed, the emission was focused onto an analog photodetector (Thorlabs APD410a). Before the photodetector, a 532 nm notch filter (Semrock NF03-532E-25) and a 647 nm long pass filter (Semrock BLP01-647R-25) removed remaining green excitation and filtered out $NV^0$ emission. The output of the photodetector was connected to an analog input of a DAQ (National Instruments, PXIe-6363) for readout. The readout approach and optimization followed the methodology outlined in previous widefield experiments[24].

Microwave spin manipulation as well as experiment timing and control was accomplished using the same hardware used in single NV measurements.

Ensemble coherence time measurements were acquired at a bias magnetic field around 41 mT. The dataset shown in Figure 4C used the Hahn echo CPMG sequence for one π pulse and the XY-8 sequence for (8 to 240) π pulse measurements. $^{19}$F NMR sensing experiments were performed between 35 mT and 42 mT. Lifetime ($T_1$) measurements were conducted at low field (approximately 3 mT).

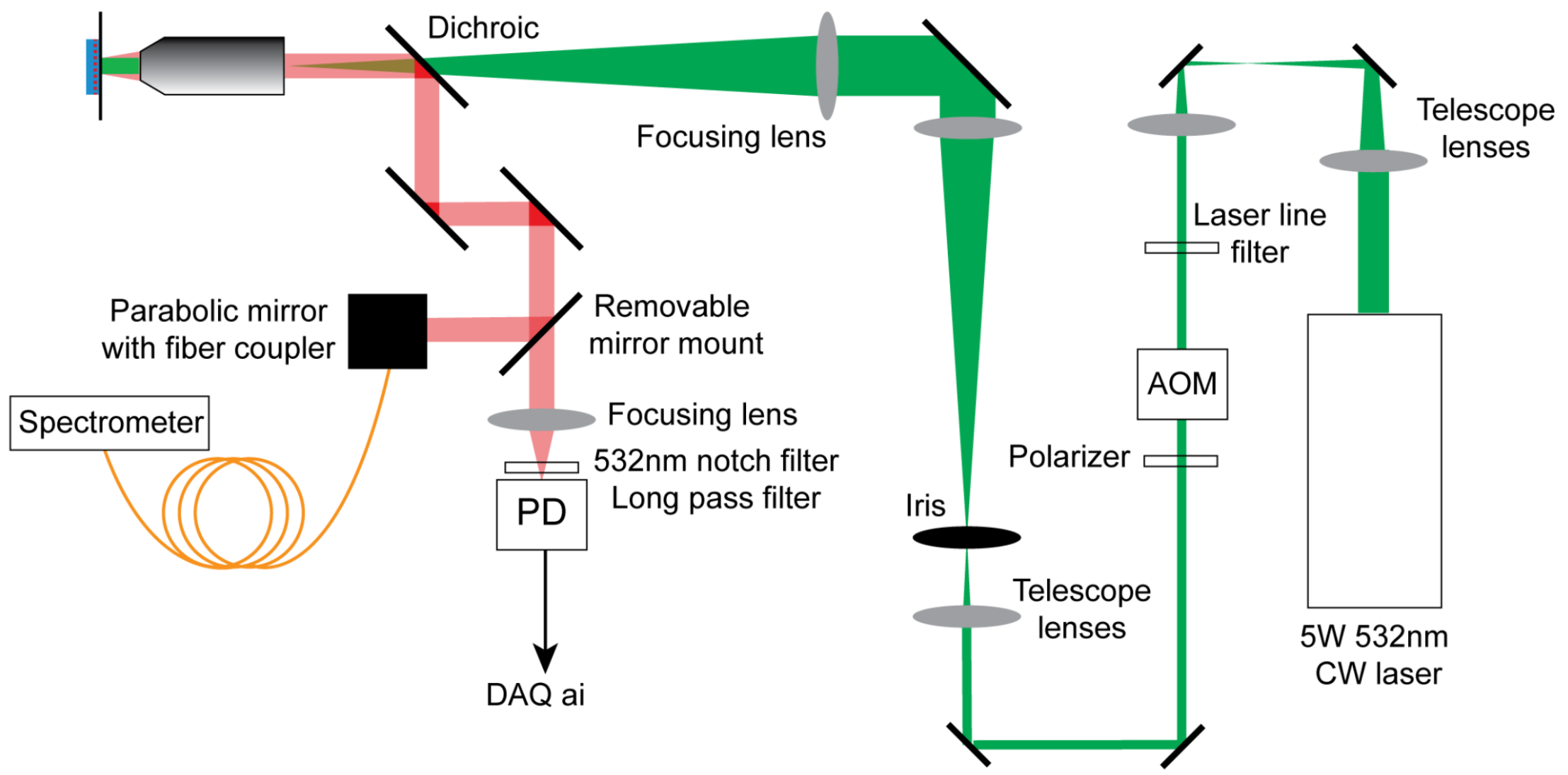

**Figure S34.** Ensemble NV measurement setup diagram.

### *NV ensemble sample preparation*

All samples were purchased or obtained from Element Six.

**Sample 3**. An ELSC sample that was cut from the same diamond as Sample 2 after low density ion implantation and activation annealing. Next, the sample was implanted with higher dose (2.5 keV, $2 \times 10^{12}/cm^2$, 7° tilt) and annealed to 800 °C in a vacuum tube furnace to form NV centers.

The sample was then oxygen annealed following the procedure established in previous work[1], hydrogen annealed, and subjected to the direct fluorination reaction.

**Sample 4.** An ELSC grade sample with an isotopically purified $^{12}C$ layer. The surface was 'as grown' (i.e., was not subjected to polishing or etching). This sample was implanted with low dose $^{15}N$ ions to form single NV centers (1.5 keV, $10^9/cm^2$, 0° tilt) and annealed to form NV centers. Next, the sample was implanted at a higher dose (2.5 keV, $2 \times 10^{12}/cm^2$, 7° tilt) and annealed to 800 °C in a vacuum tube furnace to form NV centers.

*NV ensemble charge state properties*

We studied the NV center charge state properties under green laser illumination and present the empirical results here. In these experiments, we measured the fluorescence spectra of our ensemble samples on a grating spectrometer, allowing us to see how the NV center charge state changes over time. The $NV^0$ zero phonon line (ZPL) is at 575 nm and the $NV^-$ ZPL is at 637 nm. We also observed a clear peak at 532 nm from the green laser and at 572 nm from the diamond Raman line. In all figures showing NV charge state spectra, the color transition of purple to red shows increasing time under green laser illumination. Purple traces were acquired when the sample was freshly illuminated and red traces were acquired at the end of the measurement.

First, we studied the functionalized sample charge state behavior under green laser illumination in different environments. When we mounted a functionalized sample (Sample 3) in 'air' on the stripline (i.e., no liquid between sample and stripline, as depicted in Figure S35A), we initially did not observe a clear $NV^-$ signal. After some time under green laser illumination, the $NV^0$ signal appeared to slightly increase and then, with more time, the $NV^0$ signal decreased and the $NV^-$ signal increased (Figure S35(b)). When we mounted the same sample in deuterated propylene carbonate, we also observed that the $NV^-$ signal increased under green laser illumination (Figure S35(d)). We note that this sample was subjected to several acid cleaning procedures in between these two measurements which could have impacted the difference in starting $NV^-$ population.

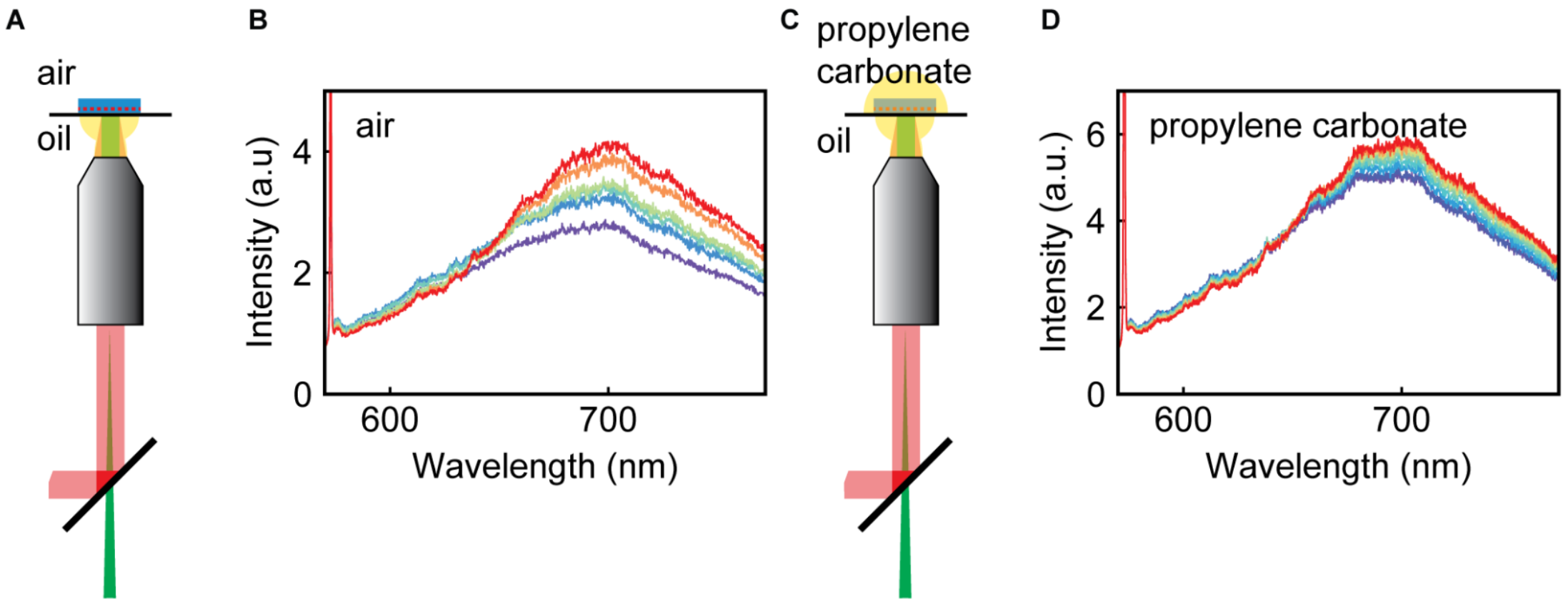

**Figure S35.** Fluorescence spectra of Sample 3 in different chemical environments. (**A**) Cartoon of sample on stripline without any liquid between stripline and sample. This will be referred to as the sample mounted in air. (**B**) Fluorescence spectra of Sample 3 in air. Purple to red traces show

spectra over time under continuous wave (CW) green laser illumination over the course of approximately12 min. (**C**) Cartoon of sample on stripline mounted in propylene carbonate. (**D**) Fluorescence spectra of Sample 3 in propylene carbonate under CW green laser illumination over the course of approximately 30 min.

### *Additional NV ensemble characterization after functionalization*

Since NV-NMR sensitivity depends on the NV OD-ESR readout contrast, we benchmarked the Rabi contrast of Sample 3 after the initial activation anneal to form NV centers, an oxygen anneal[1], and photochemical fluorination (Figure S36). The measured Rabi contrasts range from 10% to 13%, where the highest Rabi contrast was observed after oxygen annealing. We note that this contrast depends sensitively on time under green laser illumination and the chemical environment of the sample (Figure S35).

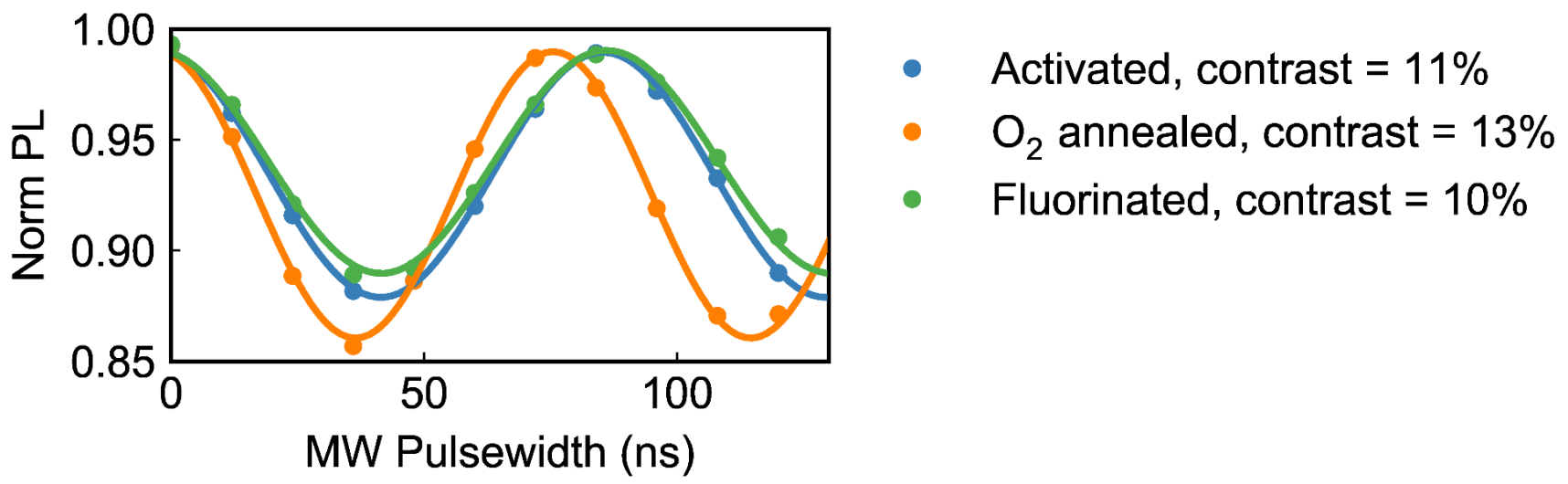

**Figure S36.** OD-ESR contrast for Sample 3 after activation anneal, oxygen anneal, and photochemical fluorination. These measurements were acquired with the sample mounted in air, as depicted in Figure S35(a).

Additionally, we measured the single quantum (SQ) and double quantum (DQ) lifetimes before and after functionalization and observed a slight reduction in DQ lifetime (Figure S37).

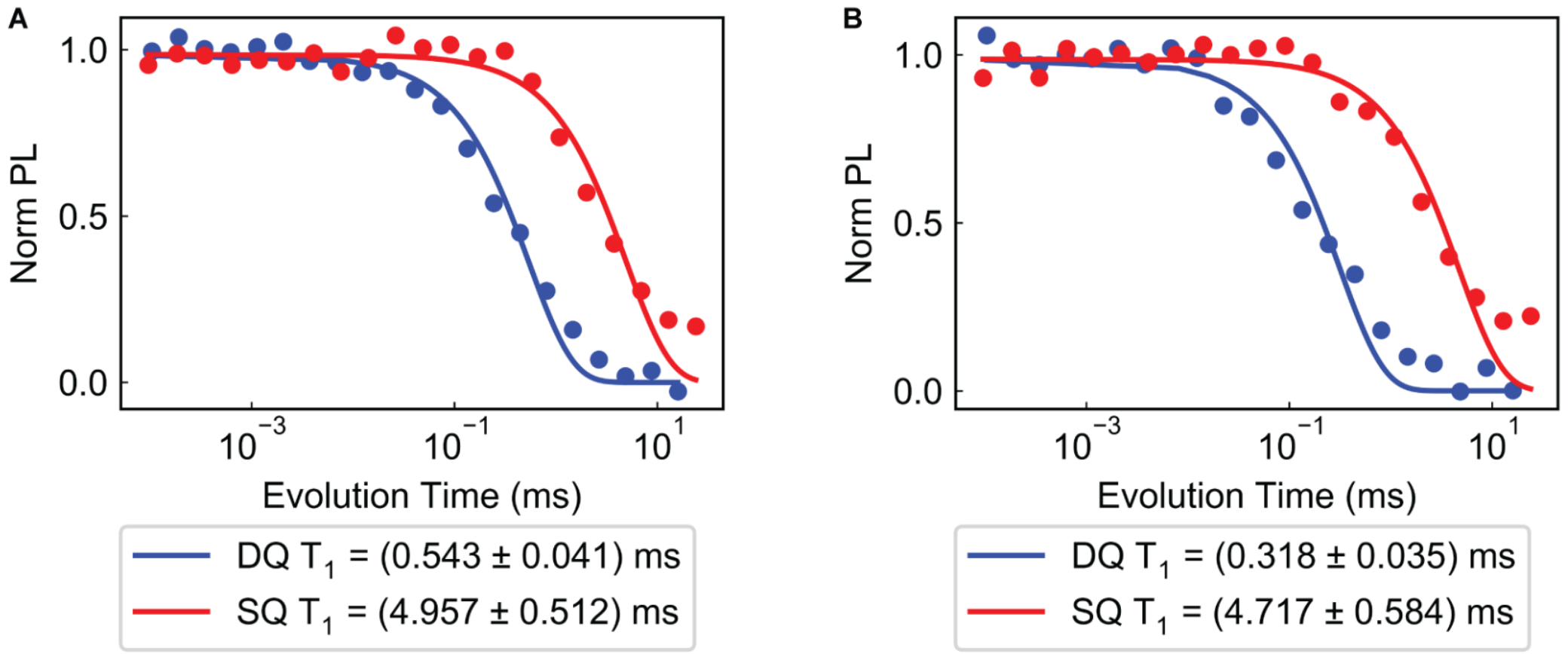

**Figure S37.** SQ and DQ lifetimes for Sample 3 after oxygen annealing (**A**) and after photochemical fluorination (**B**).

***Additional NV-NMR data***

NV-NMR experiments were performed with the sample mounted in deuterated propylene carbonate (Cambridge Isotopes DLM-1279-1). This was chosen because it has a similar index of refraction to oil, led to favorable NV charge state properties (Figure S35), did not cause fluorescence features that overlapped with the NV spectrum of interest, and is commercially available in deuterated form. Furthermore, the deuterated solvent helps to reduce the background proton signal. Since proton and fluorine nuclear spins have similar gyromagnetic ratios, their signals appear near each other in an NMR spectrum. In this experiment, we were trying to detect a small fluorine signal from our functionalized layer, so it was helpful to minimize the background proton signal.

Data were collected by scanning the free precession time ($\boldsymbol{\tau}_{\mathrm{p}}$) from 256 ns to 32.256 μs in 800 steps. Example time domain data are shown in Figure S38. For the data shown in the main text (Fig. 4E), the data were zero filled out to 96.256 μs prior to taking the fast Fourier transform (FFT). The time in between π pulses within XY-8 blocks for the data shown in main text Figure 4E was 296 ns.

The signal-to-noise ratio (SNR) is calculated by finding the amplitude of the datapoint nearest to the expected fluorine frequency, subtracting the baseline, and dividing by the standard deviation of the noise floor. The NMR dataset shown in the main text (Fig. 4E) had an SNR of 13 before zero filling and 9 after zero filling.

To verify that the candidate signal does indeed arise from the nuclear spin of interest, we swept the magnetic field and observed a shift in $^{19}$F signal frequency. Figure S39 shows NMR spectra from three different fields. The lines for $^{19}$F and $^{1}$H show the expected frequencies for these nuclear spin species at the three fields. Each magnetic field contains data from 3 to 4 spots averaged together.

We studied the homogeneity of the signal across the sample surface by interrogating 8 spots while sweeping the magnetic field (Figure S40). We observed a signal at the fluorine frequency with an SNR of at least 2 for 8 out of 8 spots. Each spot was averaged for 1 hour.

Finally, we tested to see if our candidate fluorine signal could arise from contamination with sample handling, mounting, or measurement. We used Sample 4 which was implanted and prepared identically to Sample 3 but was not subjected to the functionalization procedures. After performing an identical measurement, we did not observe a peak at the fluorine frequency, demonstrating that the $^{19}$F signal detected with Sample 3 originated from the functionalized layer at the surface (Figure S41).

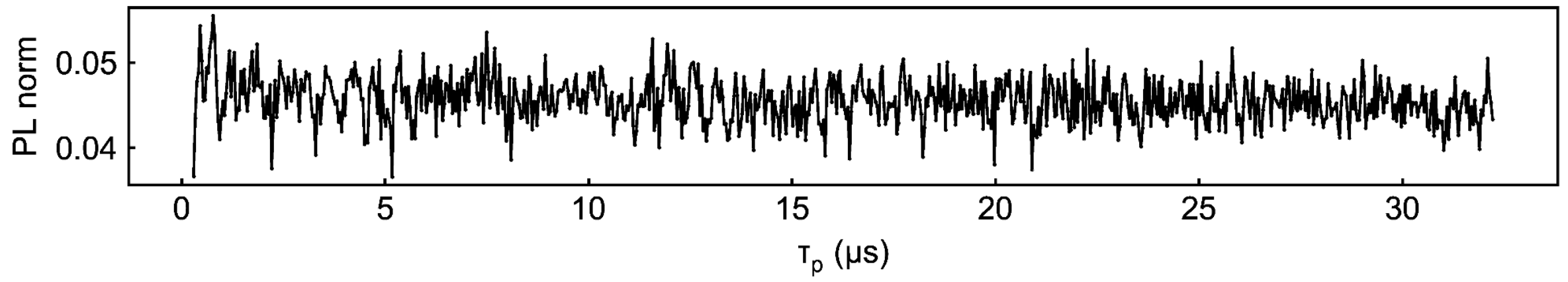

**Figure S38.** Time domain trace of NV-NMR experiment. The spectrum in Figure 4E is the FFT of this dataset after zero-filling.

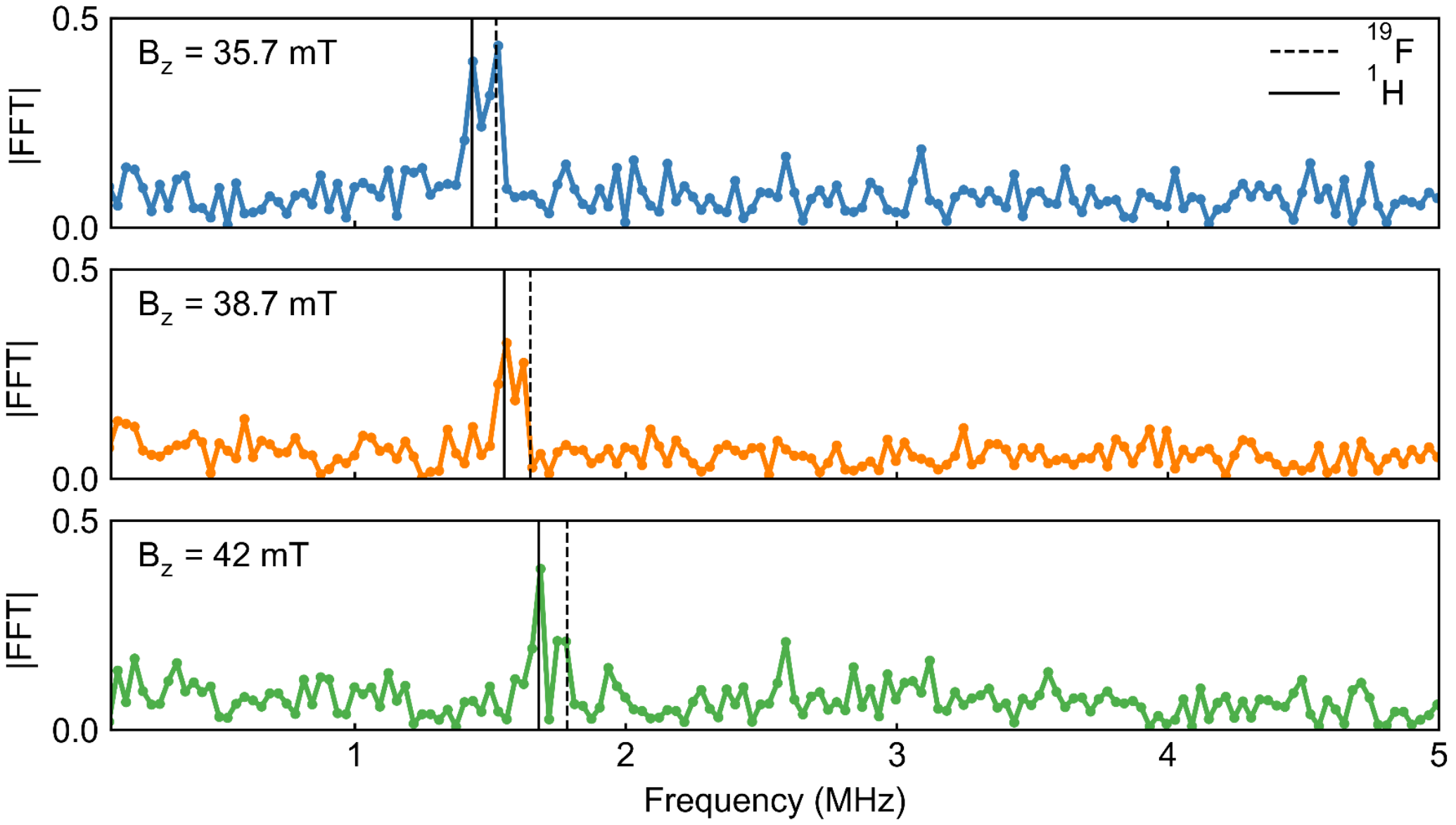

**Figure S39.** Magnetic field dependence of the $^{19}F$ signal. Each magnetic field dataset includes 3 to 4 spots averaged together.

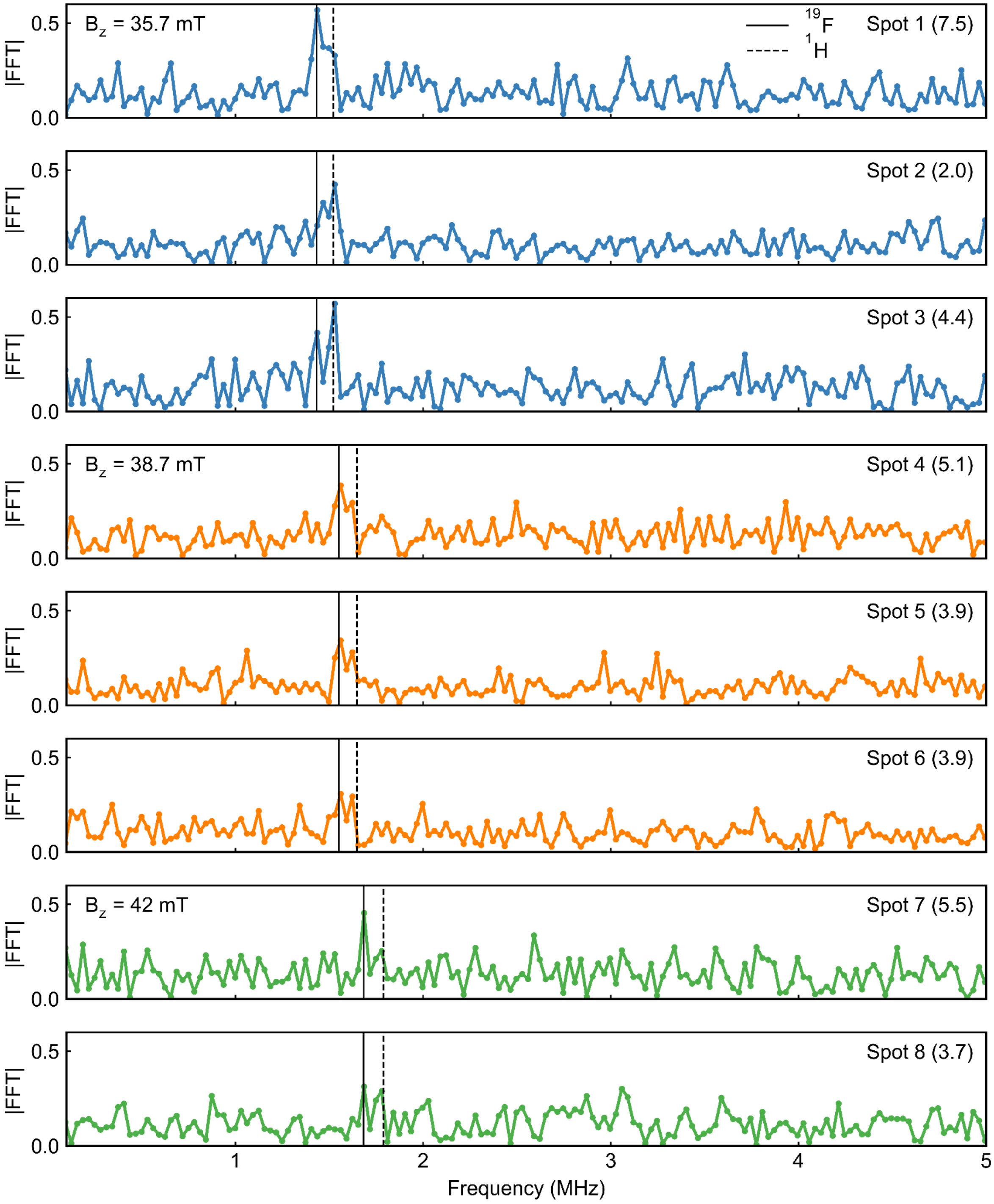

**Figure S40.** Spatial homogeneity studies of $^{19}F$ NMR signal. Each trace is from a different spot on the diamond. Different colors indicate different magnetic fields. Dashed lines show the expected frequency for $^{19}F$ and $^{1}H$ for each magnetic field. The SNR of the $^{19}F$ peak for each spectrum is given in parenthesis following the spot number. Each spot was measured for approximately 1 hour.

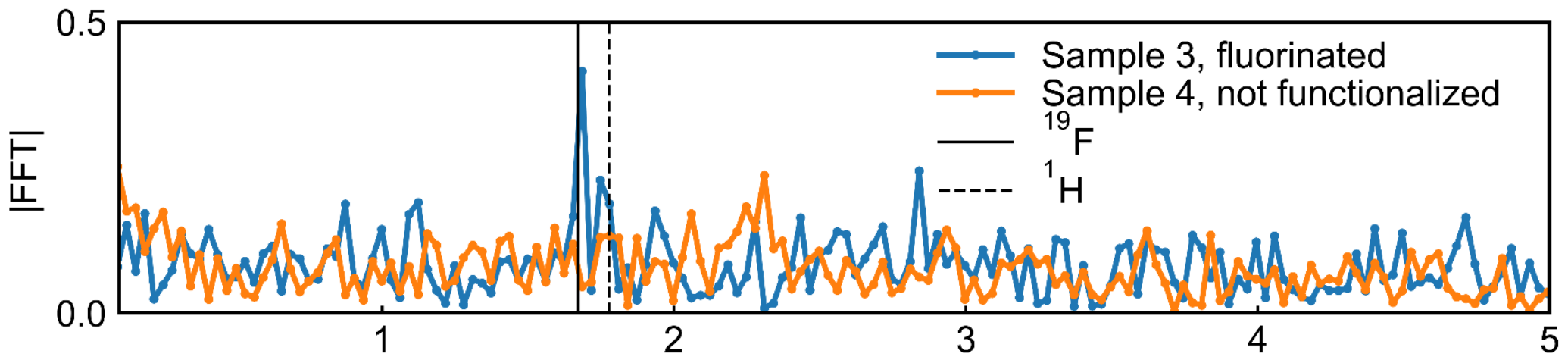

**Figure S41.** Contamination test for spurious $^{19}$F signals. A non-functionalized sample (Sample 4) was measured in the NV-NMR setup following identical sample preparation (except for functionalization), handling, and mounting procedures. No $^{19}$F signal was detected. The $^{1}$H signal arises from a combination of protons covalently bound to the surface and adventitious hydrocarbons and is expected to vary across samples.

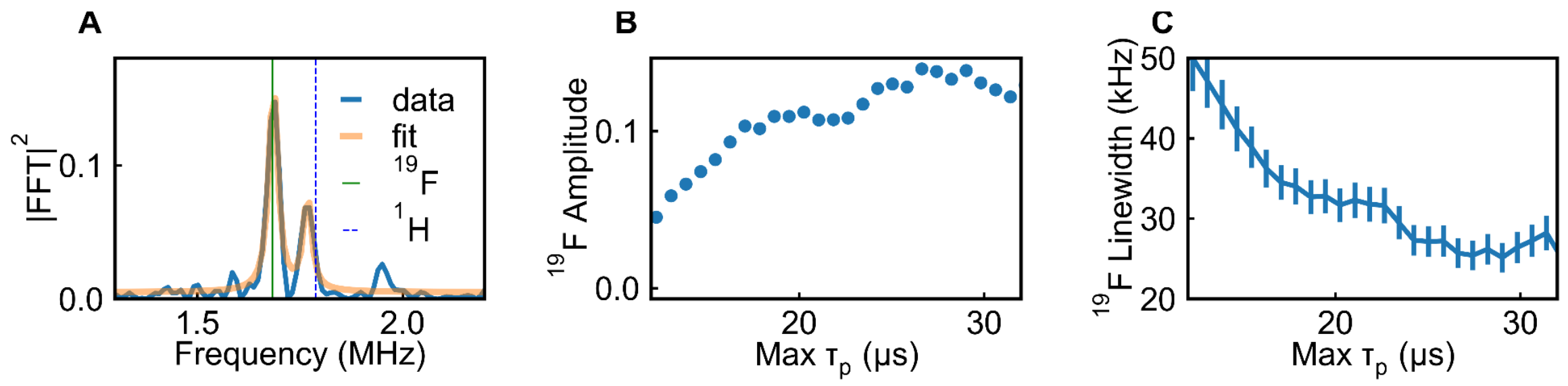

**Figure S42.** $^{19}$F linewidth analysis of NV-NMR power spectrum. (**A**) Power spectrum of data shown in the main text Figure 4E overlaid with a fit to two Lorentzians corresponding to $^{19}$F and $^{1}$H signals. The extracted linewidth for the $^{19}$F peak is 25.44 kHz ± 2.02 kHz. (**B**) $^{19}$F amplitude evolution as a function of free precession time (max $\tau_p$). (**C**) $^{19}$F linewidth evolution as a function of free precession time (max $\tau_p$).

As described in the main text, we fit the power spectrum of the NV-NMR data after zero filling and extract a linewidth of 25.44 kHz ± 2.02 kHz for the peak at the $^{19}$F frequency (Figure S42A). To investigate whether this linewidth is limited by the correlation pulse sequence or the apparent dephasing of the nuclear spins, we plot the evolution of the amplitude and linewidth of the $^{19}$F signal during the correlation sequence by zero filling after increasing maximum free precession time ($\tau_p$) values (Figure S42B, C). The amplitude of the $^{19}$F signal of the resulting spectra was found by taking the amplitude of the datapoint nearest to the $^{19}$F frequency and subtracting the average value of the baseline. Next, the power spectrum was fitted to two Lorentzian peaks to extract a linewidth. This analysis shows that the amplitude of the $^{19}$F signal saturates after a maximum $\tau_p$ of roughly 25 μs (Figure S42B). Similarly, the linewidth decreases with increasing $\tau_p$ and then saturates after roughly 25 μs of free evolution time (Figure S39C). Possible contributions to this apparent nuclear spin dephasing time include dipolar interactions with nuclear spins or electron spins at the diamond surface and changes in the applied magnetic field ($B_z$) during long experiment averaging. As a reference, the dipolar coupling between two nearest neighbor fluorine spins for a fully fluorinated (100) diamond surface is calculated to be 6.8 kHz[25]. We also note that this linewidth of approximately 25 kHz is consistent with previous work using ensembles of shallow NV centers to detect the nuclear spin signal from protons in polymer samples[26] and surface-bound fluorine groups[27]. Previous work[27] investigating the NV-NMR signal of fluorine groups contained in chains covalently bound to a functionalized alumina surface showed narrower linewidths for fluorine groups attached to longer chains, plausibly due to increased rotational freedom. An interesting direction for future work is to investigate if the same trend holds true for functional groups bound directly to the diamond surface and to quantitatively study the scaling with molecular length. Other potential strategies to further decrease the NV-NMR linewidth for future experiments with covalently bound groups include homonuclear and heteronuclear decoupling sequences[26].

## References

1. Sangtawesin, S. *et al.* Origins of Diamond Surface Noise Probed by Correlating Single-Spin Measurements with Surface Spectroscopy. *Phys. Rev. X* **9**, 031052 (2019).
2. Stacey, A. *et al.* Evidence for Primal sp $^2$ Defects at the Diamond Surface: Candidates for Electron Trapping and Noise Sources. *Adv. Mater. Interfaces* **6**, 1801449 (2019).
3. Seshan, V. *et al.* Hydrogen termination of CVD diamond films by high-temperature annealing at atmospheric pressure. *J. Chem. Phys.* **138**, 234707 (2013).
4. Zhang, Z.-H. *et al.* Neutral Silicon Vacancy Centers in Undoped Diamond via Surface Control. *Phys. Rev. Lett.* **130**, 166902 (2023).
5. Ristein, J. Diamond surfaces: familiar and amazing. *Appl. Phys. A* **82**, 377–384 (2006).
6. Stacey, A. *et al.* Depletion of nitrogen-vacancy color centers in diamond via hydrogen passivation. *Appl. Phys. Lett.* **100**, 071902 (2012).
7. Czaplyski, W. L., Na, C. G. & Alexanian, E. J. C–H Xanthylation: A Synthetic Platform for Alkane Functionalization. *J. Am. Chem. Soc.* **138**, 13854–13857 (2016).
8. Quinn, R. K. *et al.* Site-Selective Aliphatic C–H Chlorination Using *N* -Chloroamides Enables a Synthesis of Chlorolissoclimide. *J. Am. Chem. Soc.* **138**, 696–702 (2016).
9. Tzirakis, M. D., Lykakis, I. N. & Orfanopoulos, M. Decatungstate as an efficient photocatalyst in organic chemistry. *Chem. Soc. Rev.* **38**, 2609 (2009).
10. Seki, K. *et al.* High-Energy Spectroscopic Studies of the Electronic Structures of Organic Systems Formed from Carbon and Fluorine by UPS, Vacuum-UV Optical Spectroscopy, and NEXAFS: Poly(hexafluoro-1,3-butadiene) $[C(CF_3) = C(CF_3)]_n$ , Fluorinated Graphites (CF, $C_2F$, and $C_6F$), Perfluoroalkanes $n$-$C_nF_{2n+2}$ , Poly(tetrafluoroethylene) $(CF_2)_n$ , and Fluorinated Fullerenes ($C_{60}F_x$ and $C_{70}F_x$ ). *Mol. Cryst. Liq. Cryst. Sci. Technol. Sect.*

*Mol. Cryst. Liq. Cryst.* **355**, 247–274 (2001).

11. Zhang, X., Guo, S. & Tang, P. Transition-metal free oxidative aliphatic C–H fluorination. *Org. Chem. Front.* **2**, 806–810 (2015).
12. Tierney, M. M., Crespi, S., Ravelli, D. & Alexanian, E. J. Identifying Amidyl Radicals for Intermolecular C–H Functionalizations. *J. Org. Chem.* **84**, 12983–12991 (2019).
13. Chambers, R. D., Parsons, M. & Sandford, G. SELECTIVE NITROGEN FUNCTIONALISATION OF ORGANIC COMPOUNDS. (2002).
14. Latham, K. G., Dose, W. M., Allen, J. A. & Donne, S. W. Nitrogen doped heat treated and activated hydrothermal carbon: NEXAFS examination of the carbon surface at different temperatures. *Carbon* **128**, 179–190 (2018).
15. Zubavichus, Y., Shaporenko, A., Grunze, M. & Zharnikov, M. Innershell Absorption Spectroscopy of Amino Acids at All Relevant Absorption Edges. *J. Phys. Chem. A* **109**, 6998–7000 (2005).
16. Kresse, G. & Furthmüller, J. Efficiency of ab-initio total energy calculations for metals and semiconductors using a plane-wave basis set. *Comput. Mater. Sci.* **6**, 15–50 (1996).
17. Heyd, J., Scuseria, G. E. & Ernzerhof, M. Hybrid functionals based on a screened Coulomb potential. *J. Chem. Phys.* **118**, 8207–8215 (2003).
18. Kaviani, M. *et al.* Proper Surface Termination for Luminescent Near-Surface NV Centers in Diamond. *Nano Lett.* **14**, 4772–4777 (2014).
19. Maier, F., Ristein, J. & Ley, L. Electron affinity of plasma-hydrogenated and chemically oxidized diamond (100) surfaces. *Phys. Rev. B* **64**, 165411 (2001).
20. Rietwyk, K. J. *et al.* Work function and electron affinity of the fluorine-terminated (100) diamond surface. *Appl. Phys. Lett.* **102**, 091604 (2013).

21. Hauf, M. V. *et al.* Chemical control of the charge state of nitrogen-vacancy centers in diamond. *Phys. Rev. B* **83**, 081304 (2011).

22. Dwyer, B. L. *et al.* Probing Spin Dynamics on Diamond Surfaces Using a Single Quantum Sensor. *PRX Quantum* **3**, 040328 (2022).

23. Pham, L. M. *et al.* NMR technique for determining the depth of shallow nitrogen-vacancy centers in diamond. *Phys. Rev. B* **93**, 045425 (2016).

24. Bucher, D. B. *et al.* Quantum diamond spectrometer for nanoscale NMR and ESR spectroscopy. *Nat. Protoc.* **14**, 2707–2747 (2019).

25. Cai, J., Retzker, A., Jelezko, F. & Plenio, M. B. A large-scale quantum simulator on a diamond surface at room temperature. *Nat. Phys.* **9**, 168–173 (2013).

26. Aslam, N. *et al.* Nanoscale nuclear magnetic resonance with chemical resolution. *Science* **357**, 67–71 (2017).

27. Liu, K. S. *et al.* Surface NMR using quantum sensors in diamond. *Proc. Natl. Acad. Sci.* **119**, e2111607119 (2022).